\documentclass[reprint,amsmath,amssymb, aps,pre,floatfix,superscriptaddress,longbibliography]{revtex4-1}

\usepackage{graphicx} 
\usepackage[separate-uncertainty=true]{siunitx}
\usepackage[version=4]{mhchem}
\usepackage{comment}
\usepackage[x11names]{xcolor}
\usepackage{tikz}
\usepackage{scalerel}
\usepackage{xifthen}
\usepackage{chngcntr}
\usetikzlibrary{svg.path}
\usepackage[colorlinks,allcolors=blue]{hyperref}

\DeclareSIUnit\molar{M}
\DeclareSIUnit{\wtpc}{wt\%}

\newcommand\Pen{\mbox{\textit{Pe}}} 
\newcommand\Ray{\mbox{\textit{Ra}}} 

\AtBeginDocument{%
    \newwrite\bibnotes
    \def\bibnotesext{Notes.bib}
    \immediate\openout\bibnotes=\jobname\bibnotesext
    \immediate\write\bibnotes{@CONTROL{REVTEX41Control}}
    \immediate\write\bibnotes{@CONTROL{%
    apsrev41Control,author="08",editor="1",pages="1",title="0",year="1"}}
     \if@filesw
     \immediate\write\@auxout{\string\citation{apsrev41Control}}%
    \fi
}%

\newcommand{\figref}[2][{}]{Fig.\ \ref{#2}\ifthenelse{\isempty{#1}}{}{\,(#1)}}
\newcommand{\SIfigref}[2][{}]{Supplement Fig.\ \ref{#2}\ifthenelse{\isempty{#1}}{}{\,(#1)}}

\renewcommand{\vec}{\mathbf}

\definecolor{orcidlogocol}{HTML}{A6CE39}
\tikzset{
  orcidlogo/.pic={
    \fill[orcidlogocol] svg{M256,128c0,70.7-57.3,128-128,128C57.3,256,0,198.7,0,128C0,57.3,57.3,0,128,0C198.7,0,256,57.3,256,128z};
    \fill[white] svg{M86.3,186.2H70.9V79.1h15.4v48.4V186.2z}
                 svg{M108.9,79.1h41.6c39.6,0,57,28.3,57,53.6c0,27.5-21.5,53.6-56.8,53.6h-41.8V79.1z M124.3,172.4h24.5c34.9,0,42.9-26.5,42.9-39.7c0-21.5-13.7-39.7-43.7-39.7h-23.7V172.4z}
                 svg{M88.7,56.8c0,5.5-4.5,10.1-10.1,10.1c-5.6,0-10.1-4.6-10.1-10.1c0-5.6,4.5-10.1,10.1-10.1C84.2,46.7,88.7,51.3,88.7,56.8z};
  }
}

\newcommand\orcid[1]{\href{https://orcid.org/#1}{\mbox{\scalerel*{
\begin{tikzpicture}[yscale=-1,transform shape]
\pic{orcidlogo};
\end{tikzpicture}
}{|}}}}

\makeatletter
\def\maketitle{
\@author@finish
\title@column\titleblock@produce
\suppressfloats[t]}
\makeatother

\newcommand{\beginsupplement
}{%
    \setcounter{table}{0}
    \renewcommand{\thetable}{S\arabic{table}}%
    \setcounter{figure}{0}
    \renewcommand{\thefigure}{S\arabic{figure}}%
    \setcounter{equation}{0}
    \renewcommand{\theequation}{S\arabic{equation}}%
    \setcounter{section}{0}
    \setcounter{subsection}{0}
    \renewcommand{\thesection}{\arabic{section}}
        \renewcommand{\thepage}{S\arabic{page}}%
    \setcounter{page}{1}
    \maketitle
     }

\begin{document}

\setlength{\unitlength}{1cm}

\newcommand{\goeaffila}{Max Planck Institute for Dynamics and Self-Organization, Am Fa\ss{}berg 17, 37077 G\"ottingen, Germany}
\newcommand{\goeaffilb}{Institute for the Dynamics of Complex Systems, Georg August Universit\"at G\"ottingen, Germany}
\newcommand{\twaffil}{Physics of Fluids Group, Max Planck Center for Complex Fluid Dynamics and J. M. Burgers Center for Fluid Dynamics, University of Twente, PO Box 217,7500AE Enschede, Netherlands}
\newcommand{\amsaffil}{University of Amsterdam, Science Park 904, 1098 XH Amsterdam, Netherlands}
\newcommand{\mzaffil}{Max Planck Institute for Polymer Research, Ackermannweg 10, 55128 Mainz, Germany}
\newcommand{\praffil}{Department of Chemical and Biological Engineering, Princeton University, 41 Olden Street, Princeton NJ 08544, USA}

\title{Frozen by heating: temperature controlled dynamic states in droplet microswimmers}
\author{Prashanth Ramesh~\orcid{0000-0003-3264-5084 }}
\affiliation{\goeaffila}
\affiliation{\twaffil}
\author{Yibo Chen~\orcid{0000-0001-6786-707X}}
\affiliation{\twaffil}
\author{Petra R\"ader}
\author{Svenja Morsbach~\orcid{0000-0001-9662-8190}}
\affiliation{\mzaffil}
\author{Maziyar Jalaal~\orcid{0000-0002-5654-8505}}
\affiliation{\amsaffil}
 \author{Corinna C. Maass~\orcid{0000-0001-6287-4107}}
 \email{corinna.maass@ds.mpg.de}%
\affiliation{\goeaffila}
\affiliation{\twaffil}
\date{\today}%

 \begin{abstract}
Self-propelling active matter relies on the conversion of energy from the undirected, nanoscopic scale to directed, macroscopic motion. One of the challenges in the design of synthetic active matter lies in the control of dynamic states, or motility gaits. Here, we present an experimental system of self-propelling droplets with thermally controllable and reversible dynamic states, from unsteady over meandering to persistent to arrested motion. These states are known to depend on the Péclet number of the molecular process powering the motion, which we can now tune by using a temperature sensitive mixture of surfactants as propulsion fuel. We quantify the droplet dynamics  by analysing flow and chemical fields for the individual states, comparing them to canonical models for autophoretic particles.  In the context of these models, we experimentally observe, in situ, the fundamental first broken symmetry that translates an isotropic, immotile base state to self-propelled motility.

\end{abstract}

\maketitle
\section{Introduction}
Active matter is defined by the nonlinear conversion of free energy on the molecular scale into macroscopic dynamics~\cite{bowick2022_symmetry}: these nonlinear dynamics pose a challenge to the control of active agents, as small changes in system parameters can crucially tip the system over into a new dynamic equilibrium, for example from metastable inactivity to a state of directed motility.
More so, like a sorcerer's apprentice~\cite{goethe1798_zauberlehrling} we find that stopping this motion is not a trivial task; after all, one is required to deplete an energy reservoir or reverse a dynamic instability~\cite{obyrne2022_time}. 
This kind of control is important in the design and study of artificial and biological microswimmers, their theoretical modeling, experimental realization, and, ultimately, to provide design principles and dynamic control for technological application \cite{tu2017_motion,fusi2023_achieving}.
\par
Autophoretic particles \cite{bechinger2016_active,moran2019_microswimmers,zhang2021_chemicallypowered,zottl2023_modeling} and, particularly, their experimental counterpart, active droplets propelled by spontaneously self-generated chemophoretic flows \cite{maass2016_swimming,babu2022_motile,birrer2022_we,dwivedi2022_selfpropelled,michelin2023_selfpropulsion,hanczyc2007_fatty,thutupalli2011_swarming,peddireddy2012_solubilization,izri2014_selfpropulsion,herminghaus2014_interfacial}, 
are popular active matter models driven by purely physicochemical mechanisms, with typical sizes ranging between $\approx \SI{30}{\um}$ up to several hundred \SI{}{\um}. 
Generally, their dynamics are characterized by a dimensionless P\'eclet number $\Pen{}=U_cR/D$ representing the ratio of advective and diffusive transport of chemical fuel~\cite{michelin2013_spontaneous}, with $U_c$, $R$ and $D$ quantifying the characteristic flow speed, droplet radius and the fuel's diffusivity. With increasing \Pen{}, autophoretic particles first transition  from \textit{passive} isotropic chemical conversion to \textit{active} self-propulsion, and further from persistent to unsteady motion via a sequence of broken symmetries and interfacial flow modes of increasing complexity~\cite{michelin2013_spontaneous,izri2014_selfpropulsion,suga2018_selfpropelled,izzet2020_tunable,meredith2020_predatorprey,hokmabad2021_emergence,suda2021_straightcurvilinear}.
\par
Recent studies have investigated the control of speed and dynamic states of such microswimmers in response to externally applied stimuli such as temperature \cite{tu2017_selfpropelled,cholakova2021_rechargeable} or illumination \cite{florea2014_photochemopropulsion,kaneko2017_phototactic,xiao2018_moving,lancia2019_reorientation,alvarez2021_reconfigurable,ryabchun2022_runandhalt,benzion2022_cooperation,vankesteren2023_selfpropelling}.
On heating, physical intuition might suggest that motion should accelerate and destabilize, either by the increase of translational and rotational diffusion with decreasing viscosity, or by increased activity from the molecular thermodynamics driving the motion. 
However, the nonlinear dynamics of self-organized activity can drive counter-intuitive effects, as we have previously found for active droplets which destabilize with increasing viscosity due to an increase in \Pen{}~\cite{hokmabad2021_emergence}. 
\par
In this study, we explore another counter-intuitive response, this time to increasing temperature. We study active droplets using a temperature sensitive combination of co-surfactants in aqueous solution as a fuel medium. With  increasing ambient temperature we find a transition between distinct dynamic states from unsteady to oscillatory to steady, straight swimming to eventual arrest, which is reversible and cyclic with temperature. Notably, by this method we are able to drive \Pen{} across the critical activity threshold: we experimentally observe, live and in situ, the fundamental first symmetry breaking of the inactive isotropic base state into directed self-propulsion, analyzing both chemical and hydrodynamic fields.
 This observation requires an undisturbed, unchanged ambient fuel medium, which we engineer by exploiting temperature dependent surfactant-polymer interactions~\cite{tam2006_insights} that were previously not considered in active droplets - opening new design possibilities to control the motion of micro-droplets as active agents in smart materials driven by purely physicochemical mechanisms~\cite{zhang2021_autonomous}.

\section{Results and discussion}
\par
\subsection{Self-propelling droplets in co-surfactant solutions}
\begin{figure*}
	\includegraphics[width=\linewidth]{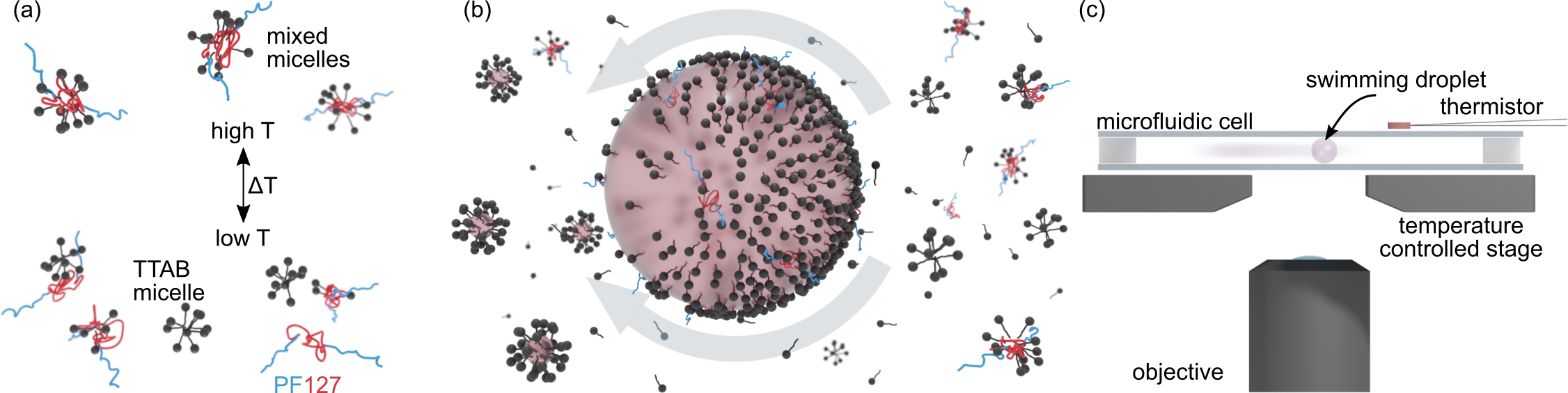}
	\caption{\textbf{Schematics of the experimental setup and droplet propulsion mechanism.} (a) With increasing/decreasing temperature, there is an increased/decreased affinity for TTAB to form mixed micelles with PF127. Simplified aggregation schematic \cite{nambam2012_effects}. (b) Droplet propulsion during solubilization: an inhomogeneous distribution of empty TTAB micelles causes a self-sustaining Marangoni gradient at the oil-water interface. (c) Setup: The droplet (diameter $d=\SI{50\pm 5}{\um}$) swims in a quasi-2D ($\SI{13}{\mm}\times\SI{8}{\mm}\times\SI{50}{\um}$) cell on a temperature controlled microscope stage. All schematics are not to scale (see SI Figure S2)\label{fig:setup}}
	
\end{figure*}
\par
Our experimental system consists of oil droplets (CB15) immersed in an aqueous solution of the ionic surfactant tetradecyltrimethylammonium bromide (TTAB) at 9-1\SI{5}{\wtpc} (267--\SI{445}{\milli\molar}) and the triblock copolymer Pluronic F127  (PF127) at \SI{4}{\wtpc}, or \SI{3}{\milli\molar}. 
\par
CB15 droplets self-propel reliably in supramicellar solutions of pure TTAB above \SI{5}{\wtpc}~\cite{jin2017_chemotaxis,hokmabad2021_emergence}. Briefly put, the swimming is due to oil diffusing from the droplet into TTAB micelles~\cite{peddireddy2012_solubilization,herminghaus2014_interfacial,izzet2020_tunable}, which removes surfactant from the droplet posterior, while the anterior is replenished by the advection of fresh surfactant (Fig.~\ref{fig:setup}b). The key point here is that the empty micelles at the anterior are less thermodynamically stable than the oil-filled ones at the posterior~\cite{izzet2020_tunable,rosen2012_surfactants}: in consequence the critical micelle concentration is higher in front of a moving droplet. The resulting self-enhancing surface tension gradient drives the droplet forward until it is dissolved. Typically, a CB15 droplet of diameter \SI{50}{\um} will swim in \SI{5}{\wtpc} TTAB for 1-2 hours.   Based on this mechanism, one can define a P\'eclet number \Pen\ of droplet activity that increases with viscosity, droplet radius  and surfactant concentration, i.e. chemical activity~\cite{izri2014_selfpropulsion,izzet2020_tunable,hokmabad2021_emergence,hokmabad2022_spontaneously,suda2021_straightcurvilinear}.  In a surfactant solution where all micelles are oil-saturated, droplets would not propel, and they are are repelled by gradients in filled and attracted by gradients in empty micelles~\cite{jin2017_chemotaxis}. We may in this sense regard empty TTAB micelles as fresh, and oil-filled ones as spent fuel.
\par
PF127 is a nonionic triblock copolymer surfactant that in a pure aqueous solution forms micelles with a hydrophobic core of propylene oxide (PPO) and an outer shell of hydrated ethylene oxide (PEO) \cite{alexandridis1994_micellization,wanka1994_phase,bohorquez1999_study,stoeber2006_passive,jalaal2016_rheology,jalaal2018_gelcontrolled} above the critical micelle temperature,  $\text{CMT}\approx\SI{21}{\celsius}$ at $\SI{4}{\wtpc}$ PF127~\cite{bohorquez1999_study}. 
In the presence of ionic co-surfactants like TTAB, which bind strongly with PF127, mixed TTAB/PF127 aggregates form (\figref{fig:setup}a, Supplement \ref{SIsec:mixedmicelles}, Figs.\ \ref{SIfig:DLS}, \ref{SIfig:Zaverage} and \ref{SIfig:DSC}), with excess TTAB forming single-species micelles \cite{nambam2012_effects,hecht1994_interaction,li2001_binding,tam2006_insights}. 
It has been found \cite{li2001_binding} that the TTAB binding capacity of PF127 is amplified with increasing temperature, due to an increasing dehydration of the PPO blocks (hydrophobic effect~\cite{butt2003_physics,alexandridis1994_micellization}).

Under our experimental conditions, we expect a significant coverage of TTAB at the interface (Supplement \figref{SIfig:anchoring}).  Furthermore, CB15 droplets are completely immotile and hardly solubilize in pure PF127 solutions (Supplement \figref{SIfig:dissolution}), while in a mixed TTAB/PF127 medium self-propelling at speeds $\approx\SI{20}{\um/\second}$, comparable to experiments in pure TTAB solutions (cf. \cite{jin2017_chemotaxis,hokmabad2021_emergence} and Supplement \figref{SIfig:ttab}). We therefore regard TTAB as the primary surfactant mediating the solubilization and the interfacial gradients driving the droplet motion. This activity is controlled by PF127 binding and releasing TTAB micelles in the bulk medium (Fig.~\ref{fig:setup}a,b).
Thus, the \Pen{} of droplet activity \textit{decreases} with \textit{increasing} temperature, as an increasing fraction of TTAB is bound in mixed micelles.

According to literature on the composition of PF127/TTAB aggregates \cite{hecht1994_interaction,li2001_binding,hecht1995_interaction}, the amount of bound TTAB in our swimming medium should exceed $\SI{1}{\wtpc}$ and further increase with temperature, which is on the order of the amounts required ($<\SI{5}{\wtpc}$) to suppress droplet motility~\cite{herminghaus2014_interfacial,hokmabad2021_emergence}, as observed in the experiments we show below (see also Supplement \figref{SIfig:swimming}). 
We further note that the swimming medium is Newtonian, with only weakly temperature dependent viscosity (Supplement \figref{SIfig:viscosity}). The Stokes-Einstein diffusion coefficient for a particle of radius $\SI{25}{\um}$ is $D<\SI{1e-2}{\um^2/s}$, such that we can neglect Brownian motion within the duration of observation. 
%
\subsection{Swimming dynamics controlled by temperature and fuel concentration}
\begin{figure*}
	\centering
	\includegraphics[width=1\linewidth]{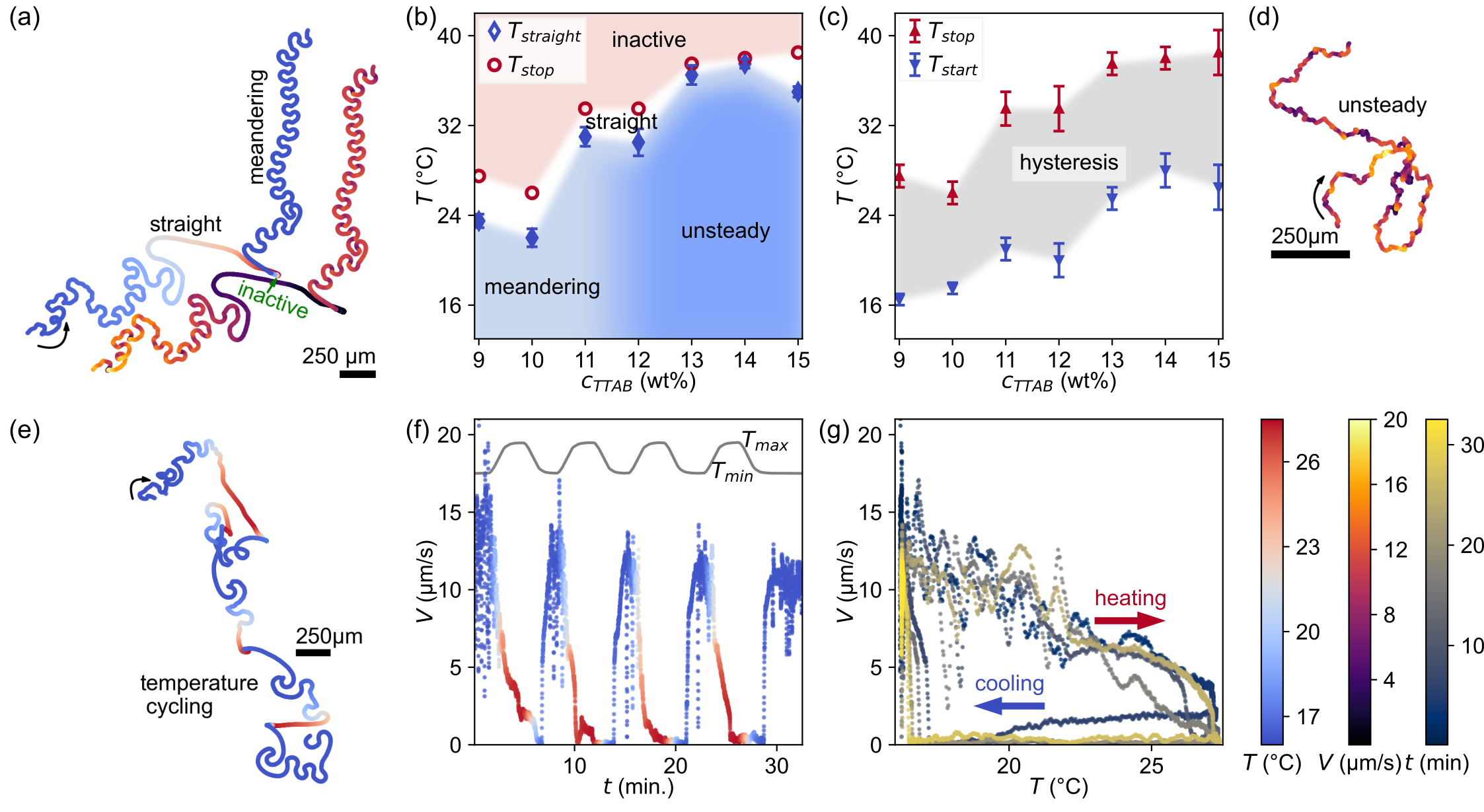}
	\caption{\textbf{Swimming dynamics controlled reversibly via temperature and fuel concentration} 
		(a) One trajectory of a droplet in a mixed surfactant swimming medium, color coded once by droplet speed and once by recorded temperature. The droplet transitions from meandering to straight swimming to arrest during a heating and subsequent cooling ramp with a set rate of \SI{1}{\kelvin/\min}. See also Movie S1. Color map and scale bars (\SI{250}{\um}) apply to all figures in the paper, and concentrations for PF127 and TTAB are always \SI{4}{\wtpc} and \SI{10}{\wtpc}, respectively, unless stated otherwise.
		(b) Map of the swimming dynamics depending on temperature and surfactant concentration. (c) Hysteresis between droplet stop and start transition temperatures. Error bars also apply to $T_\text{stop}$ in (b); experiments done in triplicate with 5--10 droplets each. (d) Example of unsteady motion at 1\SI{5}{\wtpc} TTAB and \SI{21}{\celsius} (see also Movie S2).\\
		(e) A droplet trajectory during multiple heating/cooling cycles set at \SI{10}{\kelvin/\min} and \SI{10}{\wtpc} TTAB. See also Movie S3.  
		(f) Droplet speed vs. time for (e), with an inset plot of the recorded temperature ramps (g) Droplet speed vs. recorded temperature for (e), showing a hysteresis in the re-onset of motion during cooling: the arrest during cooling is extended, with a sudden recovery of the initial speed at $T \approx \SI{17}{\celsius}$.\label{fig:traj}}	
\end{figure*}
We begin with an overview of the general swimming dynamics taken from wide-field video microscopy under changing the ambient temperature, and for a range of TTAB concentrations.
The setup contains a quasi-2D microfluidic cell on a temperature controlled stage (\figref{fig:setup}c). 
\figref{fig:traj}a (Movie S1) plots a trajectory color-coded once by speed and once by temperature, recorded at a set heating/cooling rate of \SI{1}{\kelvin/\min}, using a swimming medium containing \SI{10}{\wtpc} TTAB.
\par
We start at $T \approx \SI{15}{\celsius}< \text{CMT}$. 
Below $T_\text{straight} = \SI{22}{\celsius}$, the droplet meanders, i.e. periodically reorients. Above, the motion is straight, gradually slows down and eventually stops at $T_\text{stop} \approx \SI{27}{\celsius}$. 
During a subsequent cooling ramp, the droplet remains immotile down to a significantly lower temperature $T_\text{start} \approx \SI{17}{\celsius}$, where it abruptly starts to meander again. 
\par
As well as to temperature, the swimming dynamics are susceptible to the TTAB concentration. Previous studies on reference systems using a single surfactant species as fuel found a transition from straight to reorienting to unsteady swimming~\cite{izzet2020_tunable} with increasing fuel concentration: with activity, the flow speed $U_c$ increases, and therefore \Pen{}. Alternatively, similar transitions have been found in experiment if \Pen{} was raised via the droplet size or the viscosity of the swimming medium~\cite{suda2021_straightcurvilinear,hokmabad2021_emergence,hokmabad2022_spontaneously}.  
These dynamics fit into the canonical model for isotropic autophoretic particles~\cite{michelin2023_selfpropulsion,michelin2013_spontaneous,izri2014_selfpropulsion,morozov2019_nonlinear,morozov2020_adsorption,suda2021_straightcurvilinear}, which describes fuel consumption at a spherical interface via a hydrodynamic advection-diffusion model non-dimensionalized by \Pen{}. The solution ansatz, in the weakly non-linear limit, decomposes the flow and chemical fields into a series of higher order modes in the interfacial flow and chemical fields, with associated critical \Pen{} and growth rates. With increasing \Pen{}, starting from the isotropic base, $n=0$, higher modes like dipolar, $n=1$, and quadrupolar, $n=2$ are consecutively excited. At very high \Pen{}, numerical studies predict true chaos due to the nonlinearity of the underlying problem~\cite{li2022_swimming,hu2022_spontaneous}.
\par 
We have previously investigated the relation between these interfacial modes and straight, meandering and unsteady swimming via dual channel microscopy~\cite{hokmabad2021_emergence}: here, \textit{straight swimming} at barely supercritical $\Pen\sim 4$ corresponds to a pure dipolar interfacial flow structure, \textit{meandering} at intermediate $\Pen\sim 30$ is primarily dipolar with intermittent excitation of a quadrupolar mode that leads to smooth reorientation, while at high $\Pen\gtrsim 300$ the extensile quadrupolar mode dominates, with only short propulsive dipolar intervals, causing a characteristic \textit{unsteady} `stop-and-go' motion.

\par
Adding the PF127 co-surfactant does not appear to change these dynamics at constant low temperature qualitatively, since we find a similar transition from meandering to unsteady motion (Fig.~\ref{fig:traj}, Movie S2) with increasing TTAB concentration.  We have summarized these swimming dynamics  in a map spanned by temperature and surfactant concentration in \figref{fig:traj}b.
With increasing temperature, we observe a universal transition via straight swimming to eventual arrest. We posit that the temperature dependent TTAB depletion lowers \Pen{} below the critical thresholds of higher order interfacial modes, down to $n=1$ for straight swimming and finally $n=0$, below the fundamental advection-diffusion instability. We do not provide a quantitative estimate of \Pen{} following~\cite{hokmabad2021_emergence}, as we cannot quantify the temperature dependence of the underlying physical chemistry parameters.
Fig.~\ref{fig:traj}b also shows a general trend of the transition temperatures to increase with TTAB concentration:  
for the droplet to arrest, more TTAB needs to be removed from the swimming medium. We found the stop/start hysteresis noted above for all TTAB concentrations in use (Fig.~\ref{fig:traj}c).
\par
These state transitions are well reversible with temperature. The experiment shown in \figref{fig:traj}e-g and Movie S3 was recorded at a faster rate of \SI{10}{\kelvin/\min} to permit multiple heating and cooling cycles, with dynamics similar to the system cooled at slower rates. \figref{fig:traj}e and f show the droplet trajectory color coded by temperature and a corresponding plot of speed over time. The initial motion is recovered after each heating and cooling cycle, apart from a very gradual decrease in maximum speed which we may attribute to droplet shrinkage. The residual motion in the lab frame at peak heating, specifically during the first cycle, appears to be drift in the swimming medium (SI Fig.~S13).  We have further analyzed speed versus temperature in \figref{fig:traj}g, and found a  hysteresis cycle with a delayed re-onset of motion reproducible over multiple heating/cooling ramps. The sample in Movie S3 contained three larger droplets ($d\approx\SI{60}{\um}$): \Pen\ should increase with the droplet radius, and, correspondingly, these droplets switch from reorienting to straight motion at later times and therefore higher temperatures.
\subsection{Chemical and flow fields}
\begin{figure*}
	\centering
	\includegraphics[width=1\linewidth]{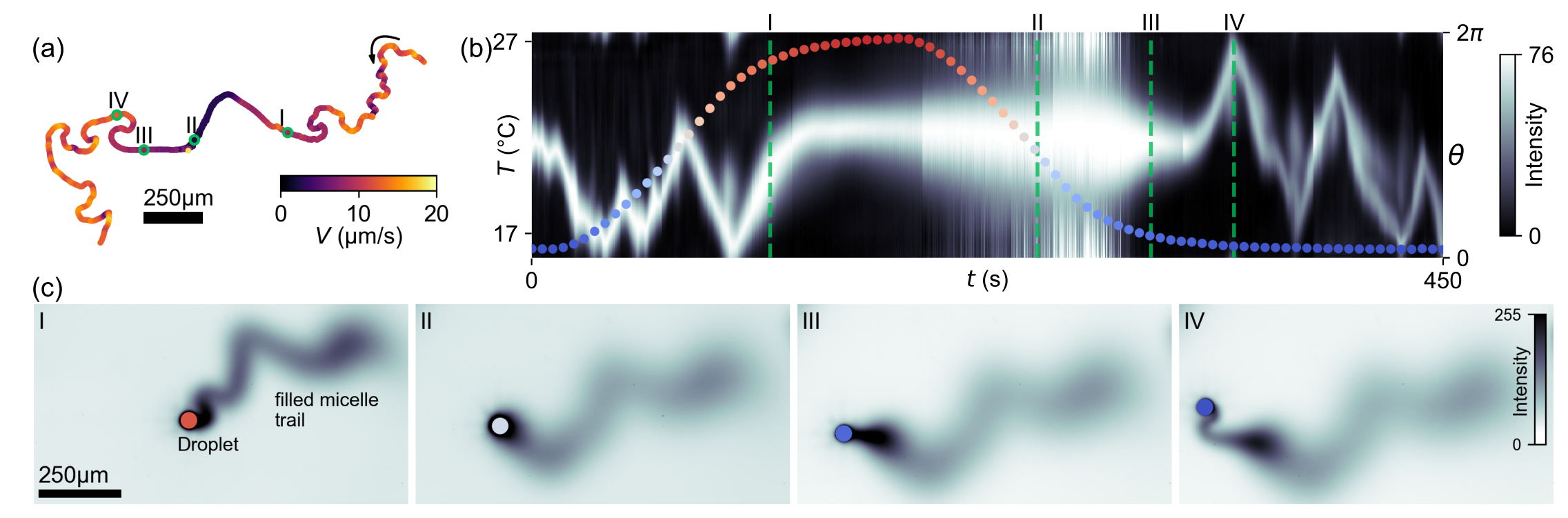}
	\caption{\textbf{The hysteresis in the re-onset of droplet motion is caused by spent fuel aggregation.} (a) Droplet trajectory during heating and subsequent cooling ramp color coded by speed. (b) Kymograph showing the evolution of chemical concentration field around the droplet interface and the recorded temperature (colored symbols). (c) Snapshots of droplet chemical trails at different temperatures as marked by I-IV in (a) and (b).
		See also Movie S4. \SI{250}{\um} scale bars and color maps as defined in Fig.~\ref{fig:traj}.\label{fig:fluo}}
	
\end{figure*}
We continue with a discussion of the chemical dynamics during the droplet arrest, to motivate the hysteresis in the re-onset of motion; and of the corresponding flow field to investigate the interfacial mode evolution.
\par
The self-generated field of spent fuel in the local environment affects the droplet motility, both via chemorepulsive gradients ~\cite{izzet2020_tunable,hokmabad2021_emergence} and via accumulation of filled micelles, which suppresses the interfacial activity~\cite{ramesh2023_interfacial}. We visualize this field by doping the droplet with the fluorescent dye Nile Red~\cite{hokmabad2022_chemotactic}, which co-moves with the oil phase into the filled micelles, and extract the fluorescence intensity $I$ from videomicroscopy data (Movie S4). In Fig.~\ref{fig:fluo}, we analyze these chemical dynamics for one droplet during a heating and cooling cycle, showing a speed-coded trajectory (a), a kymograph of $I$ around the droplet perimeter, $\theta$ vs. time and recorded temperature (b), and micrographs at the times marked I--IV (c). 
\par
During heating, the droplet transitions from unsteady to straight swimming to immotility (Fig.~\ref{fig:fluo}a, and Movie S4. We note again slight drift in the swimming medium, see SI section 10).  In the kymograph  (Fig.~\ref{fig:fluo}b), at $T < \SI{26}{\celsius}$, the band corresponding to the chemical trail translates in the angular space due to the reorientation of the droplet (I). At $T \approx \SI{26}{\celsius}$, the droplet slows down and comes to a halt. As the system is cooled down to $T \approx \SI{21}{\celsius}$, the inactive droplet still solubilizes isotropically, and oil-filled micelles accumulate around the perimeter $\theta$. Correspondingly, the band in the kymograph widens over the entire angular space (II). We know from experiments in pure TTAB media that these accumulations locally suppress the interfacial activity~\cite{ramesh2023_interfacial,morozov2020_adsorption}, as the density of empty micelle `fuel' is reduced - it follows that in the presence of oil-filled micelles even more mixed micelles need to disintegrate to restart activity. Thus, the motility transition temperature is lowered, here to $T_\text{start} = \SI{16.8\pm 0.2}{\celsius}$, where the droplet escapes the oil-filled micelle cloud (III) and swims away (IV). 
\par
\begin{figure*}
	\centering
	\includegraphics[width=1\linewidth]{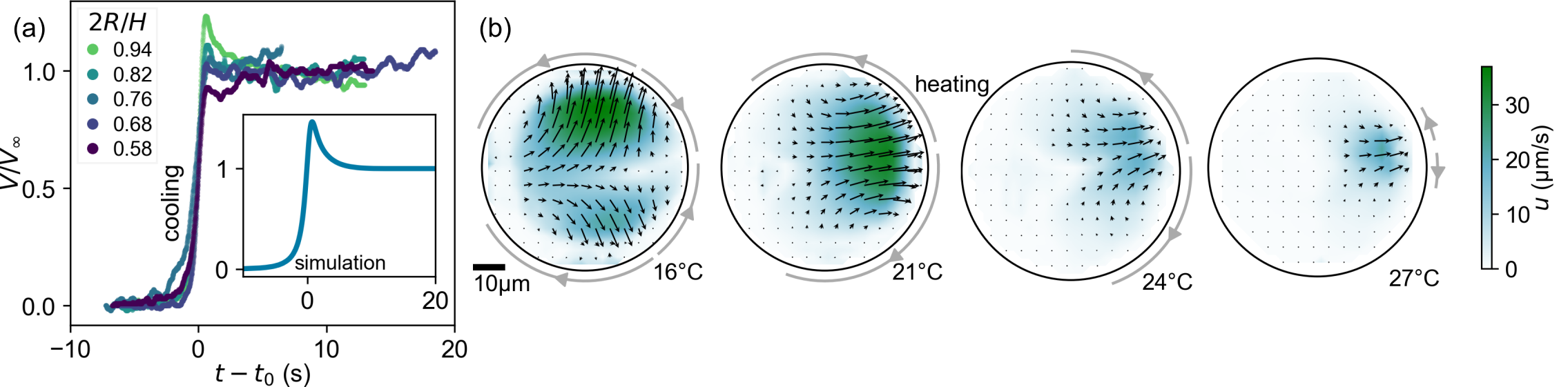}
	\caption{\textbf{Observation of the transition between passive dissolution and active propulsion.} 
		(a) Onset of motion during re-cooling, speed (normalized to steady state $V_\infty$) vs.\ time for multiple runs with $t_\text{onset}$ corresponding to $(\rm{d}V/\rm{d}t)_\text{max}$.  The initial overshoot increases with confinement $2R/H$. Inset: simulation for an isotropic autophoretic particle under comparable conditions ($H=2.2R$ confinement, $\Pen=8$), re-dimensionalized (Supplement~\ref{SIsec:numerics} and Movie S6).
		(b) Internal flow field with increasing temperature, starting at a mixed dipolar/quadrupolar mode (meandering), over a purely dipolar mode (straight) that recedes to the anterior (slowdown). 
		Vectors and color map inside the droplet indicate the velocity field $\vec{u}(x,y)$;  arrows around the perimeter mark the active regions on the droplet interface. Scale bar \SI{10}{\um}.\label{fig:piv}}	
\end{figure*}
Before discussing the flow fields, we note two more consequences of oil saturation. First, the hysteresis in \figref{fig:traj}c is reduced by several degrees if the system is not heated to full droplet arrest, but it is never entirely suppressed (Appendix Fig.\figref{SIfig:incomplete}).
This can be understood as follows:  during the late stage of the heating ramp, the droplet is already dispersing oil into its local environment by recirculation, starting from the posterior  - an effect we have also found in self-throttling pumping droplets in \cite{ramesh2023_interfacial}. During heating, the droplets will come to a stop even before the interfacial activity has fully ceded (see the discussion of \figref{fig:piv}b below), and self-propulsion would always need to restart from inside an oil-rich region as shown in \figref{fig:fluo}c-II.

Second, the regime of straight swimming appears to be highly localized on cooling (Movies S1 and S6): the droplet switches after a few seconds to a meandering motion (Movie S1). We argue here that outside the strongly localized cloud of spent fuel (see Movie S4 and Fig.~\ref{fig:fluo}c), more empty TTAB micelles have been released, such that the droplet experiences a higher \Pen{} once it escapes its self-generated local trap. We note that during this escape there is a radial gradient from filled to empty micelles, which would also locally rectify the droplet motion.

To analyze the mode evolution causing arrest and sudden onset of motion during heating and cooling (Movie S5), we added tracer colloids to the oil phase, performed high resolution bright field videomicroscopy and evaluated the internal flow field $\vec{u}(x,y)$ by particle image velocimetry (PIV) at a series of equilibrated set temperatures. \figref{fig:piv}b shows the evolution of $\vec{u}$ with increasing temperature. At $T = \SI{16}{\celsius}$, we see a  mixed  dipolar and quadrupolar flow field ($n=1,2$) corresponding to the meandering trajectory in Fig.~\ref{fig:traj}a~\cite{suda2021_straightcurvilinear,hokmabad2021_emergence}. At even higher temperature, $T = \SI{21}{\celsius}$, the droplet swims straight, \Pen{} decreases and the flow field is purely dipolar ($n=1$).  As the droplet begins to slow down, an inactive region spreads from the droplet posterior ($T = \SI{24}{\celsius}$). Finally ($T = \SI{27}{\celsius}$), just before the droplet stops ($n=0$), only a small region at the droplet anterior is active \cite{ramesh2023_interfacial}. As shown in Fig.~\ref{fig:fluo}II, the local environment isotropically saturates with spent fuel while the droplet is immotile.
\par
\subsection{Simulations}
The gradual increase of \Pen{} during cooling now allows us to directly observe the fundamental first transition from the immotile base state to self-propelled motion~\cite{michelin2013_spontaneous,michelin2023_selfpropulsion}. This can be motivated theoretically using hydrodynamic advection-diffusion models, canonically set out by Michelin et al.~\cite{michelin2013_spontaneous} as follows:

A spherical particle of radius $R$ is immersed in a fluid medium containing a chemical fuel at concentration $c$. At negligible Reynolds numbers, the flow is governed by the Stokes equations, $\mu\nabla^2\vec{u}=\nabla p$, $\nabla\cdot\vec{u}=0.$  The chemical field is coupled by an advection-diffusion equation, 

\begin{align}|\Pen|\left(\frac{\partial c}{\partial t} + \vec{u}\cdot\nabla c\right) &= \nabla^2c, &
\Pen{}&\equiv \frac{\mathcal{AM}R}{D^2},\label{eq:adv_diff}			
\end{align}
and by the particle consuming fuel at its boundary, \mbox{$\partial_t c(R) =-\mathcal{A}$}.  The P\'eclet number is set by the activity $\mathcal{A}$, mobility $\mathcal{M}$ and diffusivity $D$ of the chemical species and the particle radius $R$. Using a decomposition into squirmer modes and a linear stability analysis around the isotropic base state, $n=0$, the authors of \cite{michelin2013_spontaneous} find a transition to the propulsive dipolar state, $n=1$, above a threshold value of $\Pen{}=4$.

The mixed surfactant approach allows us to observe the growth of the dipolar mode in situ and analyzing it in the context of the canonical model. We performed a simulation of the interfacial instability, following standard protocols~\cite{michelin2013_spontaneous,picella2022_confined}, but here solving for the full 3D problem and adapted to our cell geometry (see Appendix~\ref{SIsec:numerics}). We compared our simulation to the experimental droplet speed $V$ at the onset of motion, with good agreement between re-dimensionalized numerical and experimental results (Fig.~\ref{fig:piv}a, Movie S6). For one, the timescales for the onset of motion, of $\mathcal{O}(\SI{1}{\s})$, and the evolution of the steady state at $V_\infty$, of $\mathcal{O}(\SI{5}{\s})$, are similar. Second, there is a characteristic overshoot in $V/V_\infty$ shortly after the onset of motion, common to various numerical approaches~\cite{schmitt2013_swimming,peng2023_weakly}, and already noted as `surprising'~\cite{michelin2013_spontaneous}. This initial push could be provided by the diffusive cloud of consumed or depleted fuel around the droplet (see eg. Fig.~\ref{fig:fluo}c-II), which causes a radial chemorepulsive gradient. 
Such gradients would be enhanced in confined geometries~\cite{anderson1989_colloid, picella2022_confined}: simulations have found the overshoot to increase with confinement~\cite{picella2022_confined}, and we find a similar tendency in our experimental data (Fig.\ \ref{fig:piv}(a)).

\section{Conclusions and outlook}
Tuning the dynamics of self-propelling droplets by temperature-driven surfactant interactions provides a promising framework to regulate micelle-mediated~\cite{babu2021_acceleration,wentworth2022_chemically} active droplet dynamics: we can now control self-propulsion from an unsteady or meandering state over quasi-ballistic propulsion to full arrest without needing to change the chemistry of the system. The gait control is encoded in the swimming medium and does not require complex micro-engineered swimmer design. Since the hydrophobic effect underlying the temperature dependent complex formation is entropy driven and similar aggregation effects are established for numerous surfactant-polymer combinations~\cite{nambam2012_effects,parmar2014_pluronic,li2001_binding,tam2006_insights,hecht1994_interaction}, this control method likely applies to active droplet models driven by micellar solubilization in general, and could be tested in further studies. For example, autophoretic droplet propulsion is known for SDS surfactant, DEP oil~\cite{izzet2020_tunable,thutupalli2018_flowinduced}, and a large number of oil/surfactant combinations tabulated in reviews~\cite{birrer2022_we}, and one could tune the transition temperatures by using different pluronics like PF88 or PF123~\cite{hecht1995_kinetic}. The transitions are almost fully reversible, excepting a slight reduction in peak speed that can be attributed to droplet shrinkage. 

Our hypothesis -- fuel binding by thermosensitive polymer cosurfactants --  does not account for the dynamics of adsorbed polymer at the interface, which might also be temperature dependent. However, we argue that these effects are, if present, secondary to the binding and release of TTAB in the swimming medium: generally, the desorption kinetics of large polymers are assumed to be exceedingly slow~\cite{butt2003_physics}.  
Thus, if these kinetics were the main drivers of thermoresponsive mode switching, it would not be consistent with our observations of cyclic reversibility and the fast response to changed external conditions, i.e. the instantaneous onset of motion in Fig.~\ref{fig:piv}a and particularly the fast, local adaptation to the fuel-rich medium outside the saturation area.

\par 
Our experiments fit into the framework of the  canonical theory for autophoretic particles, where the observed dynamic regimes correspond to interfacial modes becoming unstable with increasing or decreasing P\'eclet number. 
\par
While such higher order modes have been documented individually, the fundamental spontaneous transition from an isotropic zero order base state to a first order propulsion state is hard to observe experimentally, as the setup of the experiment usually provides sufficient disturbances to instantaneously set off droplet motion. By a non-invasive temperature driven crossing of the critical \Pen{} threshold, this is now experimentally observable in both chemical and flow signatures. 
As the idealized theory cited in eq.~\ref{eq:adv_diff} and its extensions are widely used in numerical work on autophoretic swimmers in complex geometries and confinement, modeling many body interactions, or chaotic dynamics in a huge parameter space, it is important to test its predictive power in an experimental context. The inactive-active transition analyzed in Fig.~\ref{fig:piv} experimentally matches  a common numerical validation case~\cite{michelin2013_spontaneous,hu2022_spontaneous,chen2021_instabilities,peng2023_weakly,wang2024_selfpropulsion} and highlights the importance of chemical history and spatial confinement in matching numerics and experiment~\cite{picella2022_confined,deblois2021_swimming,peng2023_weakly}.

The approach of depletant-mediated state control can be combined with a number of methods to guide and functionalize autophoretic swimmers in 2D and 3D~\cite{thutupalli2018_flowinduced,kruger2016_dimensionality,lancia2019_reorientation}
Solubilization based swimmers are known to exhibit chemo-, rheo-, electro-, photo- and magnetotaxis~\cite{jin2017_chemotaxis,moerman2017_solutemediated,lagzi2010_maze,dey2022_oscillatory,dwivedi2021_rheotaxis,buness2024_electrotaxis,ryabchun2022_runandhalt,xiao2018_moving,florea2014_photochemopropulsion,wagner2024_magnetotaxis,khan2024_perturbing}. Specifically, photo- and magnetotaxis are engineered by a modification of the oil phase like doping, colloidal inclusions or liquid crystalline phases, and therefore independently tunable from the fuel binding in the outer medium.

The co-surfactant based fuel depletion approach should also apply to solubilization based models that do not activate by spontaneous symmetry breaking of an isotropic base state. These could be ab initio asymmetric Janus droplets, either realized by compound droplets~\cite{balaj2020_reconfigurable}, or by adding colloidal patches to the interface~\cite{ikcheon2021_interfaciallyadsorbed}, which would also be arrested by fuel binding.

Moreover, it would be compatible with cargo functionalization - nematic self propelling droplets are stable carriers of configurable aqueous compartments which can be released by dissolution, nematic phase transition or coalescence~\cite{poulin1997_novel,hokmabad2019_topological,wang2024_selftimed}.
We also anticipate that it would inhibit activity in larger droplet-based autophoretic micropumps~\cite{ramesh2023_interfacial}.

Finally, tunable activity can be used to influence aggregation and collective behavior. Autophoretic droplets are known to show complex collective behavior depending on number densities and state of confinement~\cite{thutupalli2011_swarming,moerman2017_solutemediated,hokmabad2022_chemotactic,kruger2016_dimensionality,thutupalli2018_flowinduced,meredith2020_predatorprey}. We note that the cosurfactant mediated arrest mechanism is conceptually different from the self trapping observed previously in single surfactant systems~\cite{hokmabad2022_chemotactic}, where active droplets were trapped in self generated cages of chemorepulsive trails of oil-filled micelles, such that the swimmer ensemble is creating its own `chemical landscape'. The latter is a collective phenomenon, the trapping is transient and the $\Pen$ based state of activity is not modified. This chemorepulsive effect is also observable under PF127 addition: in samples with a higher number density, we have found multiple instances of repulsion interactions (SI section 11). Another collective effect that could be controlled by fuel depletion is the formation of `hovercraft' clusters under gravity in 3D reservoirs~\cite{kruger2016_dimensionality,thutupalli2018_flowinduced,hokmabad2022_spontaneously}: droplets self assemble into crystalline arrangements floating above the container bottom, stabilized by a balance of active upward motion and downward sedimentation. If the activity is suppressed by fuel depletion, these aggregates would sink to the bottom and disintegrate.

\par
In the context of smart, active materials, controllable dynamic states are crucial for programmable motility, function, sensing and adaptation~\cite{alvarez2021_reconfigurable,vankesteren2023_selfpropelling}. In the multi-scale and multidisciplinary study of synthetic microswimmers presented here, three
distinct macroscopic motile states emerge from the nano-scale control of molecular assembly, 
and we observe a counter-intuitive effect of dynamic arrest upon heating. These state transitions are driven by very general thermodynamic principles - our findings therefore offer a general physico-chemical design framework for the local and global control of synthetic active matter, beyond the system used here, and offer new avenues to further manipulate different types of active agents from the single to the collective scale.

\subsection*{Methods}
 All experimental and numerical methods are described  in detail in the Supporting Materials.
 
\subsection*{Author contributions}
PR designed and performed experiments, analyzed data and wrote the paper, PR\"a and SM performed experiments and analyzed data, YC designed and performed simulations, MJ designed experiments, CCM designed and performed experiments, analyzed data and wrote the paper. All authors proofread the paper.

\subsection*{Data and code availability}
 The data and Python scripts supporting the findings of this study and the numerical code underlying Fig. 4a are available at DOI:10.5281/zenodo.7818660.
 \subsection*{Competing interests}
 The authors declare no competing interests.

\subsection*{Acknowledgments}
We thank Dr.\ Babak Vajdi Hokmabad, Dr.\ Kevin Zhong and Dr.\ Stefan Karpitschka for invaluable advice and discussions, Dr.\ Stephan Weiss for providing the thermistor and Dr.\ Kristian Hantke for experimental support. We acknowledge funding from the Dutch Research Council, the German Research Foundation, grant MA 6330/1, and the Max Planck Society.


\begin{thebibliography}{88}%
\makeatletter
\providecommand \@ifxundefined [1]{%
 \@ifx{#1\undefined}
}%
\providecommand \@ifnum [1]{%
 \ifnum #1\expandafter \@firstoftwo
 \else \expandafter \@secondoftwo
 \fi
}%
\providecommand \@ifx [1]{%
 \ifx #1\expandafter \@firstoftwo
 \else \expandafter \@secondoftwo
 \fi
}%
\providecommand \natexlab [1]{#1}%
\providecommand \enquote  [1]{``#1''}%
\providecommand \bibnamefont  [1]{#1}%
\providecommand \bibfnamefont [1]{#1}%
\providecommand \citenamefont [1]{#1}%
\providecommand \href@noop [0]{\@secondoftwo}%
\providecommand \href [0]{\begingroup \@sanitize@url \@href}%
\providecommand \@href[1]{\@@startlink{#1}\@@href}%
\providecommand \@@href[1]{\endgroup#1\@@endlink}%
\providecommand \@sanitize@url [0]{\catcode `\\12\catcode `\$12\catcode
  `\&12\catcode `\#12\catcode `\^12\catcode `\_12\catcode `\%12\relax}%
\providecommand \@@startlink[1]{}%
\providecommand \@@endlink[0]{}%
\providecommand \url  [0]{\begingroup\@sanitize@url \@url }%
\providecommand \@url [1]{\endgroup\@href {#1}{\urlprefix }}%
\providecommand \urlprefix  [0]{URL }%
\providecommand \Eprint [0]{\href }%
\providecommand \doibase [0]{http://dx.doi.org/}%
\providecommand \selectlanguage [0]{\@gobble}%
\providecommand \bibinfo  [0]{\@secondoftwo}%
\providecommand \bibfield  [0]{\@secondoftwo}%
\providecommand \translation [1]{[#1]}%
\providecommand \BibitemOpen [0]{}%
\providecommand \bibitemStop [0]{}%
\providecommand \bibitemNoStop [0]{.\EOS\space}%
\providecommand \EOS [0]{\spacefactor3000\relax}%
\providecommand \BibitemShut  [1]{\csname bibitem#1\endcsname}%
\let\auto@bib@innerbib\@empty
\bibitem [{\citenamefont {Bowick}\ \emph {et~al.}(2022)\citenamefont {Bowick},
  \citenamefont {Fakhri}, \citenamefont {Marchetti},\ and\ \citenamefont
  {Ramaswamy}}]{bowick2022_symmetry}%
  \BibitemOpen
  \bibfield  {author} {\bibinfo {author} {\bibfnamefont {M.~J.}\ \bibnamefont
  {Bowick}}, \bibinfo {author} {\bibfnamefont {N.}~\bibnamefont {Fakhri}},
  \bibinfo {author} {\bibfnamefont {M.~C.}\ \bibnamefont {Marchetti}}, \ and\
  \bibinfo {author} {\bibfnamefont {S.}~\bibnamefont {Ramaswamy}},\ }\bibfield
  {title} {\enquote {\bibinfo {title} {Symmetry, {{Thermodynamics}}, and
  {{Topology}} in {{Active Matter}}},}\ }\href {\doibase
  10.1103/PhysRevX.12.010501} {\bibfield  {journal} {\bibinfo  {journal} {Phys.
  Rev. X}\ }\textbf {\bibinfo {volume} {12}},\ \bibinfo {pages} {010501}
  (\bibinfo {year} {2022})}\BibitemShut {NoStop}%
\bibitem [{\citenamefont {von Goethe}(1798)}]{goethe1798_zauberlehrling}%
  \BibitemOpen
  \bibfield  {author} {\bibinfo {author} {\bibfnamefont {J.~W.}\ \bibnamefont
  {von Goethe}},\ }\bibfield  {title} {\enquote {\bibinfo {title} {Der
  {{Zauberlehrling}}},}\ }in\ \href@noop {} {\emph {\bibinfo {booktitle}
  {Musen-{{Almanach}} F{\"u}r Das {{Jahr}} 1798}}},\ \bibinfo {editor} {edited
  by\ \bibinfo {editor} {\bibnamefont {{F. Schiller}}}}\ (\bibinfo  {publisher}
  {J.G. Cotta},\ \bibinfo {year} {1798})\ pp.\ \bibinfo {pages}
  {32--37}\BibitemShut {NoStop}%
\bibitem [{\citenamefont {O'Byrne}\ \emph {et~al.}(2022)\citenamefont
  {O'Byrne}, \citenamefont {Kafri}, \citenamefont {Tailleur},\ and\
  \citenamefont {{van Wijland}}}]{obyrne2022_time}%
  \BibitemOpen
  \bibfield  {author} {\bibinfo {author} {\bibfnamefont {J.}~\bibnamefont
  {O'Byrne}}, \bibinfo {author} {\bibfnamefont {Y.}~\bibnamefont {Kafri}},
  \bibinfo {author} {\bibfnamefont {J.}~\bibnamefont {Tailleur}}, \ and\
  \bibinfo {author} {\bibfnamefont {F.}~\bibnamefont {{van Wijland}}},\
  }\bibfield  {title} {\enquote {\bibinfo {title} {Time irreversibility in
  active matter, from micro to macro},}\ }\href {\doibase
  10.1038/s42254-021-00406-2} {\bibfield  {journal} {\bibinfo  {journal} {Nat
  Rev Phys}\ }\textbf {\bibinfo {volume} {4}},\ \bibinfo {pages} {167--183}
  (\bibinfo {year} {2022})}\BibitemShut {NoStop}%
\bibitem [{\citenamefont {Tu}\ \emph {et~al.}(2017{\natexlab{a}})\citenamefont
  {Tu}, \citenamefont {Peng},\ and\ \citenamefont {Wilson}}]{tu2017_motion}%
  \BibitemOpen
  \bibfield  {author} {\bibinfo {author} {\bibfnamefont {Y.}~\bibnamefont
  {Tu}}, \bibinfo {author} {\bibfnamefont {F.}~\bibnamefont {Peng}}, \ and\
  \bibinfo {author} {\bibfnamefont {D.~A.}\ \bibnamefont {Wilson}},\ }\bibfield
   {title} {\enquote {\bibinfo {title} {Motion {{Manipulation}} of {{Micro-}}
  and {{Nanomotors}}},}\ }\href {\doibase 10.1002/adma.201701970} {\bibfield
  {journal} {\bibinfo  {journal} {Advanced Materials}\ }\textbf {\bibinfo
  {volume} {29}},\ \bibinfo {pages} {1701970} (\bibinfo {year}
  {2017}{\natexlab{a}})}\BibitemShut {NoStop}%
\bibitem [{\citenamefont {Fusi}\ \emph {et~al.}(2023)\citenamefont {Fusi},
  \citenamefont {Li}, \citenamefont {{Llopis-Lorente}}, \citenamefont
  {Pati{\~n}o}, \citenamefont {{van Hest}},\ and\ \citenamefont
  {Abdelmohsen}}]{fusi2023_achieving}%
  \BibitemOpen
  \bibfield  {author} {\bibinfo {author} {\bibfnamefont {A.~D.}\ \bibnamefont
  {Fusi}}, \bibinfo {author} {\bibfnamefont {Y.}~\bibnamefont {Li}}, \bibinfo
  {author} {\bibfnamefont {A.}~\bibnamefont {{Llopis-Lorente}}}, \bibinfo
  {author} {\bibfnamefont {T.}~\bibnamefont {Pati{\~n}o}}, \bibinfo {author}
  {\bibfnamefont {J.~C.~M.}\ \bibnamefont {{van Hest}}}, \ and\ \bibinfo
  {author} {\bibfnamefont {L.~K. E.~A.}\ \bibnamefont {Abdelmohsen}},\
  }\bibfield  {title} {\enquote {\bibinfo {title} {Achieving {{Control}} in
  {{Micro-}}/{{Nanomotor Mobility}}},}\ }\href {\doibase
  10.1002/anie.202214754} {\bibfield  {journal} {\bibinfo  {journal}
  {Angewandte Chemie International Edition}\ }\textbf {\bibinfo {volume}
  {62}},\ \bibinfo {pages} {e202214754} (\bibinfo {year} {2023})}\BibitemShut
  {NoStop}%
\bibitem [{\citenamefont {Bechinger}\ \emph {et~al.}(2016)\citenamefont
  {Bechinger}, \citenamefont {Di~Leonardo}, \citenamefont {L{\"o}wen},
  \citenamefont {Reichhardt}, \citenamefont {Volpe},\ and\ \citenamefont
  {Volpe}}]{bechinger2016_active}%
  \BibitemOpen
  \bibfield  {author} {\bibinfo {author} {\bibfnamefont {C.}~\bibnamefont
  {Bechinger}}, \bibinfo {author} {\bibfnamefont {R.}~\bibnamefont
  {Di~Leonardo}}, \bibinfo {author} {\bibfnamefont {H.}~\bibnamefont
  {L{\"o}wen}}, \bibinfo {author} {\bibfnamefont {C.}~\bibnamefont
  {Reichhardt}}, \bibinfo {author} {\bibfnamefont {G.}~\bibnamefont {Volpe}}, \
  and\ \bibinfo {author} {\bibfnamefont {G.}~\bibnamefont {Volpe}},\ }\bibfield
   {title} {\enquote {\bibinfo {title} {Active {{Particles}} in {{Complex}} and
  {{Crowded Environments}}},}\ }\href {\doibase 10.1103/RevModPhys.88.045006}
  {\bibfield  {journal} {\bibinfo  {journal} {Reviews of Modern Physics}\
  }\textbf {\bibinfo {volume} {88}},\ \bibinfo {pages} {045006} (\bibinfo
  {year} {2016})}\BibitemShut {NoStop}%
\bibitem [{\citenamefont {Moran}\ and\ \citenamefont
  {Posner}(2019)}]{moran2019_microswimmers}%
  \BibitemOpen
  \bibfield  {author} {\bibinfo {author} {\bibfnamefont {J.}~\bibnamefont
  {Moran}}\ and\ \bibinfo {author} {\bibfnamefont {J.}~\bibnamefont {Posner}},\
  }\bibfield  {title} {\enquote {\bibinfo {title} {Microswimmers with no moving
  parts},}\ }\href {\doibase 10.1063/PT.3.4203} {\bibfield  {journal} {\bibinfo
   {journal} {Physics Today}\ }\textbf {\bibinfo {volume} {72}},\ \bibinfo
  {pages} {44--50} (\bibinfo {year} {2019})}\BibitemShut {NoStop}%
\bibitem [{\citenamefont {Zhang}\ and\ \citenamefont
  {Hess}(2021)}]{zhang2021_chemicallypowered}%
  \BibitemOpen
  \bibfield  {author} {\bibinfo {author} {\bibfnamefont {Y.}~\bibnamefont
  {Zhang}}\ and\ \bibinfo {author} {\bibfnamefont {H.}~\bibnamefont {Hess}},\
  }\bibfield  {title} {\enquote {\bibinfo {title} {Chemically-powered swimming
  and diffusion in the microscopic world},}\ }\href {\doibase
  10.1038/s41570-021-00281-6} {\bibfield  {journal} {\bibinfo  {journal} {Nat
  Rev Chem}\ }\textbf {\bibinfo {volume} {5}},\ \bibinfo {pages} {500--510}
  (\bibinfo {year} {2021})}\BibitemShut {NoStop}%
\bibitem [{\citenamefont {Z{\"o}ttl}\ and\ \citenamefont
  {Stark}(2023)}]{zottl2023_modeling}%
  \BibitemOpen
  \bibfield  {author} {\bibinfo {author} {\bibfnamefont {A.}~\bibnamefont
  {Z{\"o}ttl}}\ and\ \bibinfo {author} {\bibfnamefont {H.}~\bibnamefont
  {Stark}},\ }\bibfield  {title} {\enquote {\bibinfo {title} {Modeling {{Active
  Colloids}}: {{From Active Brownian Particles}} to {{Hydrodynamic}} and
  {{Chemical Fields}}},}\ }\href {\doibase
  10.1146/annurev-conmatphys-040821-115500} {\bibfield  {journal} {\bibinfo
  {journal} {Annual Review of Condensed Matter Physics}\ }\textbf {\bibinfo
  {volume} {14}},\ \bibinfo {pages} {109--127} (\bibinfo {year}
  {2023})}\BibitemShut {NoStop}%
\bibitem [{\citenamefont {Maass}\ \emph {et~al.}(2016)\citenamefont {Maass},
  \citenamefont {Kr{\"u}ger}, \citenamefont {Herminghaus},\ and\ \citenamefont
  {Bahr}}]{maass2016_swimming}%
  \BibitemOpen
  \bibfield  {author} {\bibinfo {author} {\bibfnamefont {C.~C.}\ \bibnamefont
  {Maass}}, \bibinfo {author} {\bibfnamefont {C.}~\bibnamefont {Kr{\"u}ger}},
  \bibinfo {author} {\bibfnamefont {S.}~\bibnamefont {Herminghaus}}, \ and\
  \bibinfo {author} {\bibfnamefont {C.}~\bibnamefont {Bahr}},\ }\bibfield
  {title} {\enquote {\bibinfo {title} {Swimming {{Droplets}}},}\ }\href
  {\doibase 10.1146/annurev-conmatphys-031115-011517} {\bibfield  {journal}
  {\bibinfo  {journal} {Annu. Rev. Condens. Matter Phys.}\ }\textbf {\bibinfo
  {volume} {7}},\ \bibinfo {pages} {171--193} (\bibinfo {year}
  {2016})}\BibitemShut {NoStop}%
\bibitem [{\citenamefont {Babu}\ \emph {et~al.}(2022)\citenamefont {Babu},
  \citenamefont {Katsonis}, \citenamefont {Lancia}, \citenamefont {Plamont},\
  and\ \citenamefont {Ryabchun}}]{babu2022_motile}%
  \BibitemOpen
  \bibfield  {author} {\bibinfo {author} {\bibfnamefont {D.}~\bibnamefont
  {Babu}}, \bibinfo {author} {\bibfnamefont {N.}~\bibnamefont {Katsonis}},
  \bibinfo {author} {\bibfnamefont {F.}~\bibnamefont {Lancia}}, \bibinfo
  {author} {\bibfnamefont {R.}~\bibnamefont {Plamont}}, \ and\ \bibinfo
  {author} {\bibfnamefont {A.}~\bibnamefont {Ryabchun}},\ }\bibfield  {title}
  {\enquote {\bibinfo {title} {Motile behaviour of droplets in lipid
  systems},}\ }\href {\doibase 10.1038/s41570-022-00392-8} {\bibfield
  {journal} {\bibinfo  {journal} {Nat. Rev. Chem.}\ }\textbf {\bibinfo {volume}
  {6}},\ \bibinfo {pages} {377--388} (\bibinfo {year} {2022})}\BibitemShut
  {NoStop}%
\bibitem [{\citenamefont {Birrer}\ \emph {et~al.}(2022)\citenamefont {Birrer},
  \citenamefont {Cheon},\ and\ \citenamefont {Zarzar}}]{birrer2022_we}%
  \BibitemOpen
  \bibfield  {author} {\bibinfo {author} {\bibfnamefont {S.}~\bibnamefont
  {Birrer}}, \bibinfo {author} {\bibfnamefont {S.~I.}\ \bibnamefont {Cheon}}, \
  and\ \bibinfo {author} {\bibfnamefont {L.~D.}\ \bibnamefont {Zarzar}},\
  }\bibfield  {title} {\enquote {\bibinfo {title} {We the droplets: {{A}}
  constitutional approach to active and self-propelled emulsions},}\ }\href
  {\doibase 10.1016/j.cocis.2022.101623} {\bibfield  {journal} {\bibinfo
  {journal} {Current Opinion in Colloid \& Interface Science}\ }\textbf
  {\bibinfo {volume} {61}},\ \bibinfo {pages} {101623} (\bibinfo {year}
  {2022})}\BibitemShut {NoStop}%
\bibitem [{\citenamefont {Dwivedi}\ \emph {et~al.}(2022)\citenamefont
  {Dwivedi}, \citenamefont {Pillai},\ and\ \citenamefont
  {Mangal}}]{dwivedi2022_selfpropelled}%
  \BibitemOpen
  \bibfield  {author} {\bibinfo {author} {\bibfnamefont {P.}~\bibnamefont
  {Dwivedi}}, \bibinfo {author} {\bibfnamefont {D.}~\bibnamefont {Pillai}}, \
  and\ \bibinfo {author} {\bibfnamefont {R.}~\bibnamefont {Mangal}},\
  }\bibfield  {title} {\enquote {\bibinfo {title} {Self-propelled swimming
  droplets},}\ }\href {\doibase 10.1016/j.cocis.2022.101614} {\bibfield
  {journal} {\bibinfo  {journal} {Current Opinion in Colloid \& Interface
  Science}\ }\textbf {\bibinfo {volume} {61}},\ \bibinfo {pages} {101614}
  (\bibinfo {year} {2022})}\BibitemShut {NoStop}%
\bibitem [{\citenamefont {Michelin}(2023)}]{michelin2023_selfpropulsion}%
  \BibitemOpen
  \bibfield  {author} {\bibinfo {author} {\bibfnamefont {S.}~\bibnamefont
  {Michelin}},\ }\bibfield  {title} {\enquote {\bibinfo {title}
  {Self-{{Propulsion}} of {{Chemically Active Droplets}}},}\ }\href {\doibase
  10.1146/annurev-fluid-120720-012204} {\bibfield  {journal} {\bibinfo
  {journal} {Annual Review of Fluid Mechanics}\ }\textbf {\bibinfo {volume}
  {55}},\ \bibinfo {pages} {77--101} (\bibinfo {year} {2023})}\BibitemShut
  {NoStop}%
\bibitem [{\citenamefont {Hanczyc}\ \emph {et~al.}(2007)\citenamefont
  {Hanczyc}, \citenamefont {Toyota}, \citenamefont {Ikegami}, \citenamefont
  {Packard},\ and\ \citenamefont {Sugawara}}]{hanczyc2007_fatty}%
  \BibitemOpen
  \bibfield  {author} {\bibinfo {author} {\bibfnamefont {M.~M.}\ \bibnamefont
  {Hanczyc}}, \bibinfo {author} {\bibfnamefont {T.}~\bibnamefont {Toyota}},
  \bibinfo {author} {\bibfnamefont {T.}~\bibnamefont {Ikegami}}, \bibinfo
  {author} {\bibfnamefont {N.}~\bibnamefont {Packard}}, \ and\ \bibinfo
  {author} {\bibfnamefont {T.}~\bibnamefont {Sugawara}},\ }\bibfield  {title}
  {\enquote {\bibinfo {title} {Fatty {{Acid Chemistry}} at the {{Oil}}-{{Water
  Interface}}:\, {{Self-Propelled Oil Droplets}}},}\ }\href {\doibase
  10.1021/ja0706955} {\bibfield  {journal} {\bibinfo  {journal} {J. Am. Chem.
  Soc.}\ }\textbf {\bibinfo {volume} {129}},\ \bibinfo {pages} {9386--9391}
  (\bibinfo {year} {2007})}\BibitemShut {NoStop}%
\bibitem [{\citenamefont {Thutupalli}\ \emph {et~al.}(2011)\citenamefont
  {Thutupalli}, \citenamefont {Seemann},\ and\ \citenamefont
  {Herminghaus}}]{thutupalli2011_swarming}%
  \BibitemOpen
  \bibfield  {author} {\bibinfo {author} {\bibfnamefont {S.}~\bibnamefont
  {Thutupalli}}, \bibinfo {author} {\bibfnamefont {R.}~\bibnamefont {Seemann}},
  \ and\ \bibinfo {author} {\bibfnamefont {S.}~\bibnamefont {Herminghaus}},\
  }\bibfield  {title} {\enquote {\bibinfo {title} {Swarming behavior of simple
  model squirmers},}\ }\href {\doibase 10.1088/1367-2630/13/7/073021}
  {\bibfield  {journal} {\bibinfo  {journal} {New J. Phys.}\ }\textbf {\bibinfo
  {volume} {13}},\ \bibinfo {pages} {073021} (\bibinfo {year}
  {2011})}\BibitemShut {NoStop}%
\bibitem [{\citenamefont {Peddireddy}\ \emph {et~al.}(2012)\citenamefont
  {Peddireddy}, \citenamefont {Kumar}, \citenamefont {Thutupalli},
  \citenamefont {Herminghaus},\ and\ \citenamefont
  {Bahr}}]{peddireddy2012_solubilization}%
  \BibitemOpen
  \bibfield  {author} {\bibinfo {author} {\bibfnamefont {K.}~\bibnamefont
  {Peddireddy}}, \bibinfo {author} {\bibfnamefont {P.}~\bibnamefont {Kumar}},
  \bibinfo {author} {\bibfnamefont {S.}~\bibnamefont {Thutupalli}}, \bibinfo
  {author} {\bibfnamefont {S.}~\bibnamefont {Herminghaus}}, \ and\ \bibinfo
  {author} {\bibfnamefont {C.}~\bibnamefont {Bahr}},\ }\bibfield  {title}
  {\enquote {\bibinfo {title} {Solubilization of thermotropic liquid crystal
  compounds in aqueous surfactant solutions.}}\ }\href {\doibase
  10.1021/la3015817} {\bibfield  {journal} {\bibinfo  {journal} {Langmuir}\
  }\textbf {\bibinfo {volume} {28}},\ \bibinfo {pages} {12426--31} (\bibinfo
  {year} {2012})}\BibitemShut {NoStop}%
\bibitem [{\citenamefont {Izri}\ \emph {et~al.}(2014)\citenamefont {Izri},
  \citenamefont {{van der Linden}}, \citenamefont {Michelin},\ and\
  \citenamefont {Dauchot}}]{izri2014_selfpropulsion}%
  \BibitemOpen
  \bibfield  {author} {\bibinfo {author} {\bibfnamefont {Z.}~\bibnamefont
  {Izri}}, \bibinfo {author} {\bibfnamefont {M.~N.}\ \bibnamefont {{van der
  Linden}}}, \bibinfo {author} {\bibfnamefont {S.}~\bibnamefont {Michelin}}, \
  and\ \bibinfo {author} {\bibfnamefont {O.}~\bibnamefont {Dauchot}},\
  }\bibfield  {title} {\enquote {\bibinfo {title} {Self-{{Propulsion}} of
  {{Pure Water Droplets}} by {{Spontaneous Marangoni-Stress-Driven Motion}}},}\
  }\href {\doibase 10.1103/PhysRevLett.113.248302} {\bibfield  {journal}
  {\bibinfo  {journal} {Phys. Rev. Lett.}\ }\textbf {\bibinfo {volume} {113}},\
  \bibinfo {pages} {248302} (\bibinfo {year} {2014})}\BibitemShut {NoStop}%
\bibitem [{\citenamefont {Herminghaus}\ \emph {et~al.}(2014)\citenamefont
  {Herminghaus}, \citenamefont {Maass}, \citenamefont {Kr{\"u}ger},
  \citenamefont {Thutupalli}, \citenamefont {Goehring},\ and\ \citenamefont
  {Bahr}}]{herminghaus2014_interfacial}%
  \BibitemOpen
  \bibfield  {author} {\bibinfo {author} {\bibfnamefont {S.}~\bibnamefont
  {Herminghaus}}, \bibinfo {author} {\bibfnamefont {C.~C.}\ \bibnamefont
  {Maass}}, \bibinfo {author} {\bibfnamefont {C.}~\bibnamefont {Kr{\"u}ger}},
  \bibinfo {author} {\bibfnamefont {S.}~\bibnamefont {Thutupalli}}, \bibinfo
  {author} {\bibfnamefont {L.}~\bibnamefont {Goehring}}, \ and\ \bibinfo
  {author} {\bibfnamefont {C.}~\bibnamefont {Bahr}},\ }\bibfield  {title}
  {\enquote {\bibinfo {title} {Interfacial mechanisms in active emulsions},}\
  }\href {\doibase 10.1039/C4SM00550C} {\bibfield  {journal} {\bibinfo
  {journal} {Soft Matter}\ }\textbf {\bibinfo {volume} {10}},\ \bibinfo {pages}
  {7008--7022} (\bibinfo {year} {2014})}\BibitemShut {NoStop}%
\bibitem [{\citenamefont {Michelin}\ \emph {et~al.}(2013)\citenamefont
  {Michelin}, \citenamefont {Lauga},\ and\ \citenamefont
  {Bartolo}}]{michelin2013_spontaneous}%
  \BibitemOpen
  \bibfield  {author} {\bibinfo {author} {\bibfnamefont {S.}~\bibnamefont
  {Michelin}}, \bibinfo {author} {\bibfnamefont {E.}~\bibnamefont {Lauga}}, \
  and\ \bibinfo {author} {\bibfnamefont {D.}~\bibnamefont {Bartolo}},\
  }\bibfield  {title} {\enquote {\bibinfo {title} {Spontaneous autophoretic
  motion of isotropic particles},}\ }\href {\doibase 10.1063/1.4810749}
  {\bibfield  {journal} {\bibinfo  {journal} {Physics of Fluids}\ }\textbf
  {\bibinfo {volume} {25}},\ \bibinfo {pages} {061701} (\bibinfo {year}
  {2013})}\BibitemShut {NoStop}%
\bibitem [{\citenamefont {Suga}\ \emph {et~al.}(2018)\citenamefont {Suga},
  \citenamefont {Suda}, \citenamefont {Ichikawa},\ and\ \citenamefont
  {Kimura}}]{suga2018_selfpropelled}%
  \BibitemOpen
  \bibfield  {author} {\bibinfo {author} {\bibfnamefont {M.}~\bibnamefont
  {Suga}}, \bibinfo {author} {\bibfnamefont {S.}~\bibnamefont {Suda}}, \bibinfo
  {author} {\bibfnamefont {M.}~\bibnamefont {Ichikawa}}, \ and\ \bibinfo
  {author} {\bibfnamefont {Y.}~\bibnamefont {Kimura}},\ }\bibfield  {title}
  {\enquote {\bibinfo {title} {Self-propelled motion switching in nematic
  liquid crystal droplets in aqueous surfactant solutions},}\ }\href {\doibase
  10.1103/PhysRevE.97.062703} {\bibfield  {journal} {\bibinfo  {journal} {Phys.
  Rev. E}\ }\textbf {\bibinfo {volume} {97}},\ \bibinfo {pages} {062703}
  (\bibinfo {year} {2018})}\BibitemShut {NoStop}%
\bibitem [{\citenamefont {Izzet}\ \emph {et~al.}(2020)\citenamefont {Izzet},
  \citenamefont {Moerman}, \citenamefont {Gross}, \citenamefont {Groenewold},
  \citenamefont {Hollingsworth}, \citenamefont {Bibette},\ and\ \citenamefont
  {Brujic}}]{izzet2020_tunable}%
  \BibitemOpen
  \bibfield  {author} {\bibinfo {author} {\bibfnamefont {A.}~\bibnamefont
  {Izzet}}, \bibinfo {author} {\bibfnamefont {P.~G.}\ \bibnamefont {Moerman}},
  \bibinfo {author} {\bibfnamefont {P.}~\bibnamefont {Gross}}, \bibinfo
  {author} {\bibfnamefont {J.}~\bibnamefont {Groenewold}}, \bibinfo {author}
  {\bibfnamefont {A.~D.}\ \bibnamefont {Hollingsworth}}, \bibinfo {author}
  {\bibfnamefont {J.}~\bibnamefont {Bibette}}, \ and\ \bibinfo {author}
  {\bibfnamefont {J.}~\bibnamefont {Brujic}},\ }\bibfield  {title} {\enquote
  {\bibinfo {title} {Tunable {{Persistent Random Walk}} in {{Swimming
  Droplets}}},}\ }\href {\doibase 10.1103/PhysRevX.10.021035} {\bibfield
  {journal} {\bibinfo  {journal} {Phys. Rev. X}\ }\textbf {\bibinfo {volume}
  {10}},\ \bibinfo {pages} {021035} (\bibinfo {year} {2020})}\BibitemShut
  {NoStop}%
\bibitem [{\citenamefont {Meredith}\ \emph {et~al.}(2020)\citenamefont
  {Meredith}, \citenamefont {Moerman}, \citenamefont {Groenewold},
  \citenamefont {Chiu}, \citenamefont {Kegel}, \citenamefont {{van
  Blaaderen}},\ and\ \citenamefont {Zarzar}}]{meredith2020_predatorprey}%
  \BibitemOpen
  \bibfield  {author} {\bibinfo {author} {\bibfnamefont {C.~H.}\ \bibnamefont
  {Meredith}}, \bibinfo {author} {\bibfnamefont {P.~G.}\ \bibnamefont
  {Moerman}}, \bibinfo {author} {\bibfnamefont {J.}~\bibnamefont {Groenewold}},
  \bibinfo {author} {\bibfnamefont {Y.-J.}\ \bibnamefont {Chiu}}, \bibinfo
  {author} {\bibfnamefont {W.~K.}\ \bibnamefont {Kegel}}, \bibinfo {author}
  {\bibfnamefont {A.}~\bibnamefont {{van Blaaderen}}}, \ and\ \bibinfo {author}
  {\bibfnamefont {L.~D.}\ \bibnamefont {Zarzar}},\ }\bibfield  {title}
  {\enquote {\bibinfo {title} {Predator-prey interactions between droplets
  driven by non-reciprocal oil exchange},}\ }\href {\doibase
  10.1038/s41557-020-00575-0} {\bibfield  {journal} {\bibinfo  {journal}
  {Nature Chemistry}\ }\textbf {\bibinfo {volume} {12}},\ \bibinfo {pages}
  {1136--1142} (\bibinfo {year} {2020})}\BibitemShut {NoStop}%
\bibitem [{\citenamefont {Hokmabad}\ \emph {et~al.}(2021)\citenamefont
  {Hokmabad}, \citenamefont {Dey}, \citenamefont {Jalaal}, \citenamefont
  {Mohanty}, \citenamefont {Almukambetova}, \citenamefont {Baldwin},
  \citenamefont {Lohse},\ and\ \citenamefont {Maass}}]{hokmabad2021_emergence}%
  \BibitemOpen
  \bibfield  {author} {\bibinfo {author} {\bibfnamefont {B.~V.}\ \bibnamefont
  {Hokmabad}}, \bibinfo {author} {\bibfnamefont {R.}~\bibnamefont {Dey}},
  \bibinfo {author} {\bibfnamefont {M.}~\bibnamefont {Jalaal}}, \bibinfo
  {author} {\bibfnamefont {D.}~\bibnamefont {Mohanty}}, \bibinfo {author}
  {\bibfnamefont {M.}~\bibnamefont {Almukambetova}}, \bibinfo {author}
  {\bibfnamefont {K.~A.}\ \bibnamefont {Baldwin}}, \bibinfo {author}
  {\bibfnamefont {D.}~\bibnamefont {Lohse}}, \ and\ \bibinfo {author}
  {\bibfnamefont {C.~C.}\ \bibnamefont {Maass}},\ }\bibfield  {title} {\enquote
  {\bibinfo {title} {Emergence of {{Bimodal Motility}} in {{Active
  Droplets}}},}\ }\href {\doibase 10.1103/PhysRevX.11.011043} {\bibfield
  {journal} {\bibinfo  {journal} {Phys. Rev. X}\ }\textbf {\bibinfo {volume}
  {11}},\ \bibinfo {pages} {011043} (\bibinfo {year} {2021})}\BibitemShut
  {NoStop}%
\bibitem [{\citenamefont {Suda}\ \emph {et~al.}(2021)\citenamefont {Suda},
  \citenamefont {Suda}, \citenamefont {Ohmura},\ and\ \citenamefont
  {Ichikawa}}]{suda2021_straightcurvilinear}%
  \BibitemOpen
  \bibfield  {author} {\bibinfo {author} {\bibfnamefont {S.}~\bibnamefont
  {Suda}}, \bibinfo {author} {\bibfnamefont {T.}~\bibnamefont {Suda}}, \bibinfo
  {author} {\bibfnamefont {T.}~\bibnamefont {Ohmura}}, \ and\ \bibinfo {author}
  {\bibfnamefont {M.}~\bibnamefont {Ichikawa}},\ }\bibfield  {title} {\enquote
  {\bibinfo {title} {Straight-to-{{Curvilinear Motion Transition}} of a
  {{Swimming Droplet Caused}} by the {{Susceptibility}} to {{Fluctuations}}},}\
  }\href {\doibase 10.1103/PhysRevLett.127.088005} {\bibfield  {journal}
  {\bibinfo  {journal} {Phys. Rev. Lett.}\ }\textbf {\bibinfo {volume} {127}},\
  \bibinfo {pages} {088005} (\bibinfo {year} {2021})}\BibitemShut {NoStop}%
\bibitem [{\citenamefont {Tu}\ \emph {et~al.}(2017{\natexlab{b}})\citenamefont
  {Tu}, \citenamefont {Peng}, \citenamefont {Sui}, \citenamefont {Men},
  \citenamefont {White}, \citenamefont {{van Hest}},\ and\ \citenamefont
  {Wilson}}]{tu2017_selfpropelled}%
  \BibitemOpen
  \bibfield  {author} {\bibinfo {author} {\bibfnamefont {Y.}~\bibnamefont
  {Tu}}, \bibinfo {author} {\bibfnamefont {F.}~\bibnamefont {Peng}}, \bibinfo
  {author} {\bibfnamefont {X.}~\bibnamefont {Sui}}, \bibinfo {author}
  {\bibfnamefont {Y.}~\bibnamefont {Men}}, \bibinfo {author} {\bibfnamefont
  {P.~B.}\ \bibnamefont {White}}, \bibinfo {author} {\bibfnamefont {J.~C.~M.}\
  \bibnamefont {{van Hest}}}, \ and\ \bibinfo {author} {\bibfnamefont {D.~A.}\
  \bibnamefont {Wilson}},\ }\bibfield  {title} {\enquote {\bibinfo {title}
  {Self-propelled supramolecular nanomotors with temperature-responsive speed
  regulation},}\ }\href {\doibase 10.1038/nchem.2674} {\bibfield  {journal}
  {\bibinfo  {journal} {Nature Chemistry}\ }\textbf {\bibinfo {volume} {9}},\
  \bibinfo {pages} {480--486} (\bibinfo {year}
  {2017}{\natexlab{b}})}\BibitemShut {NoStop}%
\bibitem [{\citenamefont {Cholakova}\ \emph {et~al.}(2021)\citenamefont
  {Cholakova}, \citenamefont {Lisicki}, \citenamefont {Smoukov}, \citenamefont
  {Tcholakova}, \citenamefont {Lin}, \citenamefont {Chen}, \citenamefont
  {De~Canio}, \citenamefont {Lauga},\ and\ \citenamefont
  {Denkov}}]{cholakova2021_rechargeable}%
  \BibitemOpen
  \bibfield  {author} {\bibinfo {author} {\bibfnamefont {D.}~\bibnamefont
  {Cholakova}}, \bibinfo {author} {\bibfnamefont {M.}~\bibnamefont {Lisicki}},
  \bibinfo {author} {\bibfnamefont {S.~K.}\ \bibnamefont {Smoukov}}, \bibinfo
  {author} {\bibfnamefont {S.}~\bibnamefont {Tcholakova}}, \bibinfo {author}
  {\bibfnamefont {E.~E.}\ \bibnamefont {Lin}}, \bibinfo {author} {\bibfnamefont
  {J.}~\bibnamefont {Chen}}, \bibinfo {author} {\bibfnamefont {G.}~\bibnamefont
  {De~Canio}}, \bibinfo {author} {\bibfnamefont {E.}~\bibnamefont {Lauga}}, \
  and\ \bibinfo {author} {\bibfnamefont {N.}~\bibnamefont {Denkov}},\
  }\bibfield  {title} {\enquote {\bibinfo {title} {Rechargeable self-assembled
  droplet microswimmers driven by surface phase transitions},}\ }\href
  {\doibase 10.1038/s41567-021-01291-3} {\bibfield  {journal} {\bibinfo
  {journal} {Nature Physics}\ }\textbf {\bibinfo {volume} {17}},\ \bibinfo
  {pages} {1050--1055} (\bibinfo {year} {2021})}\BibitemShut {NoStop}%
\bibitem [{\citenamefont {Florea}\ \emph {et~al.}(2014)\citenamefont {Florea},
  \citenamefont {Wagner}, \citenamefont {Wagner}, \citenamefont {Wallace},
  \citenamefont {Benito-Lopez}, \citenamefont {Officer},\ and\ \citenamefont
  {Diamond}}]{florea2014_photochemopropulsion}%
  \BibitemOpen
  \bibfield  {author} {\bibinfo {author} {\bibfnamefont {L.}~\bibnamefont
  {Florea}}, \bibinfo {author} {\bibfnamefont {K.}~\bibnamefont {Wagner}},
  \bibinfo {author} {\bibfnamefont {P.}~\bibnamefont {Wagner}}, \bibinfo
  {author} {\bibfnamefont {G.~G.}\ \bibnamefont {Wallace}}, \bibinfo {author}
  {\bibfnamefont {F.}~\bibnamefont {Benito-Lopez}}, \bibinfo {author}
  {\bibfnamefont {D.~L.}\ \bibnamefont {Officer}}, \ and\ \bibinfo {author}
  {\bibfnamefont {D.}~\bibnamefont {Diamond}},\ }\bibfield  {title} {\enquote
  {\bibinfo {title} {Photo-{{Chemopropulsion}} -- {{Light-Stimulated Movement}}
  of {{Microdroplets}}},}\ }\href {\doibase 10.1002/adma.201403007} {\bibfield
  {journal} {\bibinfo  {journal} {Advanced Materials}\ }\textbf {\bibinfo
  {volume} {26}},\ \bibinfo {pages} {7339--7345} (\bibinfo {year}
  {2014})}\BibitemShut {NoStop}%
\bibitem [{\citenamefont {Kaneko}\ \emph {et~al.}(2017)\citenamefont {Kaneko},
  \citenamefont {Asakura},\ and\ \citenamefont
  {Banno}}]{kaneko2017_phototactic}%
  \BibitemOpen
  \bibfield  {author} {\bibinfo {author} {\bibfnamefont {S.}~\bibnamefont
  {Kaneko}}, \bibinfo {author} {\bibfnamefont {K.}~\bibnamefont {Asakura}}, \
  and\ \bibinfo {author} {\bibfnamefont {T.}~\bibnamefont {Banno}},\ }\bibfield
   {title} {\enquote {\bibinfo {title} {Phototactic behavior of self-propelled
  micrometer-sized oil droplets in a surfactant solution},}\ }\href {\doibase
  10.1039/C6CC09236E} {\bibfield  {journal} {\bibinfo  {journal} {Chem.
  Commun.}\ }\textbf {\bibinfo {volume} {53}},\ \bibinfo {pages} {2237--2240}
  (\bibinfo {year} {2017})}\BibitemShut {NoStop}%
\bibitem [{\citenamefont {Xiao}\ \emph {et~al.}(2018)\citenamefont {Xiao},
  \citenamefont {Zarghami}, \citenamefont {Wagner}, \citenamefont {Wagner},
  \citenamefont {Gordon}, \citenamefont {Florea}, \citenamefont {Diamond},\
  and\ \citenamefont {Officer}}]{xiao2018_moving}%
  \BibitemOpen
  \bibfield  {author} {\bibinfo {author} {\bibfnamefont {Y.}~\bibnamefont
  {Xiao}}, \bibinfo {author} {\bibfnamefont {S.}~\bibnamefont {Zarghami}},
  \bibinfo {author} {\bibfnamefont {K.}~\bibnamefont {Wagner}}, \bibinfo
  {author} {\bibfnamefont {P.}~\bibnamefont {Wagner}}, \bibinfo {author}
  {\bibfnamefont {K.~C.}\ \bibnamefont {Gordon}}, \bibinfo {author}
  {\bibfnamefont {L.}~\bibnamefont {Florea}}, \bibinfo {author} {\bibfnamefont
  {D.}~\bibnamefont {Diamond}}, \ and\ \bibinfo {author} {\bibfnamefont
  {D.~L.}\ \bibnamefont {Officer}},\ }\bibfield  {title} {\enquote {\bibinfo
  {title} {Moving {{Droplets}} in {{3D Using Light}}},}\ }\href {\doibase
  10.1002/adma.201801821} {\bibfield  {journal} {\bibinfo  {journal} {Advanced
  Materials}\ }\textbf {\bibinfo {volume} {30}},\ \bibinfo {pages} {1801821}
  (\bibinfo {year} {2018})}\BibitemShut {NoStop}%
\bibitem [{\citenamefont {Lancia}\ \emph {et~al.}(2019)\citenamefont {Lancia},
  \citenamefont {Yamamoto}, \citenamefont {Ryabchun}, \citenamefont
  {Yamaguchi}, \citenamefont {Sano},\ and\ \citenamefont
  {Katsonis}}]{lancia2019_reorientation}%
  \BibitemOpen
  \bibfield  {author} {\bibinfo {author} {\bibfnamefont {F.}~\bibnamefont
  {Lancia}}, \bibinfo {author} {\bibfnamefont {T.}~\bibnamefont {Yamamoto}},
  \bibinfo {author} {\bibfnamefont {A.}~\bibnamefont {Ryabchun}}, \bibinfo
  {author} {\bibfnamefont {T.}~\bibnamefont {Yamaguchi}}, \bibinfo {author}
  {\bibfnamefont {M.}~\bibnamefont {Sano}}, \ and\ \bibinfo {author}
  {\bibfnamefont {N.}~\bibnamefont {Katsonis}},\ }\bibfield  {title} {\enquote
  {\bibinfo {title} {Reorientation behavior in the helical motility of
  light-responsive spiral droplets},}\ }\href {\doibase
  10.1038/s41467-019-13201-6} {\bibfield  {journal} {\bibinfo  {journal} {Nat.
  Commun.}\ }\textbf {\bibinfo {volume} {10}},\ \bibinfo {pages} {5238}
  (\bibinfo {year} {2019})}\BibitemShut {NoStop}%
\bibitem [{\citenamefont {Alvarez}\ \emph {et~al.}(2021)\citenamefont
  {Alvarez}, \citenamefont {{Fernandez-Rodriguez}}, \citenamefont {Alegria},
  \citenamefont {{Arrese-Igor}}, \citenamefont {Zhao}, \citenamefont
  {Kr{\"o}ger},\ and\ \citenamefont {Isa}}]{alvarez2021_reconfigurable}%
  \BibitemOpen
  \bibfield  {author} {\bibinfo {author} {\bibfnamefont {L.}~\bibnamefont
  {Alvarez}}, \bibinfo {author} {\bibfnamefont {M.~A.}\ \bibnamefont
  {{Fernandez-Rodriguez}}}, \bibinfo {author} {\bibfnamefont {A.}~\bibnamefont
  {Alegria}}, \bibinfo {author} {\bibfnamefont {S.}~\bibnamefont
  {{Arrese-Igor}}}, \bibinfo {author} {\bibfnamefont {K.}~\bibnamefont {Zhao}},
  \bibinfo {author} {\bibfnamefont {M.}~\bibnamefont {Kr{\"o}ger}}, \ and\
  \bibinfo {author} {\bibfnamefont {L.}~\bibnamefont {Isa}},\ }\bibfield
  {title} {\enquote {\bibinfo {title} {Reconfigurable artificial microswimmers
  with internal feedback},}\ }\href {\doibase 10.1038/s41467-021-25108-2}
  {\bibfield  {journal} {\bibinfo  {journal} {Nature Communications}\ }\textbf
  {\bibinfo {volume} {12}},\ \bibinfo {pages} {4762} (\bibinfo {year}
  {2021})}\BibitemShut {NoStop}%
\bibitem [{\citenamefont {Ryabchun}\ \emph {et~al.}(2022)\citenamefont
  {Ryabchun}, \citenamefont {Babu}, \citenamefont {Movilli}, \citenamefont
  {Plamont}, \citenamefont {Stuart},\ and\ \citenamefont
  {Katsonis}}]{ryabchun2022_runandhalt}%
  \BibitemOpen
  \bibfield  {author} {\bibinfo {author} {\bibfnamefont {A.}~\bibnamefont
  {Ryabchun}}, \bibinfo {author} {\bibfnamefont {D.}~\bibnamefont {Babu}},
  \bibinfo {author} {\bibfnamefont {J.}~\bibnamefont {Movilli}}, \bibinfo
  {author} {\bibfnamefont {R.}~\bibnamefont {Plamont}}, \bibinfo {author}
  {\bibfnamefont {M.~C.~A.}\ \bibnamefont {Stuart}}, \ and\ \bibinfo {author}
  {\bibfnamefont {N.}~\bibnamefont {Katsonis}},\ }\bibfield  {title} {\enquote
  {\bibinfo {title} {Run-and-halt motility of droplets in response to light},}\
  }\href {\doibase 10.1016/j.chempr.2022.06.017} {\bibfield  {journal}
  {\bibinfo  {journal} {Chem}\ }\textbf {\bibinfo {volume} {8}},\ \bibinfo
  {pages} {2290--2300} (\bibinfo {year} {2022})}\BibitemShut {NoStop}%
\bibitem [{\citenamefont {Ben~Zion}\ \emph {et~al.}(2022)\citenamefont
  {Ben~Zion}, \citenamefont {Caba}, \citenamefont {Modin},\ and\ \citenamefont
  {Chaikin}}]{benzion2022_cooperation}%
  \BibitemOpen
  \bibfield  {author} {\bibinfo {author} {\bibfnamefont {M.~Y.}\ \bibnamefont
  {Ben~Zion}}, \bibinfo {author} {\bibfnamefont {Y.}~\bibnamefont {Caba}},
  \bibinfo {author} {\bibfnamefont {A.}~\bibnamefont {Modin}}, \ and\ \bibinfo
  {author} {\bibfnamefont {P.~M.}\ \bibnamefont {Chaikin}},\ }\bibfield
  {title} {\enquote {\bibinfo {title} {Cooperation in a fluid swarm of
  fuel-free micro-swimmers},}\ }\href {\doibase 10.1038/s41467-021-27870-9}
  {\bibfield  {journal} {\bibinfo  {journal} {Nat Commun}\ }\textbf {\bibinfo
  {volume} {13}},\ \bibinfo {pages} {184} (\bibinfo {year} {2022})}\BibitemShut
  {NoStop}%
\bibitem [{\citenamefont {{van Kesteren}}\ \emph {et~al.}(2023)\citenamefont
  {{van Kesteren}}, \citenamefont {Alvarez}, \citenamefont {{Arrese-Igor}},
  \citenamefont {Alegria},\ and\ \citenamefont
  {Isa}}]{vankesteren2023_selfpropelling}%
  \BibitemOpen
  \bibfield  {author} {\bibinfo {author} {\bibfnamefont {S.}~\bibnamefont {{van
  Kesteren}}}, \bibinfo {author} {\bibfnamefont {L.}~\bibnamefont {Alvarez}},
  \bibinfo {author} {\bibfnamefont {S.}~\bibnamefont {{Arrese-Igor}}}, \bibinfo
  {author} {\bibfnamefont {A.}~\bibnamefont {Alegria}}, \ and\ \bibinfo
  {author} {\bibfnamefont {L.}~\bibnamefont {Isa}},\ }\bibfield  {title}
  {\enquote {\bibinfo {title} {Self-propelling colloids with finite state
  dynamics},}\ }\href {\doibase 10.1073/pnas.2213481120} {\bibfield  {journal}
  {\bibinfo  {journal} {Proceedings of the National Academy of Sciences}\
  }\textbf {\bibinfo {volume} {120}},\ \bibinfo {pages} {e2213481120} (\bibinfo
  {year} {2023})}\BibitemShut {NoStop}%
\bibitem [{\citenamefont {Tam}\ and\ \citenamefont
  {{Wyn-Jones}}(2006)}]{tam2006_insights}%
  \BibitemOpen
  \bibfield  {author} {\bibinfo {author} {\bibfnamefont {K.~C.}\ \bibnamefont
  {Tam}}\ and\ \bibinfo {author} {\bibfnamefont {E.}~\bibnamefont
  {{Wyn-Jones}}},\ }\bibfield  {title} {\enquote {\bibinfo {title} {Insights on
  polymer surfactant complex structures during the binding of surfactants to
  polymers as measured by equilibrium and structural techniques},}\ }\href
  {\doibase 10.1039/B415140M} {\bibfield  {journal} {\bibinfo  {journal}
  {Chemical Society Reviews}\ }\textbf {\bibinfo {volume} {35}},\ \bibinfo
  {pages} {693--709} (\bibinfo {year} {2006})}\BibitemShut {NoStop}%
\bibitem [{\citenamefont {Zhang}\ \emph {et~al.}(2021)\citenamefont {Zhang},
  \citenamefont {Mozaffari},\ and\ \citenamefont {{de
  Pablo}}}]{zhang2021_autonomous}%
  \BibitemOpen
  \bibfield  {author} {\bibinfo {author} {\bibfnamefont {R.}~\bibnamefont
  {Zhang}}, \bibinfo {author} {\bibfnamefont {A.}~\bibnamefont {Mozaffari}}, \
  and\ \bibinfo {author} {\bibfnamefont {J.~J.}\ \bibnamefont {{de Pablo}}},\
  }\bibfield  {title} {\enquote {\bibinfo {title} {Autonomous materials systems
  from active liquid crystals},}\ }\href {\doibase 10.1038/s41578-020-00272-x}
  {\bibfield  {journal} {\bibinfo  {journal} {Nature Reviews Materials}\ ,\
  \bibinfo {pages} {1--17}} (\bibinfo {year} {2021})}\BibitemShut {NoStop}%
\bibitem [{\citenamefont {Nambam}\ and\ \citenamefont
  {Philip}(2012)}]{nambam2012_effects}%
  \BibitemOpen
  \bibfield  {author} {\bibinfo {author} {\bibfnamefont {J.~S.}\ \bibnamefont
  {Nambam}}\ and\ \bibinfo {author} {\bibfnamefont {J.}~\bibnamefont
  {Philip}},\ }\bibfield  {title} {\enquote {\bibinfo {title} {Effects of
  {{Interaction}} of {{Ionic}} and {{Nonionic Surfactants}} on
  {{Self-Assembly}} of {{PEO}}--{{PPO}}--{{PEO Triblock Copolymer}} in
  {{Aqueous Solution}}},}\ }\href {\doibase 10.1021/jp208902a} {\bibfield
  {journal} {\bibinfo  {journal} {The Journal of Physical Chemistry B}\
  }\textbf {\bibinfo {volume} {116}},\ \bibinfo {pages} {1499--1507} (\bibinfo
  {year} {2012})}\BibitemShut {NoStop}%
\bibitem [{\citenamefont {Jin}\ \emph {et~al.}(2017)\citenamefont {Jin},
  \citenamefont {Kr{\"u}ger},\ and\ \citenamefont
  {Maass}}]{jin2017_chemotaxis}%
  \BibitemOpen
  \bibfield  {author} {\bibinfo {author} {\bibfnamefont {C.}~\bibnamefont
  {Jin}}, \bibinfo {author} {\bibfnamefont {C.}~\bibnamefont {Kr{\"u}ger}}, \
  and\ \bibinfo {author} {\bibfnamefont {C.~C.}\ \bibnamefont {Maass}},\
  }\bibfield  {title} {\enquote {\bibinfo {title} {Chemotaxis and
  autochemotaxis of self-propelling droplet swimmers},}\ }\href {\doibase
  10.1073/pnas.1619783114} {\bibfield  {journal} {\bibinfo  {journal} {PNAS}\
  }\textbf {\bibinfo {volume} {114}},\ \bibinfo {pages} {5089--5094} (\bibinfo
  {year} {2017})}\BibitemShut {NoStop}%
\bibitem [{\citenamefont {Rosen}\ and\ \citenamefont
  {Kunjappu}(2012)}]{rosen2012_surfactants}%
  \BibitemOpen
  \bibfield  {author} {\bibinfo {author} {\bibfnamefont {M.~J.}\ \bibnamefont
  {Rosen}}\ and\ \bibinfo {author} {\bibfnamefont {J.~T.}\ \bibnamefont
  {Kunjappu}},\ }\href@noop {} {\emph {\bibinfo {title} {Surfactants and
  Interfacial Phenomena}}},\ \bibinfo {edition} {4th}\ ed.\ (\bibinfo
  {publisher} {Wiley},\ \bibinfo {address} {Hoboken, N.J.},\ \bibinfo {year}
  {2012})\BibitemShut {NoStop}%
\bibitem [{\citenamefont {Hokmabad}\ \emph
  {et~al.}(2022{\natexlab{a}})\citenamefont {Hokmabad}, \citenamefont
  {Nishide}, \citenamefont {Ramesh}, \citenamefont {Kr{\"u}ger},\ and\
  \citenamefont {Maass}}]{hokmabad2022_spontaneously}%
  \BibitemOpen
  \bibfield  {author} {\bibinfo {author} {\bibfnamefont {B.~V.}\ \bibnamefont
  {Hokmabad}}, \bibinfo {author} {\bibfnamefont {A.}~\bibnamefont {Nishide}},
  \bibinfo {author} {\bibfnamefont {P.}~\bibnamefont {Ramesh}}, \bibinfo
  {author} {\bibfnamefont {C.}~\bibnamefont {Kr{\"u}ger}}, \ and\ \bibinfo
  {author} {\bibfnamefont {C.~C.}\ \bibnamefont {Maass}},\ }\bibfield  {title}
  {\enquote {\bibinfo {title} {Spontaneously rotating clusters of active
  droplets},}\ }\href {\doibase 10.1039/D1SM01795K} {\bibfield  {journal}
  {\bibinfo  {journal} {Soft Matter}\ }\textbf {\bibinfo {volume} {18}},\
  \bibinfo {pages} {2731--2741} (\bibinfo {year}
  {2022}{\natexlab{a}})}\BibitemShut {NoStop}%
\bibitem [{\citenamefont {Alexandridis}\ \emph {et~al.}(1994)\citenamefont
  {Alexandridis}, \citenamefont {Holzwarth},\ and\ \citenamefont
  {Hatton}}]{alexandridis1994_micellization}%
  \BibitemOpen
  \bibfield  {author} {\bibinfo {author} {\bibfnamefont {P.}~\bibnamefont
  {Alexandridis}}, \bibinfo {author} {\bibfnamefont {J.~F.}\ \bibnamefont
  {Holzwarth}}, \ and\ \bibinfo {author} {\bibfnamefont {T.~A.}\ \bibnamefont
  {Hatton}},\ }\bibfield  {title} {\enquote {\bibinfo {title} {Micellization of
  {{Poly}}(ethylene oxide)-{{Poly}}(propylene oxide)-{{Poly}}(ethylene oxide)
  {{Triblock Copolymers}} in {{Aqueous Solutions}}: {{Thermodynamics}} of
  {{Copolymer Association}}},}\ }\href {\doibase 10.1021/ma00087a009}
  {\bibfield  {journal} {\bibinfo  {journal} {Macromolecules}\ }\textbf
  {\bibinfo {volume} {27}},\ \bibinfo {pages} {2414--2425} (\bibinfo {year}
  {1994})}\BibitemShut {NoStop}%
\bibitem [{\citenamefont {Wanka}\ \emph {et~al.}(1994)\citenamefont {Wanka},
  \citenamefont {Hoffmann},\ and\ \citenamefont {Ulbricht}}]{wanka1994_phase}%
  \BibitemOpen
  \bibfield  {author} {\bibinfo {author} {\bibfnamefont {G.}~\bibnamefont
  {Wanka}}, \bibinfo {author} {\bibfnamefont {H.}~\bibnamefont {Hoffmann}}, \
  and\ \bibinfo {author} {\bibfnamefont {W.}~\bibnamefont {Ulbricht}},\
  }\bibfield  {title} {\enquote {\bibinfo {title} {Phase {{Diagrams}} and
  {{Aggregation Behavior}} of
  {{Poly}}(oxyethylene)-{{Poly}}(oxypropylene)-{{Poly}}(oxyethylene) {{Triblock
  Copolymers}} in {{Aqueous Solutions}}},}\ }\href {\doibase
  10.1021/ma00093a016} {\bibfield  {journal} {\bibinfo  {journal}
  {Macromolecules}\ }\textbf {\bibinfo {volume} {27}},\ \bibinfo {pages}
  {4145--4159} (\bibinfo {year} {1994})}\BibitemShut {NoStop}%
\bibitem [{\citenamefont {Bohorquez}\ \emph {et~al.}(1999)\citenamefont
  {Bohorquez}, \citenamefont {Koch}, \citenamefont {Trygstad},\ and\
  \citenamefont {Pandit}}]{bohorquez1999_study}%
  \BibitemOpen
  \bibfield  {author} {\bibinfo {author} {\bibfnamefont {M.}~\bibnamefont
  {Bohorquez}}, \bibinfo {author} {\bibfnamefont {C.}~\bibnamefont {Koch}},
  \bibinfo {author} {\bibfnamefont {T.}~\bibnamefont {Trygstad}}, \ and\
  \bibinfo {author} {\bibfnamefont {N.}~\bibnamefont {Pandit}},\ }\bibfield
  {title} {\enquote {\bibinfo {title} {A {{Study}} of the
  {{Temperature-Dependent Micellization}} of {{Pluronic F127}}},}\ }\href
  {\doibase 10.1006/jcis.1999.6273} {\bibfield  {journal} {\bibinfo  {journal}
  {Journal of Colloid and Interface Science}\ }\textbf {\bibinfo {volume}
  {216}},\ \bibinfo {pages} {34--40} (\bibinfo {year} {1999})}\BibitemShut
  {NoStop}%
\bibitem [{\citenamefont {Stoeber}\ \emph {et~al.}(2006)\citenamefont
  {Stoeber}, \citenamefont {Hu}, \citenamefont {Liepmann},\ and\ \citenamefont
  {Muller}}]{stoeber2006_passive}%
  \BibitemOpen
  \bibfield  {author} {\bibinfo {author} {\bibfnamefont {B.}~\bibnamefont
  {Stoeber}}, \bibinfo {author} {\bibfnamefont {C.-M.~J.}\ \bibnamefont {Hu}},
  \bibinfo {author} {\bibfnamefont {D.}~\bibnamefont {Liepmann}}, \ and\
  \bibinfo {author} {\bibfnamefont {S.~J.}\ \bibnamefont {Muller}},\ }\bibfield
   {title} {\enquote {\bibinfo {title} {Passive flow control in microdevices
  using thermally responsive polymer solutions},}\ }\href {\doibase
  10.1063/1.2204077} {\bibfield  {journal} {\bibinfo  {journal} {Physics of
  Fluids}\ }\textbf {\bibinfo {volume} {18}},\ \bibinfo {pages} {053103}
  (\bibinfo {year} {2006})}\BibitemShut {NoStop}%
\bibitem [{\citenamefont {Jalaal}\ \emph {et~al.}(2016)\citenamefont {Jalaal},
  \citenamefont {Cottrell}, \citenamefont {Balmforth},\ and\ \citenamefont
  {Stoeber}}]{jalaal2016_rheology}%
  \BibitemOpen
  \bibfield  {author} {\bibinfo {author} {\bibfnamefont {M.}~\bibnamefont
  {Jalaal}}, \bibinfo {author} {\bibfnamefont {G.}~\bibnamefont {Cottrell}},
  \bibinfo {author} {\bibfnamefont {N.}~\bibnamefont {Balmforth}}, \ and\
  \bibinfo {author} {\bibfnamefont {B.}~\bibnamefont {Stoeber}},\ }\bibfield
  {title} {\enquote {\bibinfo {title} {On the rheology of {{Pluronic F127}}
  aqueous solutions},}\ }\href {\doibase 10.1122/1.4971992} {\bibfield
  {journal} {\bibinfo  {journal} {Journal of Rheology}\ }\textbf {\bibinfo
  {volume} {61}},\ \bibinfo {pages} {139--146} (\bibinfo {year}
  {2016})}\BibitemShut {NoStop}%
\bibitem [{\citenamefont {Jalaal}\ \emph {et~al.}(2018)\citenamefont {Jalaal},
  \citenamefont {Seyfert}, \citenamefont {Stoeber},\ and\ \citenamefont
  {Balmforth}}]{jalaal2018_gelcontrolled}%
  \BibitemOpen
  \bibfield  {author} {\bibinfo {author} {\bibfnamefont {M.}~\bibnamefont
  {Jalaal}}, \bibinfo {author} {\bibfnamefont {C.}~\bibnamefont {Seyfert}},
  \bibinfo {author} {\bibfnamefont {B.}~\bibnamefont {Stoeber}}, \ and\
  \bibinfo {author} {\bibfnamefont {N.~J.}\ \bibnamefont {Balmforth}},\
  }\bibfield  {title} {\enquote {\bibinfo {title} {Gel-controlled droplet
  spreading},}\ }\href {\doibase 10.1017/jfm.2017.844} {\bibfield  {journal}
  {\bibinfo  {journal} {Journal of Fluid Mechanics}\ }\textbf {\bibinfo
  {volume} {837}},\ \bibinfo {pages} {115--128} (\bibinfo {year}
  {2018})}\BibitemShut {NoStop}%
\bibitem [{\citenamefont {Hecht}\ and\ \citenamefont
  {Hoffmann}(1994)}]{hecht1994_interaction}%
  \BibitemOpen
  \bibfield  {author} {\bibinfo {author} {\bibfnamefont {E.}~\bibnamefont
  {Hecht}}\ and\ \bibinfo {author} {\bibfnamefont {H.}~\bibnamefont
  {Hoffmann}},\ }\bibfield  {title} {\enquote {\bibinfo {title} {Interaction of
  {{ABA}} block copolymers with ionic surfactants in aqueous solution},}\
  }\href {\doibase 10.1021/la00013a013} {\bibfield  {journal} {\bibinfo
  {journal} {Langmuir}\ }\textbf {\bibinfo {volume} {10}},\ \bibinfo {pages}
  {86--91} (\bibinfo {year} {1994})}\BibitemShut {NoStop}%
\bibitem [{\citenamefont {Li}\ \emph {et~al.}(2001)\citenamefont {Li},
  \citenamefont {Xu}, \citenamefont {Couderc}, \citenamefont {Bloor},
  \citenamefont {Holzwarth},\ and\ \citenamefont
  {{Wyn-Jones}}}]{li2001_binding}%
  \BibitemOpen
  \bibfield  {author} {\bibinfo {author} {\bibfnamefont {Y.}~\bibnamefont
  {Li}}, \bibinfo {author} {\bibfnamefont {R.}~\bibnamefont {Xu}}, \bibinfo
  {author} {\bibfnamefont {S.}~\bibnamefont {Couderc}}, \bibinfo {author}
  {\bibfnamefont {D.~M.}\ \bibnamefont {Bloor}}, \bibinfo {author}
  {\bibfnamefont {J.~F.}\ \bibnamefont {Holzwarth}}, \ and\ \bibinfo {author}
  {\bibfnamefont {E.}~\bibnamefont {{Wyn-Jones}}},\ }\bibfield  {title}
  {\enquote {\bibinfo {title} {Binding of {{Tetradecyltrimethylammonium
  Bromide}} to the {{ABA Block Copolymer Pluronic F127}} ({{EO97 PO69
  EO97}}):\, {{Electromotive Force}}, {{Microcalorimetry}}, and {{Light
  Scattering Studies}}},}\ }\href {\doibase 10.1021/la010004x} {\bibfield
  {journal} {\bibinfo  {journal} {Langmuir}\ }\textbf {\bibinfo {volume}
  {17}},\ \bibinfo {pages} {5742--5747} (\bibinfo {year} {2001})}\BibitemShut
  {NoStop}%
\bibitem [{\citenamefont {Butt}\ \emph {et~al.}(2003)\citenamefont {Butt},
  \citenamefont {Graf},\ and\ \citenamefont {Kappl}}]{butt2003_physics}%
  \BibitemOpen
  \bibfield  {author} {\bibinfo {author} {\bibfnamefont {H.-J.}\ \bibnamefont
  {Butt}}, \bibinfo {author} {\bibfnamefont {K.}~\bibnamefont {Graf}}, \ and\
  \bibinfo {author} {\bibfnamefont {M.}~\bibnamefont {Kappl}},\ }\href@noop {}
  {\emph {\bibinfo {title} {Physics and {{Chemistry}} of {{Interfaces}}}}}\
  (\bibinfo  {publisher} {John Wiley \& Sons},\ \bibinfo {year}
  {2003})\BibitemShut {NoStop}%
\bibitem [{\citenamefont {Hecht}\ \emph {et~al.}(1995)\citenamefont {Hecht},
  \citenamefont {Mortensen}, \citenamefont {Gradzielski},\ and\ \citenamefont
  {Hoffmann}}]{hecht1995_interaction}%
  \BibitemOpen
  \bibfield  {author} {\bibinfo {author} {\bibfnamefont {E.}~\bibnamefont
  {Hecht}}, \bibinfo {author} {\bibfnamefont {K.}~\bibnamefont {Mortensen}},
  \bibinfo {author} {\bibfnamefont {M.}~\bibnamefont {Gradzielski}}, \ and\
  \bibinfo {author} {\bibfnamefont {H.}~\bibnamefont {Hoffmann}},\ }\bibfield
  {title} {\enquote {\bibinfo {title} {Interaction of {{ABA Block Copolymers}}
  with {{Ionic Surfactants}}: {{Influence}} on {{Micellization}} and
  {{Gelation}}},}\ }\href {\doibase 10.1021/j100013a068} {\bibfield  {journal}
  {\bibinfo  {journal} {J. Phys. Chem.}\ }\textbf {\bibinfo {volume} {99}},\
  \bibinfo {pages} {4866--4874} (\bibinfo {year} {1995})}\BibitemShut {NoStop}%
\bibitem [{\citenamefont {Morozov}\ and\ \citenamefont
  {Michelin}(2019)}]{morozov2019_nonlinear}%
  \BibitemOpen
  \bibfield  {author} {\bibinfo {author} {\bibfnamefont {M.}~\bibnamefont
  {Morozov}}\ and\ \bibinfo {author} {\bibfnamefont {S.}~\bibnamefont
  {Michelin}},\ }\bibfield  {title} {\enquote {\bibinfo {title} {Nonlinear
  dynamics of a chemically-active drop: {{From}} steady to chaotic
  self-propulsion},}\ }\href {\doibase 10.1063/1.5080539} {\bibfield  {journal}
  {\bibinfo  {journal} {J. Chem. Phys.}\ }\textbf {\bibinfo {volume} {150}},\
  \bibinfo {pages} {044110} (\bibinfo {year} {2019})}\BibitemShut {NoStop}%
\bibitem [{\citenamefont {Morozov}(2020)}]{morozov2020_adsorption}%
  \BibitemOpen
  \bibfield  {author} {\bibinfo {author} {\bibfnamefont {M.}~\bibnamefont
  {Morozov}},\ }\bibfield  {title} {\enquote {\bibinfo {title} {Adsorption
  inhibition by swollen micelles may cause multistability in active
  droplets},}\ }\href {\doibase 10.1039/D0SM00662A} {\bibfield  {journal}
  {\bibinfo  {journal} {Soft Matter}\ }\textbf {\bibinfo {volume} {16}},\
  \bibinfo {pages} {5624--5632} (\bibinfo {year} {2020})}\BibitemShut {NoStop}%
\bibitem [{\citenamefont {Li}(2022)}]{li2022_swimming}%
  \BibitemOpen
  \bibfield  {author} {\bibinfo {author} {\bibfnamefont {G.}~\bibnamefont
  {Li}},\ }\bibfield  {title} {\enquote {\bibinfo {title} {Swimming dynamics of
  a self-propelled droplet},}\ }\href {\doibase 10.1017/jfm.2021.1154}
  {\bibfield  {journal} {\bibinfo  {journal} {Journal of Fluid Mechanics}\
  }\textbf {\bibinfo {volume} {934}} (\bibinfo {year} {2022}),\
  10.1017/jfm.2021.1154}\BibitemShut {NoStop}%
\bibitem [{\citenamefont {Hu}\ \emph {et~al.}(2022)\citenamefont {Hu},
  \citenamefont {Lin}, \citenamefont {Rafai},\ and\ \citenamefont
  {Misbah}}]{hu2022_spontaneous}%
  \BibitemOpen
  \bibfield  {author} {\bibinfo {author} {\bibfnamefont {W.-F.}\ \bibnamefont
  {Hu}}, \bibinfo {author} {\bibfnamefont {T.-S.}\ \bibnamefont {Lin}},
  \bibinfo {author} {\bibfnamefont {S.}~\bibnamefont {Rafai}}, \ and\ \bibinfo
  {author} {\bibfnamefont {C.}~\bibnamefont {Misbah}},\ }\bibfield  {title}
  {\enquote {\bibinfo {title} {Spontaneous locomotion of phoretic particles in
  three dimensions},}\ }\href {\doibase 10.1103/PhysRevFluids.7.034003}
  {\bibfield  {journal} {\bibinfo  {journal} {Phys. Rev. Fluids}\ }\textbf
  {\bibinfo {volume} {7}},\ \bibinfo {pages} {034003} (\bibinfo {year}
  {2022})}\BibitemShut {NoStop}%
\bibitem [{\citenamefont {Ramesh}\ \emph {et~al.}(2023)\citenamefont {Ramesh},
  \citenamefont {Hokmabad}, \citenamefont {Pushkin}, \citenamefont
  {Mathijssen},\ and\ \citenamefont {Maass}}]{ramesh2023_interfacial}%
  \BibitemOpen
  \bibfield  {author} {\bibinfo {author} {\bibfnamefont {P.}~\bibnamefont
  {Ramesh}}, \bibinfo {author} {\bibfnamefont {B.~V.}\ \bibnamefont
  {Hokmabad}}, \bibinfo {author} {\bibfnamefont {D.~O.}\ \bibnamefont
  {Pushkin}}, \bibinfo {author} {\bibfnamefont {A.~J. T.~M.}\ \bibnamefont
  {Mathijssen}}, \ and\ \bibinfo {author} {\bibfnamefont {C.~C.}\ \bibnamefont
  {Maass}},\ }\bibfield  {title} {\enquote {\bibinfo {title} {Interfacial
  activity dynamics of confined active droplets},}\ }\href {\doibase
  10.1017/jfm.2023.411} {\bibfield  {journal} {\bibinfo  {journal} {Journal of
  Fluid Mechanics}\ }\textbf {\bibinfo {volume} {966}},\ \bibinfo {pages} {A29}
  (\bibinfo {year} {2023})}\BibitemShut {NoStop}%
\bibitem [{\citenamefont {Hokmabad}\ \emph
  {et~al.}(2022{\natexlab{b}})\citenamefont {Hokmabad}, \citenamefont
  {{Agudo-Canalejo}}, \citenamefont {Saha}, \citenamefont {Golestanian},\ and\
  \citenamefont {Maass}}]{hokmabad2022_chemotactic}%
  \BibitemOpen
  \bibfield  {author} {\bibinfo {author} {\bibfnamefont {B.~V.}\ \bibnamefont
  {Hokmabad}}, \bibinfo {author} {\bibfnamefont {J.}~\bibnamefont
  {{Agudo-Canalejo}}}, \bibinfo {author} {\bibfnamefont {S.}~\bibnamefont
  {Saha}}, \bibinfo {author} {\bibfnamefont {R.}~\bibnamefont {Golestanian}}, \
  and\ \bibinfo {author} {\bibfnamefont {C.~C.}\ \bibnamefont {Maass}},\
  }\bibfield  {title} {\enquote {\bibinfo {title} {Chemotactic self-caging in
  active emulsions},}\ }\href {\doibase 10.1073/pnas.2122269119} {\bibfield
  {journal} {\bibinfo  {journal} {PNAS}\ }\textbf {\bibinfo {volume} {119}},\
  \bibinfo {pages} {e2122269119} (\bibinfo {year}
  {2022}{\natexlab{b}})}\BibitemShut {NoStop}%
\bibitem [{\citenamefont {Picella}\ and\ \citenamefont
  {Michelin}(2022)}]{picella2022_confined}%
  \BibitemOpen
  \bibfield  {author} {\bibinfo {author} {\bibfnamefont {F.}~\bibnamefont
  {Picella}}\ and\ \bibinfo {author} {\bibfnamefont {S.}~\bibnamefont
  {Michelin}},\ }\bibfield  {title} {\enquote {\bibinfo {title} {Confined
  self-propulsion of an isotropic active colloid},}\ }\href {\doibase
  10.1017/jfm.2021.1081} {\bibfield  {journal} {\bibinfo  {journal} {Journal of
  Fluid Mechanics}\ }\textbf {\bibinfo {volume} {933}} (\bibinfo {year}
  {2022}),\ 10.1017/jfm.2021.1081}\BibitemShut {NoStop}%
\bibitem [{\citenamefont {Schmitt}\ and\ \citenamefont
  {Stark}(2013)}]{schmitt2013_swimming}%
  \BibitemOpen
  \bibfield  {author} {\bibinfo {author} {\bibfnamefont {M.}~\bibnamefont
  {Schmitt}}\ and\ \bibinfo {author} {\bibfnamefont {H.}~\bibnamefont
  {Stark}},\ }\bibfield  {title} {\enquote {\bibinfo {title} {Swimming active
  droplet: {{A}} theoretical analysis},}\ }\href {\doibase
  10.1209/0295-5075/101/44008} {\bibfield  {journal} {\bibinfo  {journal}
  {EPL}\ }\textbf {\bibinfo {volume} {101}},\ \bibinfo {pages} {44008}
  (\bibinfo {year} {2013})}\BibitemShut {NoStop}%
\bibitem [{\citenamefont {Peng}\ and\ \citenamefont
  {Schnitzer}(2023)}]{peng2023_weakly}%
  \BibitemOpen
  \bibfield  {author} {\bibinfo {author} {\bibfnamefont {G.~G.}\ \bibnamefont
  {Peng}}\ and\ \bibinfo {author} {\bibfnamefont {O.}~\bibnamefont
  {Schnitzer}},\ }\bibfield  {title} {\enquote {\bibinfo {title} {Weakly
  nonlinear dynamics of a chemically active particle near the threshold for
  spontaneous motion. {{II}}. {{History-dependent}} motion},}\ }\href {\doibase
  10.1103/PhysRevFluids.8.033602} {\bibfield  {journal} {\bibinfo  {journal}
  {Phys. Rev. Fluids}\ }\textbf {\bibinfo {volume} {8}},\ \bibinfo {pages}
  {033602} (\bibinfo {year} {2023})}\BibitemShut {NoStop}%
\bibitem [{\citenamefont {Anderson}(1989)}]{anderson1989_colloid}%
  \BibitemOpen
  \bibfield  {author} {\bibinfo {author} {\bibfnamefont {J.~L.}\ \bibnamefont
  {Anderson}},\ }\bibfield  {title} {\enquote {\bibinfo {title} {Colloid
  transport by interfacial forces},}\ }\href {\doibase
  10.1146/annurev.fluid.21.1.61} {\bibfield  {journal} {\bibinfo  {journal}
  {Annual Review of Fluid Mechanics}\ }\textbf {\bibinfo {volume} {21}},\
  \bibinfo {pages} {61--99} (\bibinfo {year} {1989})}\BibitemShut {NoStop}%
\bibitem [{\citenamefont {Babu}\ \emph {et~al.}(2021)\citenamefont {Babu},
  \citenamefont {Scanes}, \citenamefont {Plamont}, \citenamefont {Ryabchun},
  \citenamefont {Lancia}, \citenamefont {Kudernac}, \citenamefont {Fletcher},\
  and\ \citenamefont {Katsonis}}]{babu2021_acceleration}%
  \BibitemOpen
  \bibfield  {author} {\bibinfo {author} {\bibfnamefont {D.}~\bibnamefont
  {Babu}}, \bibinfo {author} {\bibfnamefont {R.~J.~H.}\ \bibnamefont {Scanes}},
  \bibinfo {author} {\bibfnamefont {R.}~\bibnamefont {Plamont}}, \bibinfo
  {author} {\bibfnamefont {A.}~\bibnamefont {Ryabchun}}, \bibinfo {author}
  {\bibfnamefont {F.}~\bibnamefont {Lancia}}, \bibinfo {author} {\bibfnamefont
  {T.}~\bibnamefont {Kudernac}}, \bibinfo {author} {\bibfnamefont {S.~P.}\
  \bibnamefont {Fletcher}}, \ and\ \bibinfo {author} {\bibfnamefont
  {N.}~\bibnamefont {Katsonis}},\ }\bibfield  {title} {\enquote {\bibinfo
  {title} {Acceleration of lipid reproduction by emergence of microscopic
  motion},}\ }\href {\doibase 10.1038/s41467-021-23022-1} {\bibfield  {journal}
  {\bibinfo  {journal} {Nat Commun}\ }\textbf {\bibinfo {volume} {12}},\
  \bibinfo {pages} {2959} (\bibinfo {year} {2021})}\BibitemShut {NoStop}%
\bibitem [{\citenamefont {Wentworth}\ \emph {et~al.}(2022)\citenamefont
  {Wentworth}, \citenamefont {Castonguay}, \citenamefont {Moerman},
  \citenamefont {Meredith}, \citenamefont {Balaj}, \citenamefont {Cheon},\ and\
  \citenamefont {Zarzar}}]{wentworth2022_chemically}%
  \BibitemOpen
  \bibfield  {author} {\bibinfo {author} {\bibfnamefont {C.~M.}\ \bibnamefont
  {Wentworth}}, \bibinfo {author} {\bibfnamefont {A.~C.}\ \bibnamefont
  {Castonguay}}, \bibinfo {author} {\bibfnamefont {P.~G.}\ \bibnamefont
  {Moerman}}, \bibinfo {author} {\bibfnamefont {C.~H.}\ \bibnamefont
  {Meredith}}, \bibinfo {author} {\bibfnamefont {R.~V.}\ \bibnamefont {Balaj}},
  \bibinfo {author} {\bibfnamefont {S.~I.}\ \bibnamefont {Cheon}}, \ and\
  \bibinfo {author} {\bibfnamefont {L.~D.}\ \bibnamefont {Zarzar}},\ }\bibfield
   {title} {\enquote {\bibinfo {title} {Chemically {{Tuning Attractive}} and
  {{Repulsive Interactions}} between {{Solubilizing Oil Droplets}}},}\ }\href
  {\doibase 10.1002/anie.202204510} {\bibfield  {journal} {\bibinfo  {journal}
  {Angewandte Chemie International Edition}\ }\textbf {\bibinfo {volume}
  {61}},\ \bibinfo {pages} {e202204510} (\bibinfo {year} {2022})}\BibitemShut
  {NoStop}%
\bibitem [{\citenamefont {Parmar}\ \emph {et~al.}(2014)\citenamefont {Parmar},
  \citenamefont {Chavda},\ and\ \citenamefont {Bahadur}}]{parmar2014_pluronic}%
  \BibitemOpen
  \bibfield  {author} {\bibinfo {author} {\bibfnamefont {A.}~\bibnamefont
  {Parmar}}, \bibinfo {author} {\bibfnamefont {S.}~\bibnamefont {Chavda}}, \
  and\ \bibinfo {author} {\bibfnamefont {P.}~\bibnamefont {Bahadur}},\
  }\bibfield  {title} {\enquote {\bibinfo {title} {Pluronic--cationic
  surfactant mixed micelles: {{Solubilization}} and release of the drug
  hydrochlorothiazide},}\ }\href {\doibase 10.1016/j.colsurfa.2013.09.018}
  {\bibfield  {journal} {\bibinfo  {journal} {Colloids and Surfaces A:
  Physicochemical and Engineering Aspects}\ }\textbf {\bibinfo {volume}
  {441}},\ \bibinfo {pages} {389--397} (\bibinfo {year} {2014})}\BibitemShut
  {NoStop}%
\bibitem [{\citenamefont {Thutupalli}\ \emph {et~al.}(2018)\citenamefont
  {Thutupalli}, \citenamefont {Geyer}, \citenamefont {Singh}, \citenamefont
  {Adhikari},\ and\ \citenamefont {Stone}}]{thutupalli2018_flowinduced}%
  \BibitemOpen
  \bibfield  {author} {\bibinfo {author} {\bibfnamefont {S.}~\bibnamefont
  {Thutupalli}}, \bibinfo {author} {\bibfnamefont {D.}~\bibnamefont {Geyer}},
  \bibinfo {author} {\bibfnamefont {R.}~\bibnamefont {Singh}}, \bibinfo
  {author} {\bibfnamefont {R.}~\bibnamefont {Adhikari}}, \ and\ \bibinfo
  {author} {\bibfnamefont {H.~A.}\ \bibnamefont {Stone}},\ }\bibfield  {title}
  {\enquote {\bibinfo {title} {Flow-induced phase separation of active
  particles is controlled by boundary conditions},}\ }\href {\doibase
  10.1073/pnas.1718807115} {\bibfield  {journal} {\bibinfo  {journal} {PNAS}\
  }\textbf {\bibinfo {volume} {115}},\ \bibinfo {pages} {5403--5408} (\bibinfo
  {year} {2018})}\BibitemShut {NoStop}%
\bibitem [{\citenamefont {Hecht}\ and\ \citenamefont
  {Hoffmann}(1995)}]{hecht1995_kinetic}%
  \BibitemOpen
  \bibfield  {author} {\bibinfo {author} {\bibfnamefont {E.}~\bibnamefont
  {Hecht}}\ and\ \bibinfo {author} {\bibfnamefont {H.}~\bibnamefont
  {Hoffmann}},\ }\bibfield  {title} {\enquote {\bibinfo {title} {Kinetic and
  calorimetric investigations on micelle formation of block copolymers of the
  poloxamer type},}\ }\href {\doibase 10.1016/0927-7757(94)03044-Z} {\bibfield
  {journal} {\bibinfo  {journal} {Colloids and Surfaces A: Physicochemical and
  Engineering Aspects}\ }\textbf {\bibinfo {volume} {96}},\ \bibinfo {pages}
  {181--197} (\bibinfo {year} {1995})}\BibitemShut {NoStop}%
\bibitem [{\citenamefont {Chen}\ \emph {et~al.}(2021)\citenamefont {Chen},
  \citenamefont {Chong}, \citenamefont {Liu}, \citenamefont {Verzicco},\ and\
  \citenamefont {Lohse}}]{chen2021_instabilities}%
  \BibitemOpen
  \bibfield  {author} {\bibinfo {author} {\bibfnamefont {Y.}~\bibnamefont
  {Chen}}, \bibinfo {author} {\bibfnamefont {K.~L.}\ \bibnamefont {Chong}},
  \bibinfo {author} {\bibfnamefont {L.}~\bibnamefont {Liu}}, \bibinfo {author}
  {\bibfnamefont {R.}~\bibnamefont {Verzicco}}, \ and\ \bibinfo {author}
  {\bibfnamefont {D.}~\bibnamefont {Lohse}},\ }\bibfield  {title} {\enquote
  {\bibinfo {title} {Instabilities driven by diffusiophoretic flow on catalytic
  surfaces},}\ }\href {\doibase 10.1017/jfm.2021.370} {\bibfield  {journal}
  {\bibinfo  {journal} {Journal of Fluid Mechanics}\ }\textbf {\bibinfo
  {volume} {919}},\ \bibinfo {pages} {A10} (\bibinfo {year}
  {2021})}\BibitemShut {NoStop}%
\bibitem [{\citenamefont {Wang}\ \emph
  {et~al.}(2024{\natexlab{a}})\citenamefont {Wang}, \citenamefont {Zheng},\
  and\ \citenamefont {Li}}]{wang2024_selfpropulsion}%
  \BibitemOpen
  \bibfield  {author} {\bibinfo {author} {\bibfnamefont {Y.}~\bibnamefont
  {Wang}}, \bibinfo {author} {\bibfnamefont {L.}~\bibnamefont {Zheng}}, \ and\
  \bibinfo {author} {\bibfnamefont {G.}~\bibnamefont {Li}},\ }\bibfield
  {title} {\enquote {\bibinfo {title} {Self-propulsion of a droplet induced by
  combined diffusiophoresis and {{Marangoni}} effects},}\ }\href {\doibase
  10.1002/elps.202400005} {\bibfield  {journal} {\bibinfo  {journal}
  {ELECTROPHORESIS}\ } (\bibinfo {year} {2024}{\natexlab{a}}),\
  10.1002/elps.202400005}\BibitemShut {NoStop}%
\bibitem [{\citenamefont {{de Blois}}\ \emph {et~al.}(2021)\citenamefont {{de
  Blois}}, \citenamefont {Bertin}, \citenamefont {Suda}, \citenamefont
  {Ichikawa}, \citenamefont {Reyssat},\ and\ \citenamefont
  {Dauchot}}]{deblois2021_swimming}%
  \BibitemOpen
  \bibfield  {author} {\bibinfo {author} {\bibfnamefont {C.}~\bibnamefont {{de
  Blois}}}, \bibinfo {author} {\bibfnamefont {V.}~\bibnamefont {Bertin}},
  \bibinfo {author} {\bibfnamefont {S.}~\bibnamefont {Suda}}, \bibinfo {author}
  {\bibfnamefont {M.}~\bibnamefont {Ichikawa}}, \bibinfo {author}
  {\bibfnamefont {M.}~\bibnamefont {Reyssat}}, \ and\ \bibinfo {author}
  {\bibfnamefont {O.}~\bibnamefont {Dauchot}},\ }\bibfield  {title} {\enquote
  {\bibinfo {title} {Swimming droplets in {{1D}} geometries: An active
  {{Bretherton}} problem},}\ }\href {\doibase 10.1039/D1SM00387A} {\bibfield
  {journal} {\bibinfo  {journal} {Soft Matter}\ }\textbf {\bibinfo {volume}
  {17}},\ \bibinfo {pages} {6646--6660} (\bibinfo {year} {2021})},\ \Eprint
  {http://arxiv.org/abs/2103.09513} {arXiv:2103.09513} \BibitemShut {NoStop}%
\bibitem [{\citenamefont {Kr{\"u}ger}\ \emph {et~al.}(2016)\citenamefont
  {Kr{\"u}ger}, \citenamefont {Bahr}, \citenamefont {Herminghaus},\ and\
  \citenamefont {Maass}}]{kruger2016_dimensionality}%
  \BibitemOpen
  \bibfield  {author} {\bibinfo {author} {\bibfnamefont {C.}~\bibnamefont
  {Kr{\"u}ger}}, \bibinfo {author} {\bibfnamefont {C.}~\bibnamefont {Bahr}},
  \bibinfo {author} {\bibfnamefont {S.}~\bibnamefont {Herminghaus}}, \ and\
  \bibinfo {author} {\bibfnamefont {C.~C.}\ \bibnamefont {Maass}},\ }\bibfield
  {title} {\enquote {\bibinfo {title} {Dimensionality matters in the collective
  behaviour of active emulsions},}\ }\href {\doibase
  10.1140/epje/i2016-16064-y} {\bibfield  {journal} {\bibinfo  {journal} {Eur.
  Phys. J. E}\ }\textbf {\bibinfo {volume} {39}},\ \bibinfo {pages} {64}
  (\bibinfo {year} {2016})}\BibitemShut {NoStop}%
\bibitem [{\citenamefont {Moerman}\ \emph {et~al.}(2017)\citenamefont
  {Moerman}, \citenamefont {Moyses}, \citenamefont {{van der Wee}},
  \citenamefont {Grier}, \citenamefont {{van Blaaderen}}, \citenamefont
  {Kegel}, \citenamefont {Groenewold},\ and\ \citenamefont
  {Brujic}}]{moerman2017_solutemediated}%
  \BibitemOpen
  \bibfield  {author} {\bibinfo {author} {\bibfnamefont {P.~G.}\ \bibnamefont
  {Moerman}}, \bibinfo {author} {\bibfnamefont {H.~W.}\ \bibnamefont {Moyses}},
  \bibinfo {author} {\bibfnamefont {E.~B.}\ \bibnamefont {{van der Wee}}},
  \bibinfo {author} {\bibfnamefont {D.~G.}\ \bibnamefont {Grier}}, \bibinfo
  {author} {\bibfnamefont {A.}~\bibnamefont {{van Blaaderen}}}, \bibinfo
  {author} {\bibfnamefont {W.~K.}\ \bibnamefont {Kegel}}, \bibinfo {author}
  {\bibfnamefont {J.}~\bibnamefont {Groenewold}}, \ and\ \bibinfo {author}
  {\bibfnamefont {J.}~\bibnamefont {Brujic}},\ }\bibfield  {title} {\enquote
  {\bibinfo {title} {Solute-mediated interactions between active droplets},}\
  }\href {\doibase 10.1103/PhysRevE.96.032607} {\bibfield  {journal} {\bibinfo
  {journal} {Phys. Rev. E}\ }\textbf {\bibinfo {volume} {96}},\ \bibinfo
  {pages} {032607} (\bibinfo {year} {2017})}\BibitemShut {NoStop}%
\bibitem [{\citenamefont {Lagzi}\ \emph {et~al.}(2010)\citenamefont {Lagzi},
  \citenamefont {Soh}, \citenamefont {Wesson}, \citenamefont {Browne},\ and\
  \citenamefont {Grzybowski}}]{lagzi2010_maze}%
  \BibitemOpen
  \bibfield  {author} {\bibinfo {author} {\bibfnamefont {I.}~\bibnamefont
  {Lagzi}}, \bibinfo {author} {\bibfnamefont {S.}~\bibnamefont {Soh}}, \bibinfo
  {author} {\bibfnamefont {P.~J.}\ \bibnamefont {Wesson}}, \bibinfo {author}
  {\bibfnamefont {K.~P.}\ \bibnamefont {Browne}}, \ and\ \bibinfo {author}
  {\bibfnamefont {B.~A.}\ \bibnamefont {Grzybowski}},\ }\bibfield  {title}
  {\enquote {\bibinfo {title} {Maze solving by chemotactic droplets},}\
  }\href@noop {} {\bibfield  {journal} {\bibinfo  {journal} {Journal of the
  American Chemical Society}\ }\textbf {\bibinfo {volume} {132}},\ \bibinfo
  {pages} {1198--1199} (\bibinfo {year} {2010})}\BibitemShut {NoStop}%
\bibitem [{\citenamefont {Dey}\ \emph {et~al.}(2022)\citenamefont {Dey},
  \citenamefont {Buness}, \citenamefont {Hokmabad}, \citenamefont {Jin},\ and\
  \citenamefont {Maass}}]{dey2022_oscillatory}%
  \BibitemOpen
  \bibfield  {author} {\bibinfo {author} {\bibfnamefont {R.}~\bibnamefont
  {Dey}}, \bibinfo {author} {\bibfnamefont {C.~M.}\ \bibnamefont {Buness}},
  \bibinfo {author} {\bibfnamefont {B.~V.}\ \bibnamefont {Hokmabad}}, \bibinfo
  {author} {\bibfnamefont {C.}~\bibnamefont {Jin}}, \ and\ \bibinfo {author}
  {\bibfnamefont {C.~C.}\ \bibnamefont {Maass}},\ }\bibfield  {title} {\enquote
  {\bibinfo {title} {Oscillatory rheotaxis of artificial swimmers in
  microchannels},}\ }\href {\doibase 10.1038/s41467-022-30611-1} {\bibfield
  {journal} {\bibinfo  {journal} {Nat Commun}\ }\textbf {\bibinfo {volume}
  {13}},\ \bibinfo {pages} {2952} (\bibinfo {year} {2022})}\BibitemShut
  {NoStop}%
\bibitem [{\citenamefont {Dwivedi}\ \emph {et~al.}(2021)\citenamefont
  {Dwivedi}, \citenamefont {Shrivastava}, \citenamefont {Pillai},\ and\
  \citenamefont {Mangal}}]{dwivedi2021_rheotaxis}%
  \BibitemOpen
  \bibfield  {author} {\bibinfo {author} {\bibfnamefont {P.}~\bibnamefont
  {Dwivedi}}, \bibinfo {author} {\bibfnamefont {A.}~\bibnamefont
  {Shrivastava}}, \bibinfo {author} {\bibfnamefont {D.}~\bibnamefont {Pillai}},
  \ and\ \bibinfo {author} {\bibfnamefont {R.}~\bibnamefont {Mangal}},\
  }\bibfield  {title} {\enquote {\bibinfo {title} {Rheotaxis of active
  droplets},}\ }\href {\doibase 10.1063/5.0060952} {\bibfield  {journal}
  {\bibinfo  {journal} {Physics of Fluids}\ }\textbf {\bibinfo {volume} {33}},\
  \bibinfo {pages} {082108} (\bibinfo {year} {2021})}\BibitemShut {NoStop}%
\bibitem [{\citenamefont {Buness}\ \emph {et~al.}(2024)\citenamefont {Buness},
  \citenamefont {Rana}, \citenamefont {Maass},\ and\ \citenamefont
  {Dey}}]{buness2024_electrotaxis}%
  \BibitemOpen
  \bibfield  {author} {\bibinfo {author} {\bibfnamefont {C.~M.}\ \bibnamefont
  {Buness}}, \bibinfo {author} {\bibfnamefont {A.}~\bibnamefont {Rana}},
  \bibinfo {author} {\bibfnamefont {C.~C.}\ \bibnamefont {Maass}}, \ and\
  \bibinfo {author} {\bibfnamefont {R.}~\bibnamefont {Dey}},\ }\bibfield
  {title} {\enquote {\bibinfo {title} {Electrotaxis of {{Self-Propelling
  Artificial Swimmers}} in {{Microchannels}}},}\ }\href {\doibase
  10.1103/PhysRevLett.133.158301} {\bibfield  {journal} {\bibinfo  {journal}
  {Phys. Rev. Lett.}\ }\textbf {\bibinfo {volume} {133}},\ \bibinfo {pages}
  {158301} (\bibinfo {year} {2024})}\BibitemShut {NoStop}%
\bibitem [{\citenamefont {Wagner}\ \emph {et~al.}(2024)\citenamefont {Wagner},
  \citenamefont {Domburg}, \citenamefont {Kr{\"u}ger}, \citenamefont {Meyer},
  \citenamefont {Zhang}, \citenamefont {Ramesh},\ and\ \citenamefont
  {Maass}}]{wagner2024_magnetotaxis}%
  \BibitemOpen
  \bibfield  {author} {\bibinfo {author} {\bibfnamefont {M.~W.}\ \bibnamefont
  {Wagner}}, \bibinfo {author} {\bibfnamefont {F.}~\bibnamefont {Domburg}},
  \bibinfo {author} {\bibfnamefont {C.}~\bibnamefont {Kr{\"u}ger}}, \bibinfo
  {author} {\bibfnamefont {J.}~\bibnamefont {Meyer}}, \bibinfo {author}
  {\bibfnamefont {J.}~\bibnamefont {Zhang}}, \bibinfo {author} {\bibfnamefont
  {P.}~\bibnamefont {Ramesh}}, \ and\ \bibinfo {author} {\bibfnamefont {C.~C.}\
  \bibnamefont {Maass}},\ }\href {\doibase 10.48550/arXiv.2409.14558} {\enquote
  {\bibinfo {title} {Magnetotaxis in droplet microswimmers},}\ } (\bibinfo
  {year} {2024}),\ \Eprint {http://arxiv.org/abs/2409.14558} {arXiv:2409.14558
  [cond-mat]} \BibitemShut {NoStop}%
\bibitem [{\citenamefont {Khan}\ \emph {et~al.}(2024)\citenamefont {Khan},
  \citenamefont {Gardi}, \citenamefont {Soon}, \citenamefont {Zhang},\ and\
  \citenamefont {Sitti}}]{khan2024_perturbing}%
  \BibitemOpen
  \bibfield  {author} {\bibinfo {author} {\bibfnamefont {M.~T.~A.}\
  \bibnamefont {Khan}}, \bibinfo {author} {\bibfnamefont {G.}~\bibnamefont
  {Gardi}}, \bibinfo {author} {\bibfnamefont {R.~H.}\ \bibnamefont {Soon}},
  \bibinfo {author} {\bibfnamefont {M.}~\bibnamefont {Zhang}}, \ and\ \bibinfo
  {author} {\bibfnamefont {M.}~\bibnamefont {Sitti}},\ }\bibfield  {title}
  {\enquote {\bibinfo {title} {Perturbing {{Dynamics}} of {{Active Emulsions}}
  and {{Their Collectives}}},}\ }\href@noop {} {\bibfield  {journal} {\bibinfo
  {journal} {arXiv:2405.05889}\ } (\bibinfo {year} {2024})},\ \Eprint
  {http://arxiv.org/abs/2405.05889} {arXiv:2405.05889 [cond-mat]} \BibitemShut
  {NoStop}%
\bibitem [{\citenamefont {Balaj}\ and\ \citenamefont
  {Zarzar}(2020)}]{balaj2020_reconfigurable}%
  \BibitemOpen
  \bibfield  {author} {\bibinfo {author} {\bibfnamefont {R.~V.}\ \bibnamefont
  {Balaj}}\ and\ \bibinfo {author} {\bibfnamefont {L.~D.}\ \bibnamefont
  {Zarzar}},\ }\bibfield  {title} {\enquote {\bibinfo {title} {Reconfigurable
  complex emulsions: {{Design}}, properties, and applications},}\ }\href
  {\doibase 10.1063/5.0028606} {\bibfield  {journal} {\bibinfo  {journal}
  {Chem. Phys. Rev.}\ }\textbf {\bibinfo {volume} {1}},\ \bibinfo {pages}
  {011301} (\bibinfo {year} {2020})}\BibitemShut {NoStop}%
\bibitem [{\citenamefont {Ik~Cheon}\ \emph {et~al.}(2021)\citenamefont
  {Ik~Cheon}, \citenamefont {Capaverde~Silva}, \citenamefont {S.~Khair},\ and\
  \citenamefont {D.~Zarzar}}]{ikcheon2021_interfaciallyadsorbed}%
  \BibitemOpen
  \bibfield  {author} {\bibinfo {author} {\bibfnamefont {S.}~\bibnamefont
  {Ik~Cheon}}, \bibinfo {author} {\bibfnamefont {L.~B.}\ \bibnamefont
  {Capaverde~Silva}}, \bibinfo {author} {\bibfnamefont {A.}~\bibnamefont
  {S.~Khair}}, \ and\ \bibinfo {author} {\bibfnamefont {L.}~\bibnamefont
  {D.~Zarzar}},\ }\bibfield  {title} {\enquote {\bibinfo {title}
  {Interfacially-adsorbed particles enhance the self-propulsion of oil droplets
  in aqueous surfactant},}\ }\href {\doibase 10.1039/D0SM02234A} {\bibfield
  {journal} {\bibinfo  {journal} {Soft Matter}\ }\textbf {\bibinfo {volume}
  {17}},\ \bibinfo {pages} {6742--6750} (\bibinfo {year} {2021})}\BibitemShut
  {NoStop}%
\bibitem [{\citenamefont {Poulin}\ \emph {et~al.}(1997)\citenamefont {Poulin},
  \citenamefont {Stark}, \citenamefont {Lubensky},\ and\ \citenamefont
  {Weitz}}]{poulin1997_novel}%
  \BibitemOpen
  \bibfield  {author} {\bibinfo {author} {\bibfnamefont {P.}~\bibnamefont
  {Poulin}}, \bibinfo {author} {\bibfnamefont {H.}~\bibnamefont {Stark}},
  \bibinfo {author} {\bibfnamefont {T.~C.}\ \bibnamefont {Lubensky}}, \ and\
  \bibinfo {author} {\bibfnamefont {D.~A.}\ \bibnamefont {Weitz}},\ }\bibfield
  {title} {\enquote {\bibinfo {title} {Novel colloidal interactions in
  anisotropic fluids},}\ }\href {\doibase 10.1126/science.275.5307.1770}
  {\bibfield  {journal} {\bibinfo  {journal} {Science}\ }\textbf {\bibinfo
  {volume} {275}},\ \bibinfo {pages} {1770} (\bibinfo {year}
  {1997})}\BibitemShut {NoStop}%
\bibitem [{\citenamefont {Hokmabad}\ \emph {et~al.}(2019)\citenamefont
  {Hokmabad}, \citenamefont {Baldwin}, \citenamefont {Kr{\"u}ger},
  \citenamefont {Bahr},\ and\ \citenamefont
  {Maass}}]{hokmabad2019_topological}%
  \BibitemOpen
  \bibfield  {author} {\bibinfo {author} {\bibfnamefont {B.~V.}\ \bibnamefont
  {Hokmabad}}, \bibinfo {author} {\bibfnamefont {K.~A.}\ \bibnamefont
  {Baldwin}}, \bibinfo {author} {\bibfnamefont {C.}~\bibnamefont {Kr{\"u}ger}},
  \bibinfo {author} {\bibfnamefont {C.}~\bibnamefont {Bahr}}, \ and\ \bibinfo
  {author} {\bibfnamefont {C.~C.}\ \bibnamefont {Maass}},\ }\bibfield  {title}
  {\enquote {\bibinfo {title} {Topological {{Stabilization}} and {{Dynamics}}
  of {{Self-Propelling Nematic Shells}}},}\ }\href {\doibase
  10.1103/PhysRevLett.123.178003} {\bibfield  {journal} {\bibinfo  {journal}
  {Phys. Rev. Lett.}\ }\textbf {\bibinfo {volume} {123}},\ \bibinfo {pages}
  {178003} (\bibinfo {year} {2019})}\BibitemShut {NoStop}%
\bibitem [{\citenamefont {Wang}\ \emph
  {et~al.}(2024{\natexlab{b}})\citenamefont {Wang}, \citenamefont {Yang},
  \citenamefont {Roh}, \citenamefont {Hormozi}, \citenamefont {Gianneschi},\
  and\ \citenamefont {Abbott}}]{wang2024_selftimed}%
  \BibitemOpen
  \bibfield  {author} {\bibinfo {author} {\bibfnamefont {X.}~\bibnamefont
  {Wang}}, \bibinfo {author} {\bibfnamefont {Y.}~\bibnamefont {Yang}}, \bibinfo
  {author} {\bibfnamefont {S.}~\bibnamefont {Roh}}, \bibinfo {author}
  {\bibfnamefont {S.}~\bibnamefont {Hormozi}}, \bibinfo {author} {\bibfnamefont
  {N.~C.}\ \bibnamefont {Gianneschi}}, \ and\ \bibinfo {author} {\bibfnamefont
  {N.~L.}\ \bibnamefont {Abbott}},\ }\bibfield  {title} {\enquote {\bibinfo
  {title} {Self-{{Timed}} and {{Spatially Targeted Delivery}} of {{Chemical
  Cargo}} by {{Motile Liquid Crystal}}},}\ }\href {\doibase
  10.1002/adma.202311311} {\bibfield  {journal} {\bibinfo  {journal} {Advanced
  Materials}\ }\textbf {\bibinfo {volume} {36}},\ \bibinfo {pages} {2311311}
  (\bibinfo {year} {2024}{\natexlab{b}})}\BibitemShut {NoStop}%
\bibitem [{\citenamefont {{Lopez-Leon}}\ and\ \citenamefont
  {{Fernandez-Nieves}}(2011)}]{lopez-leon2011_drops}%
  \BibitemOpen
  \bibfield  {author} {\bibinfo {author} {\bibfnamefont {T.}~\bibnamefont
  {{Lopez-Leon}}}\ and\ \bibinfo {author} {\bibfnamefont {A.}~\bibnamefont
  {{Fernandez-Nieves}}},\ }\bibfield  {title} {\enquote {\bibinfo {title}
  {Drops and shells of liquid crystal},}\ }\href@noop {} {\bibfield  {journal}
  {\bibinfo  {journal} {Colloid Polym. Sci.}\ }\textbf {\bibinfo {volume}
  {289}},\ \bibinfo {pages} {345--359} (\bibinfo {year} {2011})}\BibitemShut
  {NoStop}%
\bibitem [{\citenamefont {Shechter}\ \emph {et~al.}(2020)\citenamefont
  {Shechter}, \citenamefont {Atzin}, \citenamefont {Mozaffari}, \citenamefont
  {Zhang}, \citenamefont {Zhou}, \citenamefont {Strain}, \citenamefont {Oster},
  \citenamefont {{de Pablo}},\ and\ \citenamefont
  {Ross}}]{shechter2020_direct}%
  \BibitemOpen
  \bibfield  {author} {\bibinfo {author} {\bibfnamefont {J.}~\bibnamefont
  {Shechter}}, \bibinfo {author} {\bibfnamefont {N.}~\bibnamefont {Atzin}},
  \bibinfo {author} {\bibfnamefont {A.}~\bibnamefont {Mozaffari}}, \bibinfo
  {author} {\bibfnamefont {R.}~\bibnamefont {Zhang}}, \bibinfo {author}
  {\bibfnamefont {Y.}~\bibnamefont {Zhou}}, \bibinfo {author} {\bibfnamefont
  {B.}~\bibnamefont {Strain}}, \bibinfo {author} {\bibfnamefont {L.~M.}\
  \bibnamefont {Oster}}, \bibinfo {author} {\bibfnamefont {J.~J.}\ \bibnamefont
  {{de Pablo}}}, \ and\ \bibinfo {author} {\bibfnamefont {J.~L.}\ \bibnamefont
  {Ross}},\ }\bibfield  {title} {\enquote {\bibinfo {title} {Direct
  {{Observation}} of {{Liquid Crystal Droplet Configurational Transitions}}
  using {{Optical Tweezers}}},}\ }\href {\doibase 10.1021/acs.langmuir.9b03629}
  {\bibfield  {journal} {\bibinfo  {journal} {Langmuir}\ }\textbf {\bibinfo
  {volume} {36}},\ \bibinfo {pages} {7074--7082} (\bibinfo {year}
  {2020})}\BibitemShut {NoStop}%
\bibitem [{\citenamefont {Kim}\ \emph {et~al.}(2005)\citenamefont {Kim},
  \citenamefont {Manoharan},\ and\ \citenamefont
  {Crocker}}]{kim2005_swellingbased}%
  \BibitemOpen
  \bibfield  {author} {\bibinfo {author} {\bibfnamefont {A.~J.}\ \bibnamefont
  {Kim}}, \bibinfo {author} {\bibfnamefont {V.~N.}\ \bibnamefont {Manoharan}},
  \ and\ \bibinfo {author} {\bibfnamefont {J.~C.}\ \bibnamefont {Crocker}},\
  }\bibfield  {title} {\enquote {\bibinfo {title} {Swelling-{{Based Method}}
  for {{Preparing Stable}}, {{Functionalized Polymer Colloids}}},}\ }\href
  {\doibase 10.1021/ja0450051} {\bibfield  {journal} {\bibinfo  {journal} {J.
  Am. Chem. Soc.}\ }\textbf {\bibinfo {volume} {127}},\ \bibinfo {pages}
  {1592--1593} (\bibinfo {year} {2005})}\BibitemShut {NoStop}%
\bibitem [{\citenamefont {Thielicke}\ and\ \citenamefont
  {Stamhuis}(2014)}]{thielicke2014_pivlab}%
  \BibitemOpen
  \bibfield  {author} {\bibinfo {author} {\bibfnamefont {W.}~\bibnamefont
  {Thielicke}}\ and\ \bibinfo {author} {\bibfnamefont {E.}~\bibnamefont
  {Stamhuis}},\ }\bibfield  {title} {\enquote {\bibinfo {title} {{{PIVlab}} --
  {{Towards User-friendly}}, {{Affordable}} and {{Accurate Digital Particle
  Image Velocimetry}} in {{MATLAB}}},}\ }\href {\doibase 10.5334/jors.bl}
  {\bibfield  {journal} {\bibinfo  {journal} {JORS}\ }\textbf {\bibinfo
  {volume} {2}},\ \bibinfo {pages} {e30} (\bibinfo {year} {2014})}\BibitemShut
  {NoStop}%
\bibitem [{\citenamefont {Verzicco}\ and\ \citenamefont
  {Orlandi}(1996)}]{verzicco1996_finitedifference}%
  \BibitemOpen
  \bibfield  {author} {\bibinfo {author} {\bibfnamefont {R.}~\bibnamefont
  {Verzicco}}\ and\ \bibinfo {author} {\bibfnamefont {P.}~\bibnamefont
  {Orlandi}},\ }\bibfield  {title} {\enquote {\bibinfo {title} {A
  {{Finite-Difference Scheme}} for {{Three-Dimensional Incompressible Flows}}
  in {{Cylindrical Coordinates}}},}\ }\href {\doibase 10.1006/jcph.1996.0033}
  {\bibfield  {journal} {\bibinfo  {journal} {Journal of Computational
  Physics}\ }\textbf {\bibinfo {volume} {123}},\ \bibinfo {pages} {402--414}
  (\bibinfo {year} {1996})}\BibitemShut {NoStop}%
\bibitem [{\citenamefont {Spandan}\ \emph {et~al.}(2017)\citenamefont
  {Spandan}, \citenamefont {Meschini}, \citenamefont {{Ostilla-M{\'o}nico}},
  \citenamefont {Lohse}, \citenamefont {Querzoli}, \citenamefont {{de
  Tullio}},\ and\ \citenamefont {Verzicco}}]{spandan2017_parallel}%
  \BibitemOpen
  \bibfield  {author} {\bibinfo {author} {\bibfnamefont {V.}~\bibnamefont
  {Spandan}}, \bibinfo {author} {\bibfnamefont {V.}~\bibnamefont {Meschini}},
  \bibinfo {author} {\bibfnamefont {R.}~\bibnamefont {{Ostilla-M{\'o}nico}}},
  \bibinfo {author} {\bibfnamefont {D.}~\bibnamefont {Lohse}}, \bibinfo
  {author} {\bibfnamefont {G.}~\bibnamefont {Querzoli}}, \bibinfo {author}
  {\bibfnamefont {M.~D.}\ \bibnamefont {{de Tullio}}}, \ and\ \bibinfo {author}
  {\bibfnamefont {R.}~\bibnamefont {Verzicco}},\ }\bibfield  {title} {\enquote
  {\bibinfo {title} {A parallel interaction potential approach coupled with the
  immersed boundary method for fully resolved simulations of deformable
  interfaces and membranes},}\ }\href {\doibase 10.1016/j.jcp.2017.07.036}
  {\bibfield  {journal} {\bibinfo  {journal} {Journal of Computational
  Physics}\ }\textbf {\bibinfo {volume} {348}},\ \bibinfo {pages} {567--590}
  (\bibinfo {year} {2017})}\BibitemShut {NoStop}%
\end{thebibliography}
%

\cleardoublepage
\title{Frozen by heating: temperature controlled dynamic states in droplet microswimmers. Supporting information}

\beginsupplement

\section{Materials and characterization}
\begin{figure}
    \centering
    \includegraphics[width=.9\columnwidth]{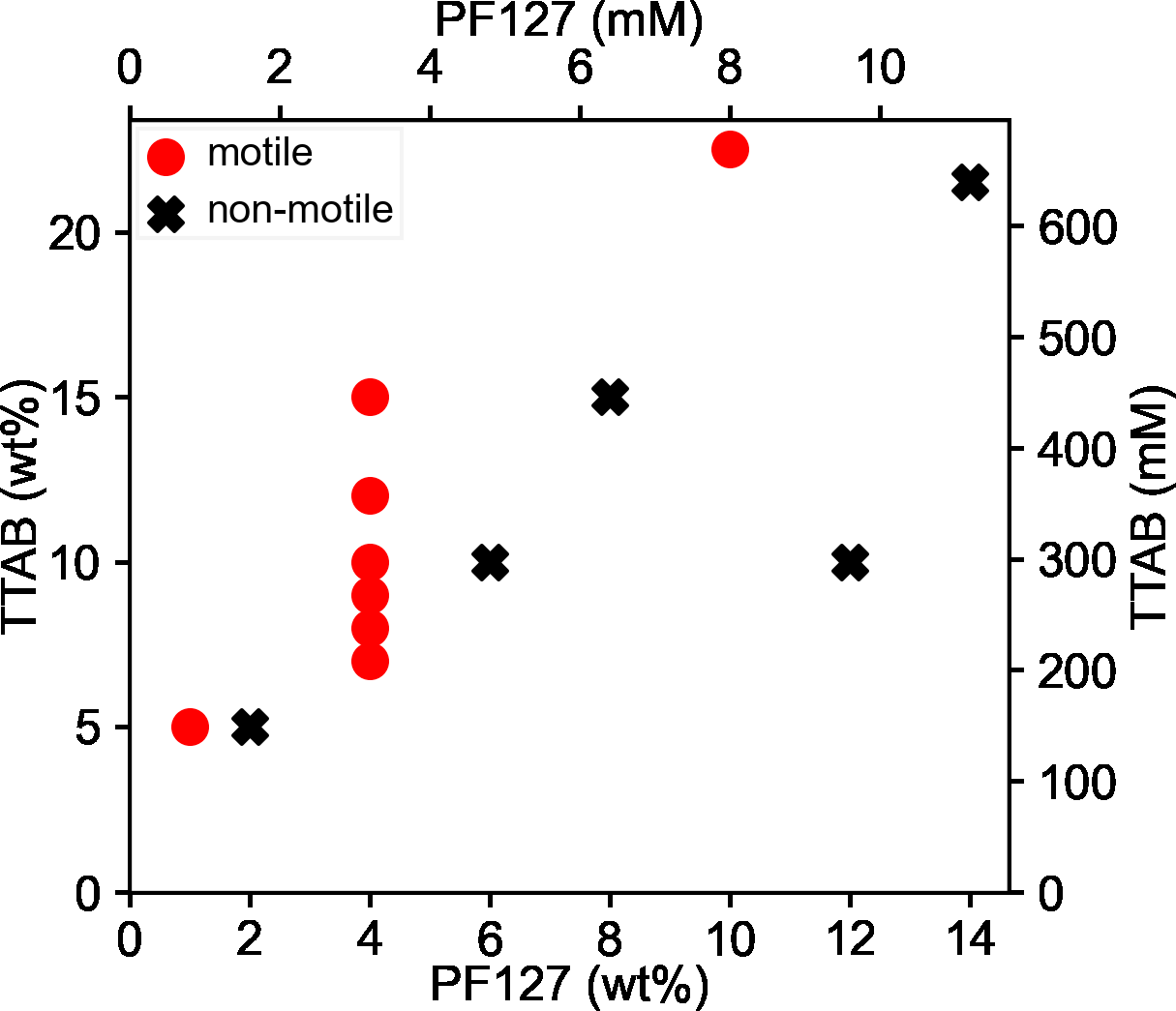}
    \caption{Motility chart for \SI{50}{\um} droplets in media of various surfactant composition, taken at room temperature.\label{SIfig:swimming}}
    
\end{figure}

We obtained CB15 (Synthon Chemicals), TTAB and PF127 (Sigma-Aldrich) and used them as-is. To study the influence of mixed micelles on active droplet motility around room temperature, we varied the TTAB concentration between \SIrange{9}{15}{\wtpc}. 
We note that a low concentration of PF127 requires heating the system to high temperatures to deplete TTAB, causing evaporation issues, while at a higher concentration of PF127 droplets would only be active at low temperatures.  This strongly limited the experimentally accessible parameter space for PF127, and we opted to keep the concentration fixed at \SI{4}{\wtpc}. An overview of the media compositions for which we observed swimming at room temperature (\SI{22}{\celsius}) is shown in Fig.\ \ref{SIfig:swimming}. Increasing the amount of PF127 suppresses motility, while an excess of TTAB promotes it. The typical molarity ratios for motility are of the order of 1:100.
\section{Microfluidic experiments}
We mass produced CB15 droplets in microfluidic flow focusing devices \cite{hokmabad2021_emergence} at a size of \SI{50\pm5}{\um}. We fabricated the experimental reservoir, a Hele-Shaw geometry, by spin coating a \SI{50}{\um} layer of SU-8 photoresist on glass, and creating a rectangular void space  of area $(13\times8)\:\rm{mm^{2}}$ by UV photolithography. For experiments,  we filled it with a dilute droplet emulsion and sealed it with a glass cover slip. The Rayleigh number, estimated for an aqueous medium, a typical length scale of $\SI{50}{\um}$ and $\Delta T=\SI{10}{\kelvin}$, is $\Ray\approx 10^{-2}$, ruling out thermal convection effects.
\par
We recorded the motion of active droplets on a  bright-field microscope (Olympus IX-81) with a temperature controlled stage (Linkam PE100) which allowed for both heating and cooling protocols. We set the reservoir temperature at the desired initial value for a period of \SI{5}{\minute}, after which we started measurements. 

\begin{figure}
    \centering
    \includegraphics[width=.7\columnwidth]{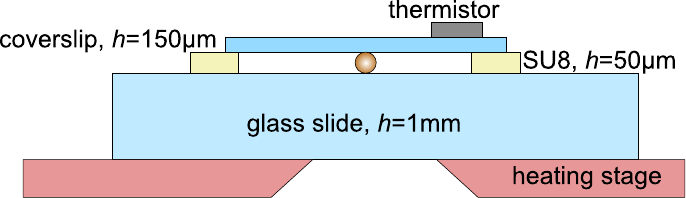}
    \caption{Cell dimensions: in addition to the heights specified in the figure, the rectangular sample reservoir in the SU8 photoresist layer had an area of $\SI{13}{\mm}\times\SI{8}{\mm}$, and the heating stage had a circular hole for observation of diameter \SI{15}{\mm}.\label{SIfig:celldetails}}
\end{figure}

We performed temperature ramp experiments with a set heating/cooling rate of \SI{1}{\kelvin/\minute}, \SI{5}{\kelvin/\minute} or \SI{10}{\kelvin/\minute}. As apparent in Fig. 2 (c), the system did not equilibrate instantaneously. We therefore recorded the temperature $T_\text{meas}\approx T_\text{sample}$ using a thermistor taped to the cover slip sealing the microfluidic reservoir (Fig. 1). 
The location of the thermistor was separated from the sample volume by a coverslip of thickness $\approx\SI{150}{\um}$, leading to a systematic deviation in the estimated sample temperature $T_\text{sample}$ due to the gradient between stage, $T_\text{stage}$, and room temperature, RT, through the cell (height \SI{1.25}{\mm}, see Fig.~\ref{SIfig:celldetails}). Here, for the movies S1 and S3, $T_\text{set}$ varied between \SI{15}{\celsius} and \SI{28}{\celsius}, and $T_\text{meas}$ between \SI{16.2\pm0.1}{\celsius} and \SI{27.3\pm0.1}{\celsius} with RT at \SI{24}{\celsius}. Assuming a linear temperature profile via conductive heat transfer, the maximum systematic error would be on the order of \SI{0.2}{\celsius}, thus within the experimental error of the transitions documented in Fig. 2. 

Movie frames were recorded at 4 frames per second (fps) using a Canon digital camera (EOS 600D) with a digital resolution of 1920 x 1080 px. 

We chose heating/cooling rates based on the experimental parameters and the required degree of quantitation. The minimum rate was limited by droplet shrinkage over time. Slower heating rates were feasible for single-cycle experiments, low TTAB concentrations and temperatures. Table \ref{SItab:rates} summarizes the protocols for the data underlying Fig. 2-4 and the corresponding supporting videos. For the best determination of $T_\text{start},T_\text{straight},$ and $T_\text{stop}$, the rate was kept as low as possible, we ensured via the recorded droplet sizes that the polydispersity was low, \SI{50\pm5}{\um}, and used small number densities ($<\SI{0.1}{mm^{-2}}$) to  exclude mutual trail interactions.
\begin{table}
    \noindent\begin{tabular}{|p{.2\columnwidth}|p{.7\columnwidth}|}
    \hline\textbf{Plot} & \textbf{Settings} \\\hline
   Fig.\ 2a      &  \SI{1}{\kelvin/\minute} \\\hline
     Fig.\ 2e-g    & \SI{10}{\kelvin/\minute}\\\hline
     Fig.\ 2b,c & \SI{5}{\kelvin/\minute} \\\hline
     Fig.\ 3 & \SI{5}{\kelvin/\minute}\\\hline
     Fig.\ 4 & stepwise heating, $\approx \SI{3}{\minute}$ equilibration before data acquisition\\\hline
    \end{tabular}
    \caption{Summary of heating rates/protocols for the data underlying the figures in the manuscript.\label{SItab:rates}}
    
\end{table}

To estimate whether the quasi 2D confinement affects the observed dynamics, we have also recorded the temperature-dependent droplet motion in reservoirs with a larger height of \SI{310}{\um} and \SI{2}{\mm}, using a swimming medium with \SI{10}{\wtpc} TTAB and \SI{4}{\wtpc} PF127. We found similar dynamic states and, within the experimental error (see Fig. 2b), the same transition temperatures ($T_\text{straight}\approx\SI{23}{\celsius}$, $T_\text{stop}\approx\SI{27}{\celsius}$) as in the case of the quasi-2D geometry, $h=\SI{50}{\um}$ (Fig.\ 2b).

\section{Investigation of TTAB/PF127 micelle formation}\label{SIsec:mixedmicelles}
To investigate the formation of individual and mixed micelles in our particular swimming medium, we recorded dynamic light scattering and calorimetry data on solutions of pure and mixed surfactants. 
The nature of these complexes is according to existing literature highly dependent on concentrations and temperature, however, at the relative concentrations exceeding 1:100 in our system,  PF127 micelles are presumably fully broken down for the entire temperature range under investigation~\cite{li2001_binding}. For this reason, we do not expect the  CMT of PF127 to be a characteristic quantity in the dynamics of our system. 
\subsection{DLS measurements}
\begin{figure*}
\centering
\includegraphics[width=1\linewidth]{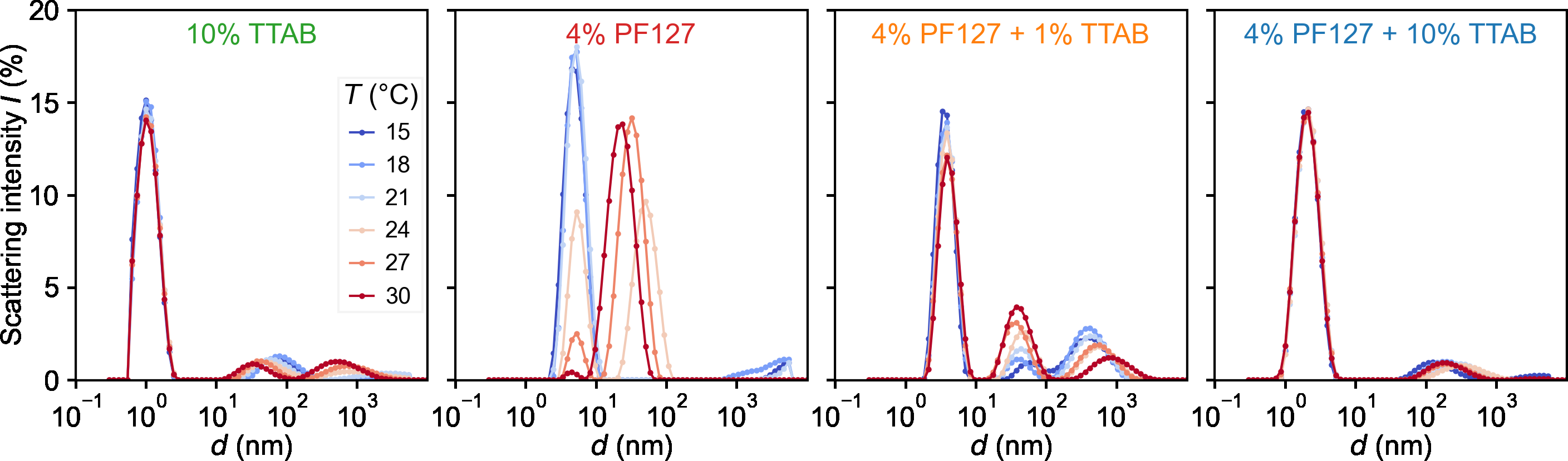}
\caption{Dynamic light scattering (DLS) measurements showing the size distributions of four aqueous solutions of PF127 and TTAB with increasing temperature, from backscattering (detector angle 173°). Scattering intensity $I$ vs.\ hydrodynamic diameter $d$. Averages over triplicate measurements.\label{SIfig:DLS}}
\end{figure*}

\begin{figure}
\centering
\includegraphics[width=1\linewidth]{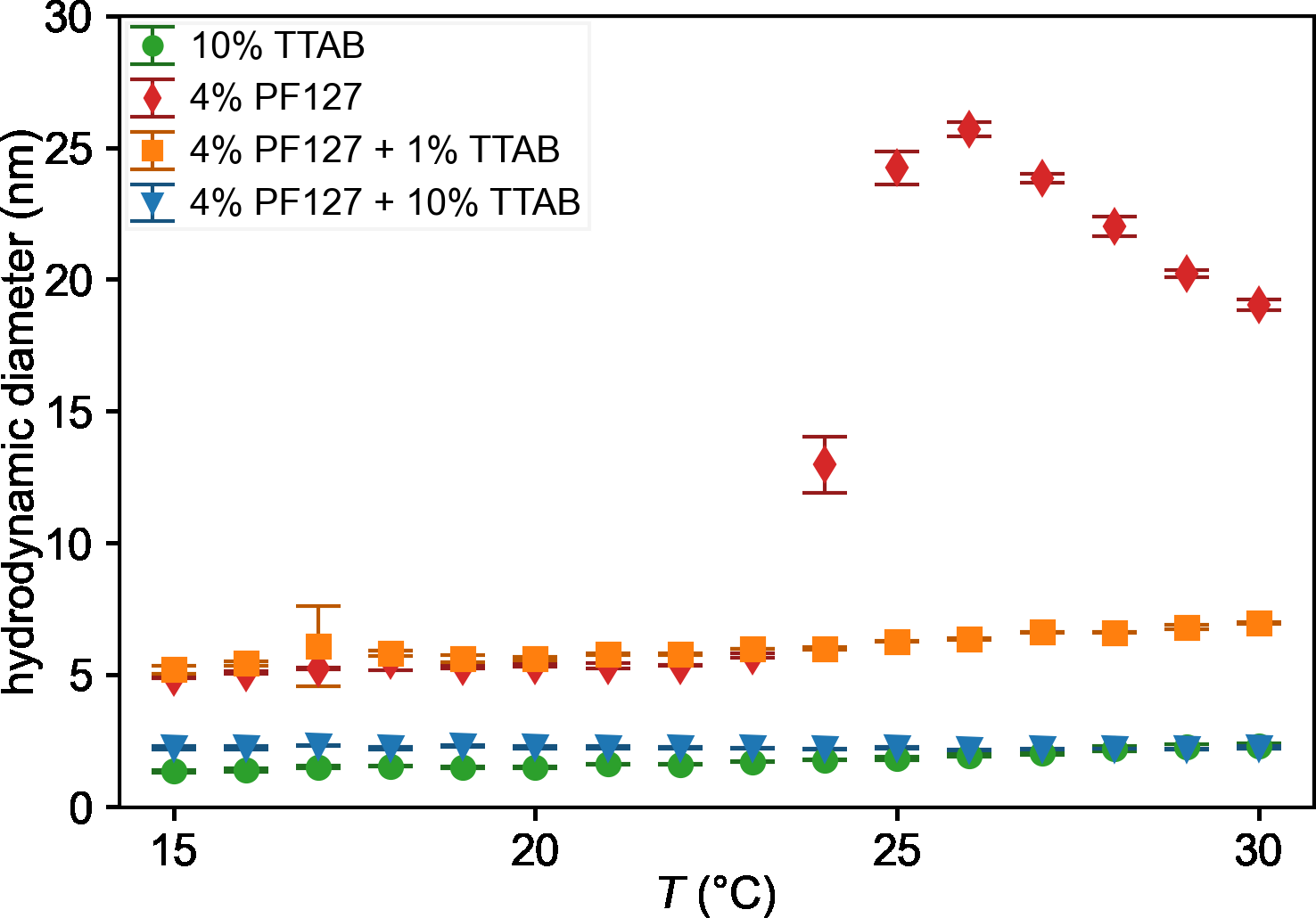}
\caption{Intensity weighted Z-average values of the hydrodynamic diameter for each measurement as shown in the DLS series in Fig.~\ref{SIfig:DLS}. The error bars use the standard deviation of the triplicate experimental runs.\label{SIfig:Zaverage}}
\end{figure}

We performed dynamic light scattering (DLS) measurements for the co-surfactant mixtures used in our experiments on a Malvern Zetasizer Ultra Red. We placed a \SI{1}{\ml} sample in a polystyrene cuvette and recorded the backscattering intensity at \SI{173}{\degree} between \SI{15}{\celsius} and  \SI{30}{\celsius}. At each temperature, the sample was allowed to equilibrate for \SI{120}{\s}, and all measurements were carried out in triplicate. We have plotted the size distributions for selected temperatures in  Fig.~\ref{SIfig:DLS} and summarized the temperature dependent behavior for all samples via the Z average hydrodynamic diameter, i.e. the  ``intensity weighted mean hydrodynamic size of the ensemble collection of particles'' (Malvern) in  Fig.~\ref{SIfig:Zaverage}. For \SI{4}{\wtpc} PF127, there is a significant increase in diameter above \SI{23}{\celsius}, corresponding to the formation of PF127 micelles (we note that this appears somewhat higher than the $\text{CMT}\approx\SI{21}{\celsius}$ literature value~\cite{bohorquez1999_study,alexandridis1994_micellization}). This increase is suppressed in the presence of TTAB, consistent with the formation of  PF127/TTAB complexes seen in literature~\cite{parmar2014_pluronic,nambam2012_effects,li2001_binding,hecht1994_interaction}. 
We note that, from the Z average, the hydrodynamic diameter of these complexes is somewhat smaller than that of a PF127 monomer, which is in line with existing studies~\cite{parmar2014_pluronic}.  Further, these complexes appear to be close in size to pure TTAB micelles in a size range $<\SI{5}{\nano\metre}$ near the lower DLS resolution limit, such that the two species probably cannot be resolved in  Fig.~\ref{SIfig:DLS}. 

\subsection{Differential scanning calorimetry}

\begin{figure}
    \centering
    \includegraphics[width=1\linewidth]{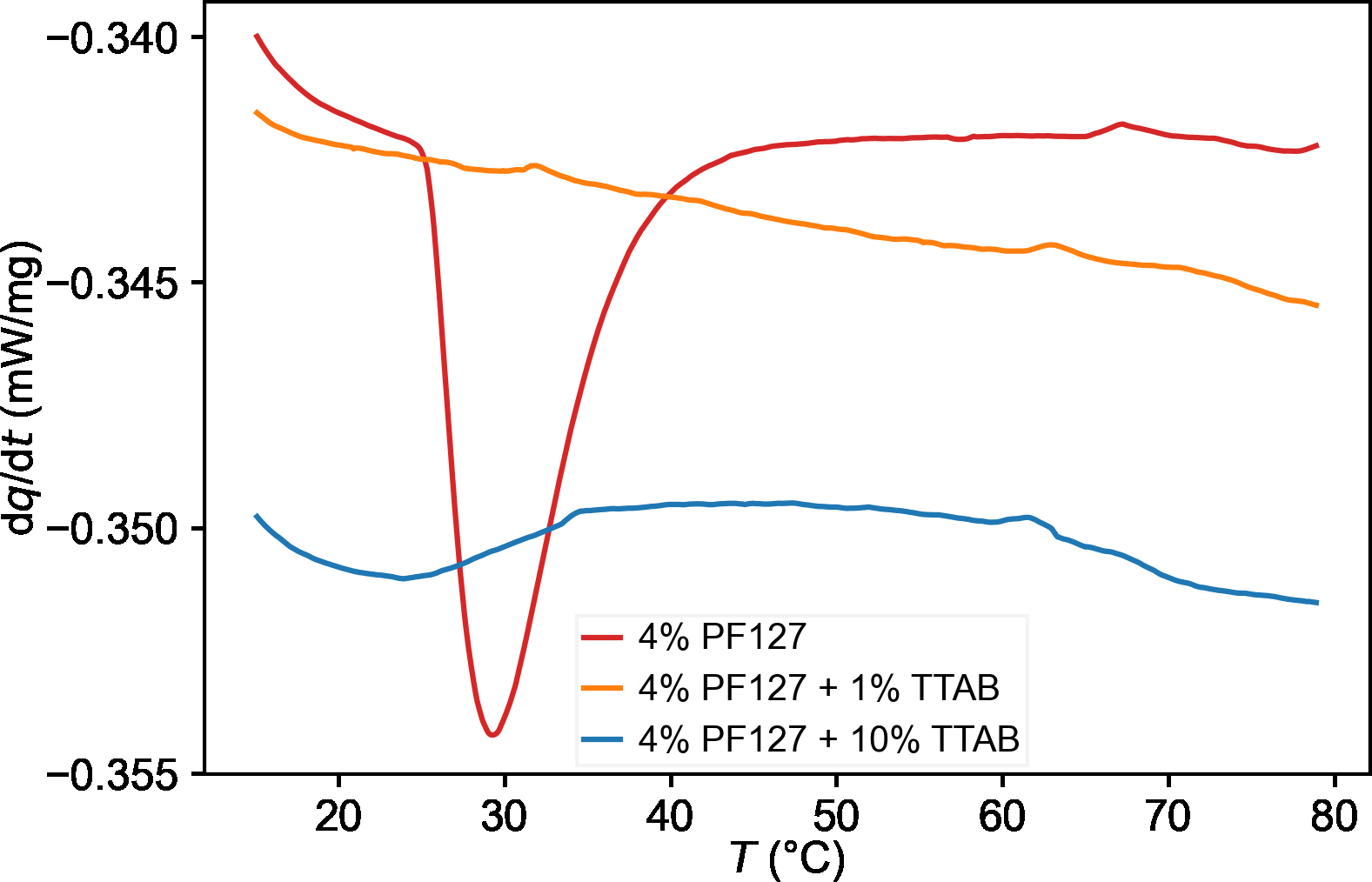}
    \caption{DSC heating curves of PF127 and TTAB solutions. Measurements were performed with a heat rate of \SI{5}{\kelvin/\min}  and normalized to sample mass.\label{SIfig:DSC}}
    
\end{figure}
We investigated the temperature dependent PF127 micellization  using differential scanning calorimetry (DSC). For the measurements, we prepared three sample types: (1) PF127 in \SI{4}{\wtpc} water, (2) PF127 \SI{4}{\wtpc} + TTAB \SI{1}{\wtpc} in water and (3) PF127 \SI{4}{\wtpc} + TTAB \SI{10}{\wtpc} in water. 
Solutions were directly transferred into DSC sample aluminum pans (volume \SI{100}{\ul}, Mettler-Toledo GmbH, Gießen, Germany). DSC pans were covered with aluminum lids. DSC measurements were performed on a DSC 823 instrument (Mettler-Toledo GmbH, Gießen, Germany). Heating-cooling cycles were recorded 
at a heating/cooling rate of \SI{5}{\kelvin/\min} between 0 and \SI{80}{\celsius}. The measurements were performed under a nitrogen atmosphere with a flow of \SI{30}{\ml/\min}. Heating curves were normalized to the sample mass. The endothermic dip in the curve for pure PF127 is consistent with the onset of micellization from the DLS results, and is similarly suppressed under the addition of TTAB.

\section{Cosurfactants on oil-water droplet interfaces}
We conducted further experiments to confirm, under our experimental conditions, (A) that TTAB is the dominant surfactant at the oil-water interface and (B) that the aqueous solubilization of CB15 is mainly TTAB mediated, as follows.

\subsection{TTAB interfacial coverage based on nematic anchoring}\label{SIsec:anchoring}
\begin{figure}
    \centering
    \includegraphics[width=1\linewidth]{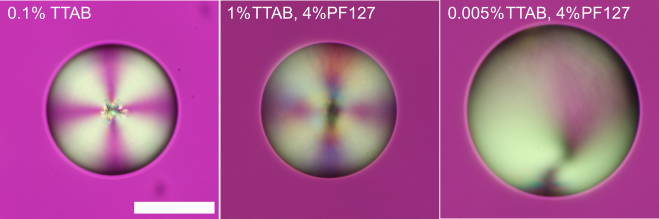}
    \caption{Polarized micrographs of 5CB droplets in cosurfactant mixtures: left, \SI{0.1}{\wtpc} TTAB, shows homeotropic; right, \SI{4}{\wtpc} PF127 and \SI{0.005}{\wtpc} TTAB, planar anchoring; middle, \SI{4}{\wtpc} PF127 and \SI{1}{\wtpc} TTAB, primarily homeotropic. Scale bar \SI{50}{\um}. \label{SIfig:anchoring}}
    
\end{figure}

In liquid crystal emulsions, the interfacial anchoring of the nematic director depends on the surfactant in use. Assuming comparable interfacial activity for both substances, we can infer the presence of TTAB at the oil-water interface from investigating the anchoring for CB15's nematic isomer 5CB (at room temperature) under polarized microscopy~\cite{lopez-leon2011_drops,shechter2020_direct}. We show this in three micrographs in Fig.\ \ref{SIfig:anchoring}: 5CB droplets (left) in $ \SI{0.1}{\wtpc}$ TTAB show a cross-shaped interference pattern and a central point defect, consistent with homeotropic (surface normal) anchoring as known for TTAB. On the right, we show a droplet in \SI{0.005}{\wtpc} TTAB and \SI{4}{\wtpc} PF127, where we observed a bipolar defect pattern, with two opposing defects at the interface (one visible in the micrograph). This pattern is typical for planar anchoring, and we associate it with the large excess of PF127. In the middle image, for \SI{1}{\wtpc} TTAB and \SI{4}{\wtpc} PF127 (middle), there is only a single point defect, indicating a transition to TTAB mediated homeotropic anchoring already far below the \SI{10}{\wtpc} TTAB used in the experiments on motile droplets. For these experiments, we could not use the surfactant conditions in the manuscript, $>\SI{5}{\wtpc}$ TTAB and \SI{4}{\wtpc} PF127, as in this case the droplets would have been motile at room temperature, distorting the birefringence patterns. However, adding more TTAB would crowd out PF127 at the droplet interface even more, such that we may assume TTAB as the dominant surfactant at the interface under experimental conditions featuring motility. 

\begin{figure}
    \centering
    \includegraphics[width=1\linewidth]{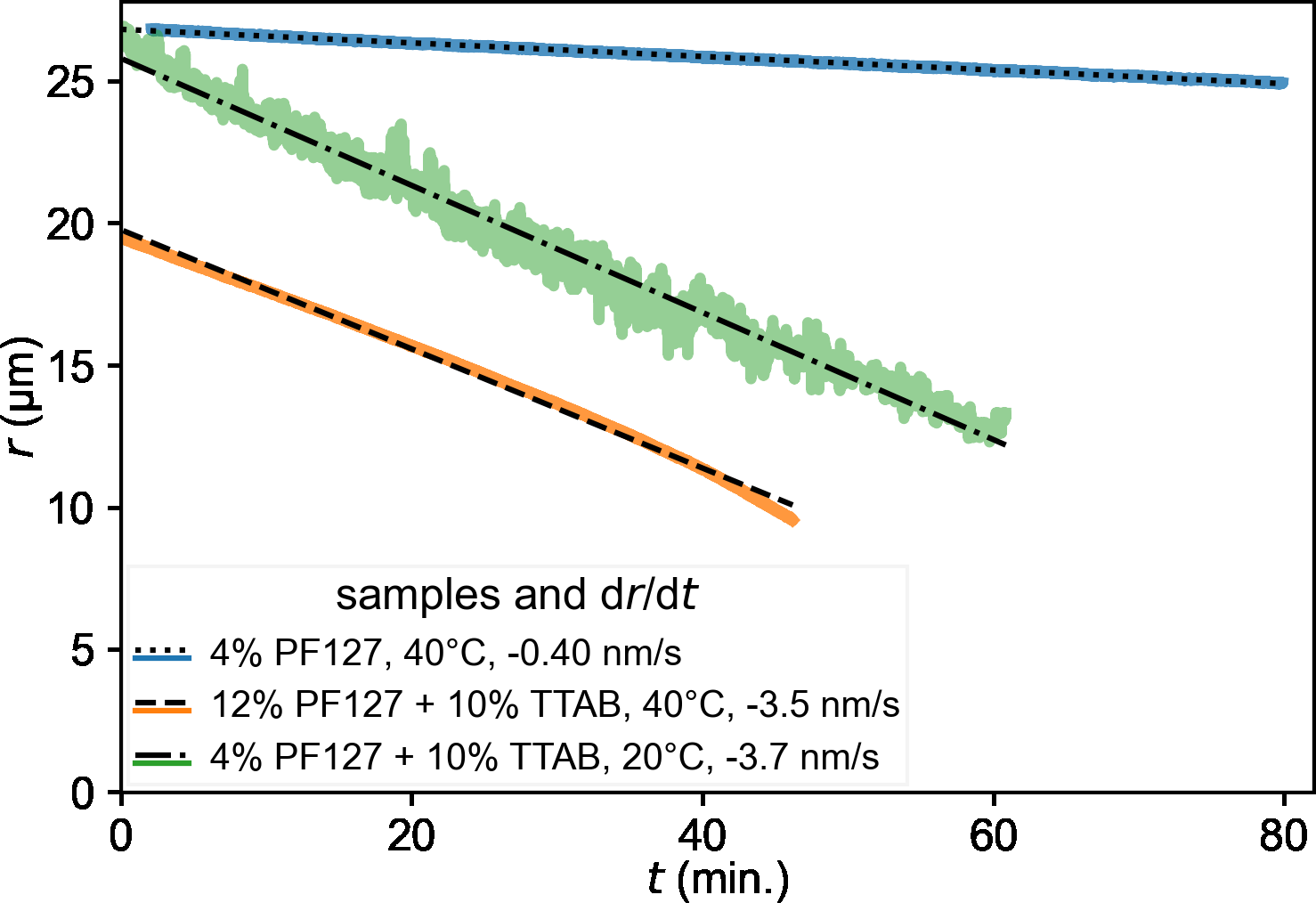}
    \caption{Decrease in CB15 droplet radius over time for non-motile droplets in \SI{4}{\wtpc} PF127 (blue) and \SI{12}{\wtpc} PF127 + \SI{10}{\wtpc} TTAB (orange) media at $T = \SI{40}{\celsius}$ (one droplet each at 40x magnification), and for motile droplets in \SI{10}{\wtpc} TTAB + \SI{4}{\wtpc} PF127 at \SI{20}{\celsius} (green, low magnification ensemble average). Dotted lines: linear regression fits to determine the shrinking rates, \SI{-0.4}{\nano\metre/\second}, \SI{-3.5}{\nano\metre/\second} and  \SI{-3.7}{\nano\metre/\second}, respectively.\label{SIfig:dissolution}}
    
\end{figure}

\subsection{Solubilization rate measurements}\label{SIsec:solubilisation}

We measured the shrinking rate of a CB15 droplet for two non-motile cases under bright field microscopy at 63x magnification at an elevated temperature of \SI{40}{\celsius} to promote dissolution by enhanced oil diffusion into the aqueous phase, using the concentrations \SI{10}{\wtpc} TTAB + \SI{12}{\wtpc} PF127 and \SI{4}{\wtpc} PF127 (note that pure PF127 at \SI{10}{\wtpc} would have formed a hydrogel~\cite{jalaal2016_rheology}).  Droplet radii and fitted shrinking rates are shown in \figref{SIfig:dissolution}. We compare these two cases to ensemble averaged data from a lower resolution experiment on motile droplets, in a medium with \SI{10}{\wtpc} TTAB + \SI{4}{\wtpc} PF127 at \SI{20}{\celsius}. Fig.~\ref{SIfig:dissolution} shows the following observations: CB15 does not significantly dissolve in pure PF127 solution (blue). If TTAB is strongly bound by PF127, i.e. at elevated temperatures and elevated PF127 concentration, CB15 droplets will solubilize, but not move (orange). At lower temperatures and with a proportionally higher concentration of TTAB, CB15 will both solubilize and move (green).

Given that (A) there appears to be a considerable fraction of TTAB at the interface (sec.~\ref{SIsec:anchoring}) and that (B) droplets neither significantly dissolve or move in pure PF127 (sec.~\ref{SIsec:solubilisation}), we conclude that the droplet motion is primarily driven by TTAB gradients in the oil-water interface, by a mechanism similar to the one driven by micellar solubilization found in pure TTAB media~
\cite{herminghaus2014_interfacial,peddireddy2012_solubilization,maass2016_swimming,izzet2020_tunable}.   We also note that the droplet speed is of a similar order of magnitude as the one in pure TTAB ($\approx\SIrange{10}{20}{\um\per\second}$, \figref{SIfig:ttab}), such that any additional adsorbed PF127 at the interface does not appear to impede the mechanism.

\section{Control experiments}
\subsection{Temperature dependent dynamics in pure TTAB solution}
\begin{figure}
    \centering
    \includegraphics[width=1\linewidth]{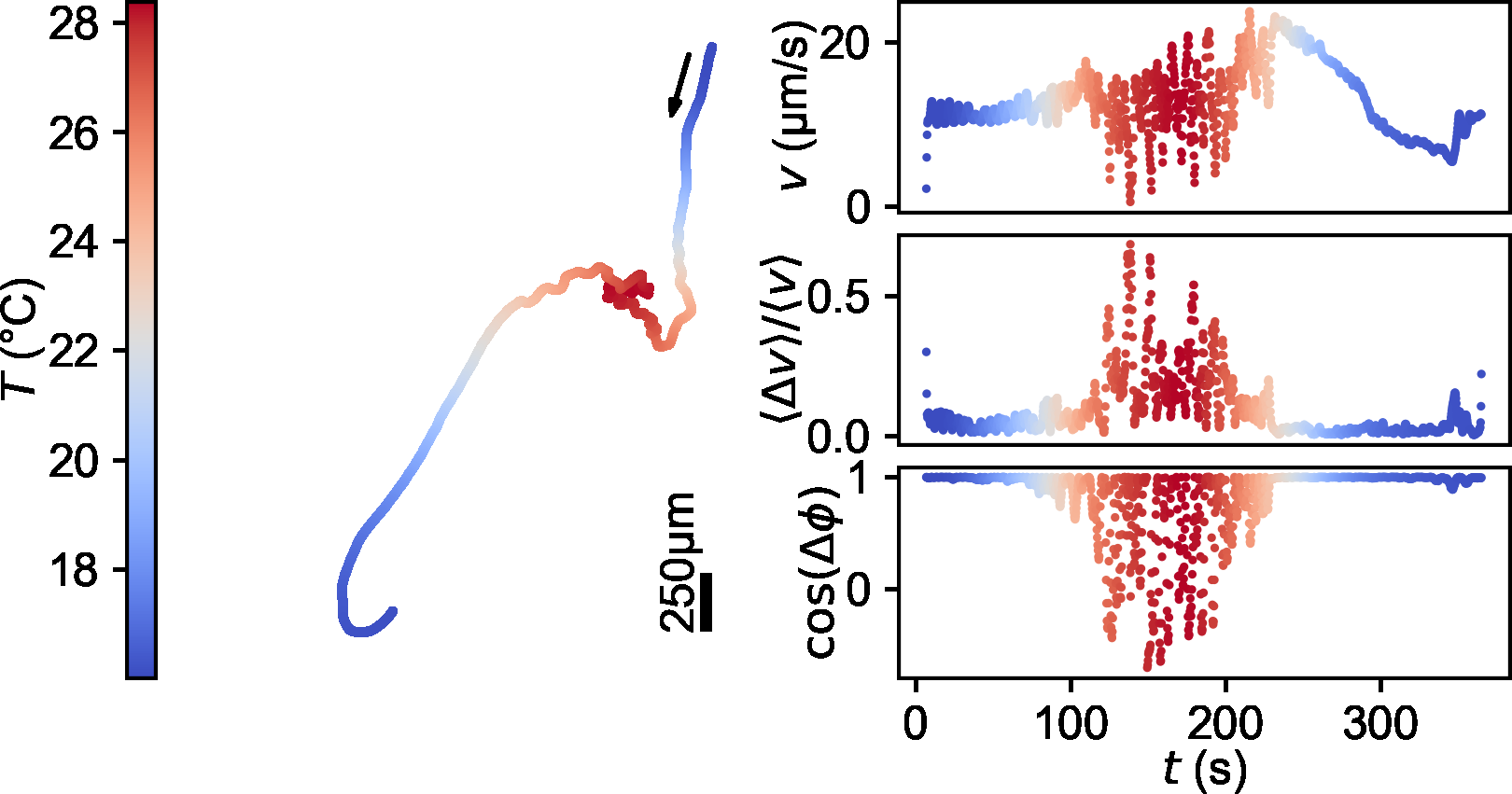}
    \caption{Temperature dependent droplet dynamics in a solution of pure TTAB at \SI{10}{\wtpc}.  (a) Droplet trajectory and (b) droplet
 speed $v$, and measures of fluctuations in speed, $\langle\Delta v\rangle/\langle v\rangle$, and orientation, $\cos(\Delta\phi)$, over time. With increasing temperature, the droplet motion accelerates and destabilizes. Scale bar \SI{250}{\um}.}
    \label{SIfig:ttab}
\end{figure}
In Fig.~\ref{SIfig:ttab}, we show a temperature-coded trajectory of a droplet studied under the same conditions as the experiments shown in the main manuscript, but in an aqueous solution of TTAB only at \SI{10}{\wtpc}. Here, the motion destabilizes with increasing temperature. We illustrate this by temperature-coded plots of three quantities: the speed (top), which increases, but also fluctuates strongly for high temperatures. Further, two simple correlation estimates, both taken over a running time window of $\tau=\SI{2}{s}$: the standard deviation over average speed, $\langle \Delta v\rangle/\langle v \rangle$ as a measure of unsteadiness in speed, and the cosine of the angle between $\vec{v}(t)$ and $\vec{v}(t+\tau)$ via their inner product as a measure of rotational fluctuation. Both indicate strong decorrelation at elevated temperatures. We note that this behavior is strongly different from the one in PF127/TTAB mixtures, where speed and unsteadiness decrease with increasing temperature.

\subsection{Temperature dependent dynamics in pure PF127 solution}
\begin{figure}
    \centering
    \includegraphics[width=\linewidth]{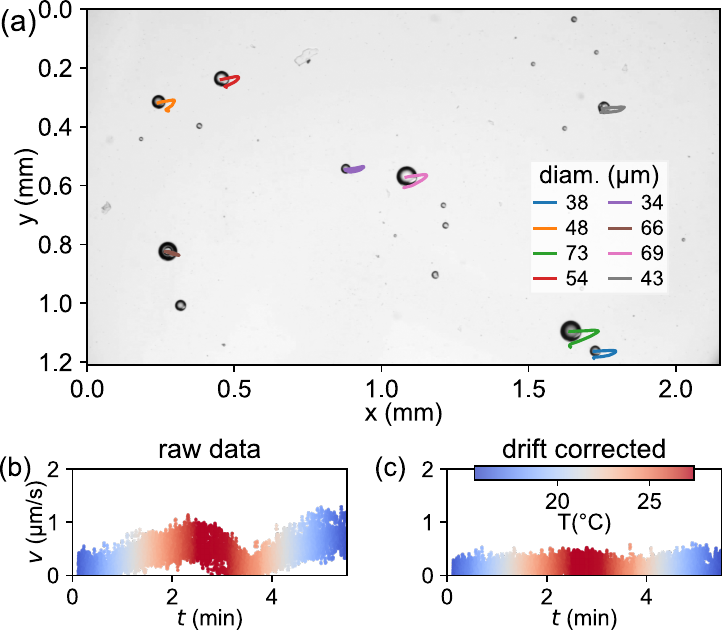}
    \caption{CB15 droplets are immotile during a temperature ramp between 17 and \SI{28}{\celsius} in a \SI{4}{\wtpc} PF127 solution. (a) snapshot with trajectories and droplet diameters. (b, c) raw and drift corrected speeds (see Sec.~\ref{SIsec:drift}).\label{SIfig:pf127dyn}}
    
\end{figure}

To further support our hypothesis that the temperature dependent arrest documented in the main manuscript requires interactions between the two cosurfactants, and not found for the surfactants individually, we show in Fig.~\ref{SIfig:pf127dyn} that there is no active motility for droplets of any relevant size in a \SI{4}{\wtpc} aqueous solution of PF127 during a temperature ramp between 17 and \SI{28}{\celsius}. 

\section{Reduced hysteresis during an incomplete heating ramp}
\begin{figure*}[t]
\centering
\includegraphics[width=\linewidth]{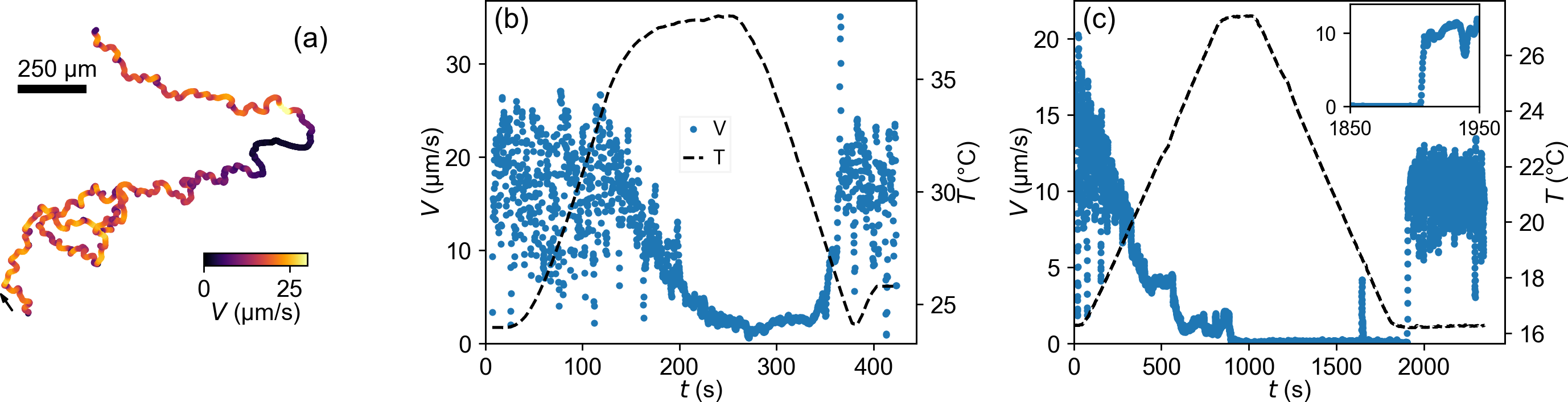}
\caption{Reduced hysteresis during an incomplete heating/cooling cycle. (a) speed coded trajectory (b) Speed and temperature versus time (c) The same quantities for a temperature ramp heating to full droplet arrest, using the data from Fig.\ 1a.}
\label{SIfig:incomplete}
\end{figure*}
We have found a reduction in hysteresis to about half the temperature difference when the droplet does not fully stop (i.e., during an incomplete heating ramp,  where it slows down to $<\SI{1}{\um/\s}$). 
\figref{SIfig:incomplete} shows an example for \SI{13}{\wtpc},  with the speed coded trajectory in (a), speed and temperature over time in (b) and in (c), for comparison, the same quantities for the trajectory plotted in Fig.\ 2a in the MS. Compared to the $T_\text{start}, T_\text{stop}$ values for $c_\text{TTAB}=\SI{13}{\wtpc}$ in Fig.\ 2c, the hysteresis is reduced almost by half, the droplet doesn't stop entirely and the re-onset of motion is far more gradual than one starting from full arrest (see  zoomed in inset in (c)). 
\section{Viscosity measurements}

\begin{figure*}
\centering
\includegraphics[width=\linewidth]{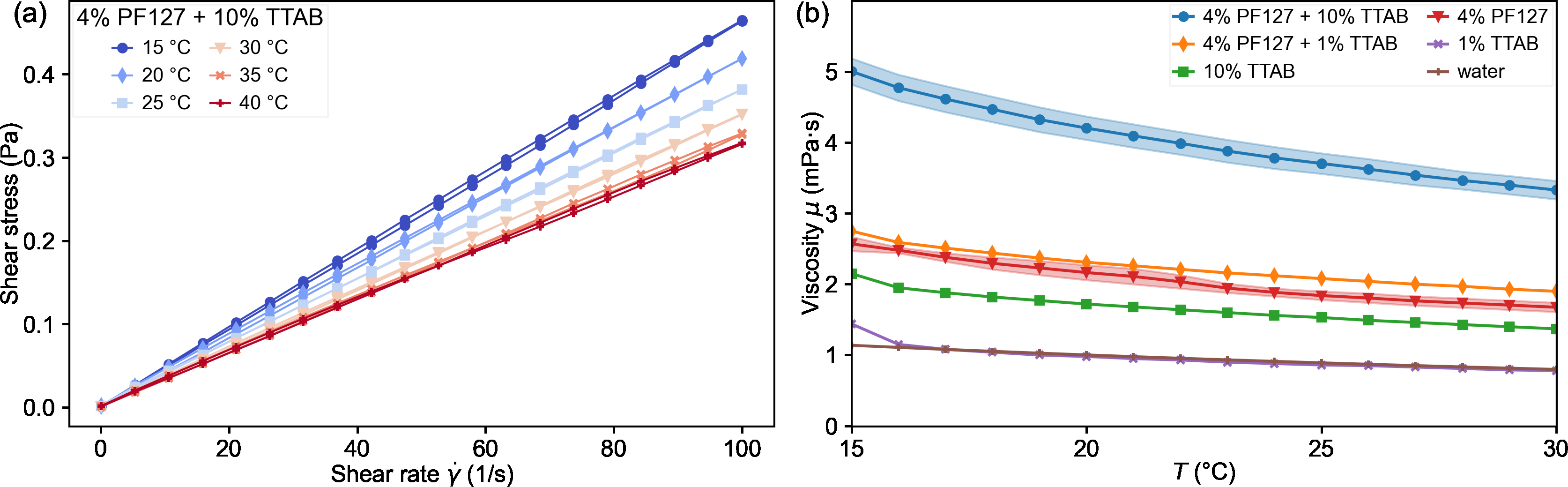}
\caption{(a) Shear stress versus shear rates at different temperatures for \SI{4}{\wtpc} PF127 + \SI{10}{\wtpc} TTAB. (b) Viscosity versus temperature for TTAB and PF127.}
\label{SIfig:viscosity}
\end{figure*}
We measured the viscosity of our swimming media on an Anton-Paar MCR 502 rheometer using a cone-plate geometry with a gap width of \SI{0.1}{\mm}. Measurements were carried out at temperatures between \SI{15}{\celsius} and \SI{40}{\celsius} and shear rates between \SI{0.1}{\s^{-1}}  and \SI{100}{\s^{-1}} (see \figref{SIfig:viscosity}). We observe Newtonian rheology and an only weakly temperature dependent viscosity. Thus, while aqueous solutions of PF127 are known to gel at high temperatures, essentially forming a network of micelles, we are still below this non-Newtonian regime at the concentration and temperature range in use.

\section{Flow and chemical field measurements}
The internal flow field was determined by adding $\rm{1\:\mu m}$ diameter tracer Silica particles (Cospheric SiO2MS-1.8) to the oil phase and analysing high magnification videomicroscopy data by particle image velocimetry (PIV). We did not measure external flow fields, since adsorbed PF127 on colloidal tracer particles \cite{kim2005_swellingbased} may cause them to aggregate and impedes accurate PIV measurements.

We recorded videomicroscopy data under a 40x objective using a 4MP camera (FLIR Grasshopper 3, GS3-U3-41C6M-C) at 40 fps at different set temperatures. 
\par
To study the sudden onset of motion from an inactive state as the system was cooled, images were recorded at a higher frame rate of 80 fps. Droplet speed over time was calculated from the recorded trajectories.
\par
To visualize the oil-filled micelle chemical trail behind the droplet, we doped the oil phase with Nile Red (Sigma-Aldrich) dye. We performed fluorescent microscopy on an Olympus IX81 microscope with a filter cube (excitation filter ET560/40x, beam splitter 585 LP and emissions filter ET630/75m, all Chroma Technology). Images were captured via a 4x objective  using a 4 MP CMOS camera (FLIR Grasshopper 3, GS3-U3-41C6M-C) at 4 fps.

\section{Digital image processing and data analysis}
\begin{figure}
    \includegraphics[width=\columnwidth]{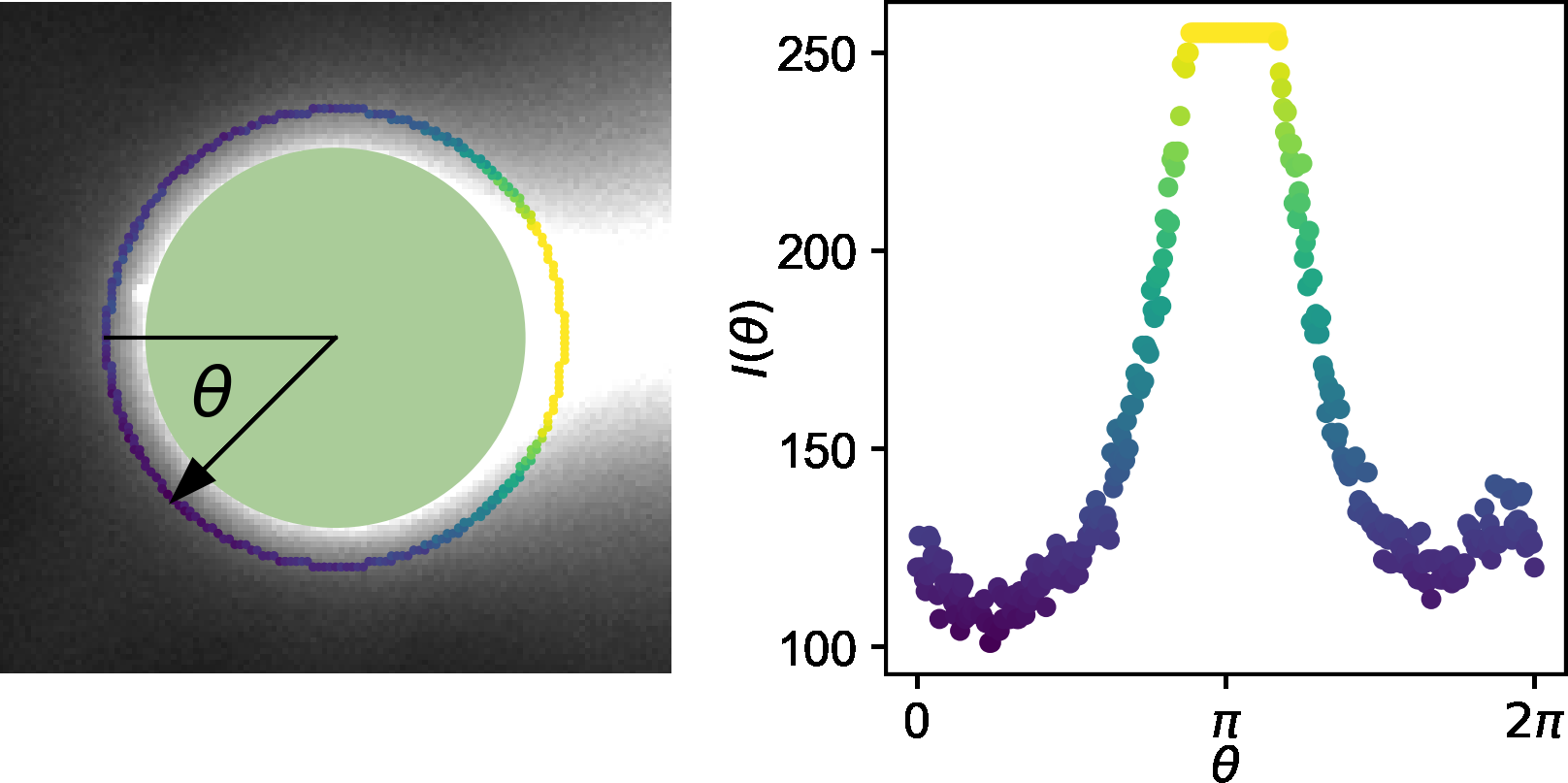}
    \caption{Protocol to extract fluorescence kymographs from microvideo data, by the example of snapshot III from Fig. 3}
    \label{SIfig:kymo}
\end{figure}

We extracted droplet coordinates from bright field microscopy using a sequence of background correction, binarization, blob
detection by contour analysis, and minimum enclosing
circle fits, and determined trajectories via a nearest-neighbor algorithm using in-house Python scripts building on numpy and opencv. For the strongly overexposed fluorescence data the droplet centroid was calculated via a distance transform algorithm on the fluorescence intensity. The polar intensity map in Fig. 3(b) was derived by taking the intensity in an annular region around the droplet at a distance of 1.2 droplet radii from the centroid~\cite{hokmabad2021_emergence,hokmabad2022_chemotactic,ramesh2023_interfacial}, as sketched in supporting figure~\ref{SIfig:kymo}.

\begin{figure}
    \centering
    \includegraphics[width=\linewidth]{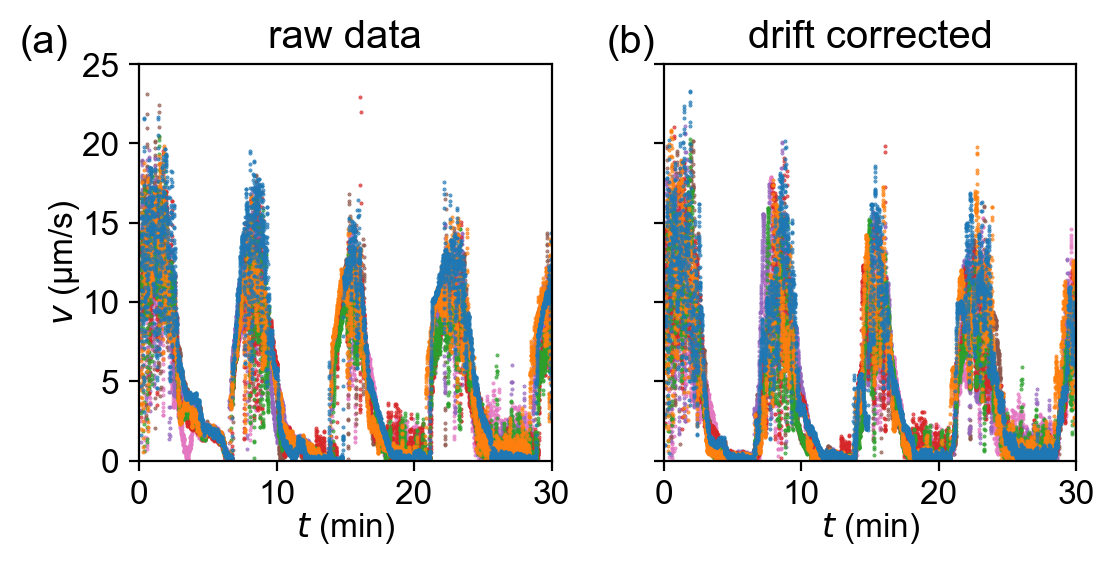}
    \caption{Drift analysis: Comparison of instantaneous speeds in the ensemble of droplets shown in Movie S3 (see also Fig. 2 e-g), once taken from the raw coordinates (a) and once with the ensemble average position subtracted. During cooling periods, the speed in (b) reverts to zero, indicating that the residual motion in (a) is uniform drift.\label{SIfig:driftanalysis}}
    
\end{figure}

Using the time-dependent droplet trajectory and temperature data, we estimated transition temperatures between dynamic states. The error bars for temperature in the regime map (Fig. 2f) represent the maximum variation within three different runs on samples containing on the order of 5--10 droplets each. It should be noted that there is some uncertainty associated with estimating $T_\text{straight}$ and $T_\text{stop}$ from low magnification droplet trajectory data. As seen in high magnification data in Movie S5, a droplet could appear to be stationary even though it shows internal activity through the motion of tracer particles. Moreover, the transition between a smooth reorientation and straight motion is somewhat gradual.

We performed PIV analyses  using the Matlab PIVlab module \cite{thielicke2014_pivlab} with a multi-pass interrogation window of 64 x 64 pixel and 32 x 32 pixel with 50\% overlap. The spatial resolution of the PIV output was \SI{4.3}{\um/px}.

\section{Drift analysis}\label{SIsec:drift}
Some experiments feature residual motion in the lab frame on the order of $\approx\SIrange{1}{2}{\um/s}$ during high temperature states that we identified as arrested and inactive. We believe this is due to drift in the swimming medium, possibly due to the response of the microfluidic cell to the thermal ramp, for the following reasons: (a) In Movie S5, there is no internal flow detectable from the colloidal tracers at high temperature. (b) The fluorescent trails in Movie S4 shift together with the droplets. (c) The residual speed in Fig. 2f is reduced to almost zero if the droplet positions are shifted by the ensemble average, i.e. removing collective translation (Fig.~\ref{SIfig:driftanalysis}). (d) The inactive droplets in Fig.~\ref{SIfig:pf127dyn} collectively drift to the right on heating and to the left on cooling. Since a correction for the ensemble average similar to Fig.~\ref{SIfig:driftanalysis} and \ref{SIfig:pf127dyn} would add considerable noise during periods of active undirected motility, it is not applied in Fig. 2f. The fluorescence kymographs in Fig.~3 and the PIV data in Fig.~4 take the droplet center as the coordinate origin, and are therefore not susceptible to drift.

\section{Additional data: trail interactions}
\begin{figure}
    \centering
    \includegraphics[width=\linewidth]{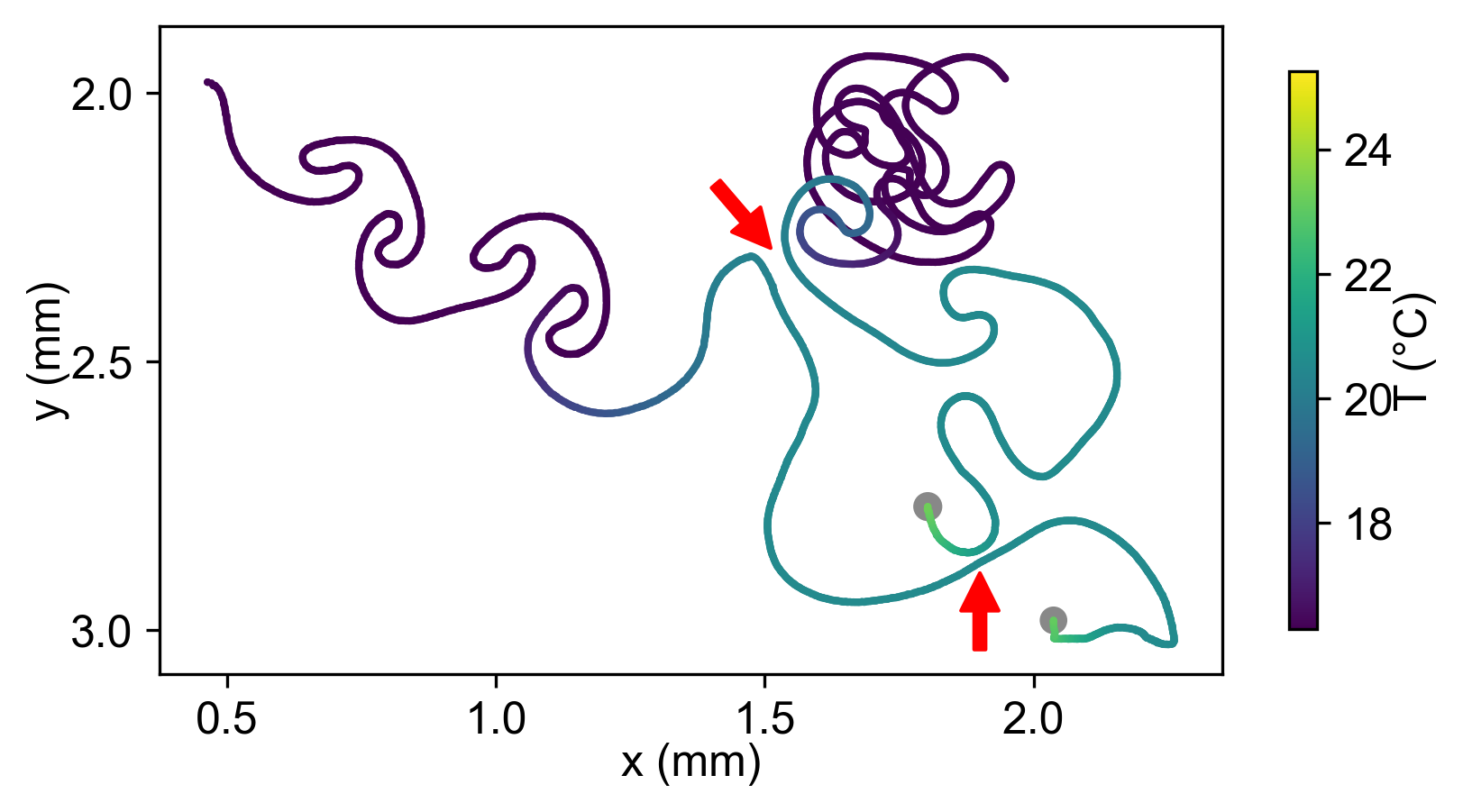}
    \caption{Collective effects: Similar to droplets swimming in pure TTAB solution, droplets are repelled from each other's trajectory in TTAB/PF127 solutions of 10 and \SI{4}{\wtpc}, respectively. Two representative trajectories, color coded by the ambient temperature, with observed collision events marked by arrows. Grey circles represent the measured droplet sizes.\label{SIfig:collective}}    
\end{figure}

Autophoretic droplets are known to show complex collective behavior depending on number densities and state of confinement. We previously observed transient trapping in pure TTAB solutions~\cite{hokmabad2022_chemotactic} at constant $\Pen\approx4$, where active droplets were trapped in self generated cages of chemorepulsive trails, such that the collective is creating its own `chemical landscape'.  Under addition of a cosurfactant, the droplets will still modify their chemical environment in a similar manner. Fig.~\ref{SIfig:collective} shows exploratory data taken from an experiment with an increased number density around \SI{4}{\per\square\mm}, at \SI{4}{\wtpc} PF127 and \SI{10}{\wtpc} TTAB, with the sample temperature increasing from 17 to \SI{25}{\celsius}.
Red arrows identify two representative trail collisions, where droplets are repelled by the persistent filled micelle signature of other droplets.

\section{Numerics}\label{SIsec:numerics}
We simulate a diffusiophoretic particle of unit radius initially located at the center of a domain of size $L_x=10$, $L_y=100$, $L_z=2.2$ with $201 \times 2001 \times 45$ grids at subcritical \Pen. The particle is propelled by diffusiophoresis, a type of microswimmer similar to active droplet, as both move in response to the forces at the surfaces, which depend on the local chemical concentration field (see also~\cite{morozov2019_nonlinear}). The chemical reaction takes place at the particle surface and whenever there is a chemical concentration difference along the surface, a flow is generated within the interaction layer near the solid surface, with thickness $\lambda$ of nanometers, which propels the particle forward. 

We use the same non-dimensional governing equations and boundary conditions at the particle interface as those in previous studies~\cite{michelin2013_spontaneous,chen2021_instabilities} (here, solving for the full 3D problem as opposed to the axisymmetric analytical approach).  
The governing equations are given as
\begin{equation}\label{eq_num1}
\frac{\partial c}{\partial t}+\boldsymbol{u}\cdot\nabla c= \frac{1}{Pe}\nabla^2 c,
\end{equation}

\begin{equation}
\gdef\thesubequation{\theequation \mbox{\textit{a}},\textit{b}}
\frac{\partial \boldsymbol{u}}{\partial t}+(\boldsymbol{u}\cdot\nabla)\boldsymbol{u}=-\nabla p+\frac{Sc}{Pe}\nabla^2 \boldsymbol{u},\quad
\boldsymbol{\nabla} \cdot \boldsymbol{u}=0,
\end{equation}
where $c$ is the concentration, $u$ the velocity. $Pe$ is the P\'eclet number, which is the ratio of advection to diffusion and $Sc$ is the Schmidt number, which is the ratio between the momentum and mass diffusivities:
\begin{equation} \label{eq_num3}
\gdef\thesubequation{\theequation \mbox{\textit{a}},\textit{b}}
Pe=\frac{M\alpha L}{D^2}, \quad
Sc=\frac{\nu}{D}.
\end{equation}
where $M$ is the mobility, $M\sim \pm k_BT\lambda^2/(\rho\nu)$ with $k_B$ the Boltzmann constant and $T$ the temperature, $\rho$ the density, $\nu$ the viscosity, and $D$ is the diffusion coefficient. 

The boundary conditions are given as:
\begin{equation} \label{eq_num2}
\gdef\thesubequation{\theequation \mbox{\textit{a}},\textit{b}}
\partial_n c=-1 \quad
u_s=\nabla_s c,
\end{equation}
where  $\partial_n c$ represents the concentration gradient at the direction normal to particle surface, $u_s$ the slip velocity and $\nabla_s$ is the tangential gradient. The top and bottom boundaries (at $z$ direction) of the domain are set as solid walls, and all other domain boundaries (at $x$ and $y$ directions) are set as periodic.

We used a central second-order finite difference scheme to spatially discretize the governing equations, with homogeneous staggered grids used in both the horizontal and vertical directions. The equations are integrated by a fractional-step method, with non-linear terms computed explicitly using a low-storage third-order Runge-Kutta scheme and the viscous terms computed implicitly by a Crank-Nicolson scheme~\cite{verzicco1996_finitedifference}.
For the particle boundary, we make use of the moving least squares (MLS) based immersed boundary (IB) method, where the particle interface is represented by a triangulated Lagrangian mesh~\cite{spandan2017_parallel}.
For the detailed numerical methods and validation, we refer to~\cite{chen2021_instabilities}.

In simulations, we observed the onset of symmetry breaking for 
 $Pe\gtrapprox 6$; the velocity profile in the inset of Fig.\ 4a is taken from a simulation with $\Pen=8$. In Fig.\ 4(a), the numerical timescale has been re-dimensionalized using the characteristic time and velocity scales $t^*$ and $V^*$ following~\cite{michelin2013_spontaneous,picella2022_confined}
 \begin{align*}
     \Pen &= \frac{R\cdot V^*}{D} & t^*&=\frac{R}{V^*}=\frac{R^2}{\Pen\cdot D}
 \end{align*}
 with $D\approx\SI{5e-10}{m^2/s}$ for the Stokes-Einstein diffusivity of a surfactant monomer~\cite{hokmabad2021_emergence}, $R=\SI{25}{\um}$ and $\Pen=8$.
\newpage

\section{Supplementary Movie captions}
\noindent\begin{minipage}{\columnwidth}
\noindent\includegraphics[width=\columnwidth]{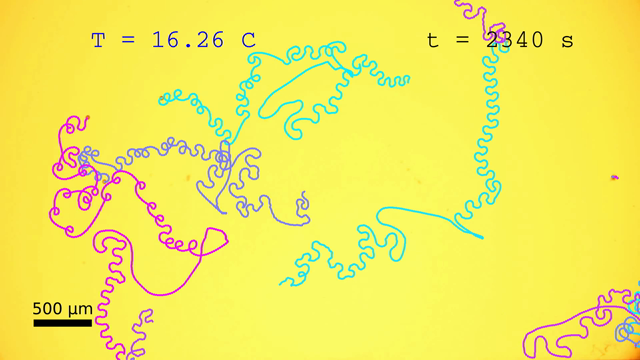}
\textbf{Movie S1.} Temperature dependent droplet dynamics at \SI{4}{\wtpc} PF127 + \SI{10}{\wtpc} TTAB, showing a reversible transition from meandering to straight swimming to arrest during a heating and subsequent cooling ramp with a set rate of \SI{1}{\kelvin/\minute}.
\end{minipage}

\noindent\begin{minipage}{\columnwidth}
\noindent\includegraphics[width=\columnwidth]{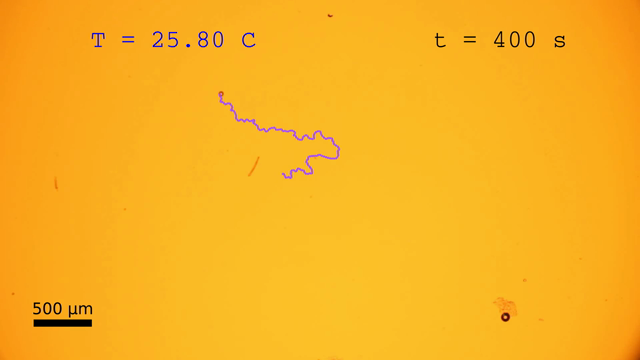}
\textbf{Movie S2.}  Temperature dependent droplet dynamics at \SI{4}{\wtpc} PF127 + \SI{13}{\wtpc} TTAB, showing a reversible transition from unsteady to straight swimming
to arrest during a heating and subsequent cooling ramp with a set rate of \SI{5}{\kelvin/\minute}\end{minipage}

\noindent\begin{minipage}{\columnwidth}
\noindent\includegraphics[width=\columnwidth]{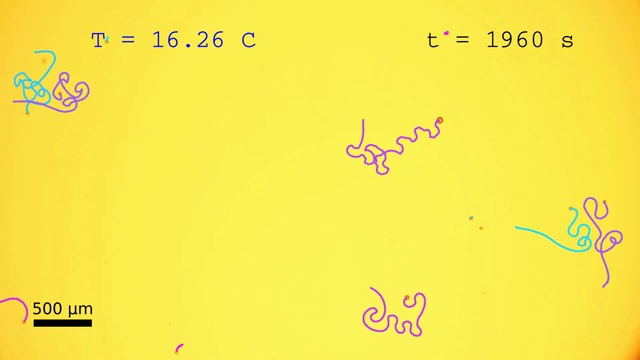}
\textbf{Movie S3.}  Droplet trajectories at \SI{4}{\wtpc} PF127 + \SI{10}{\wtpc} TTAB, during multiple heating and subsequent cooling ramps at a set rate of \SI{10}{\kelvin/\minute}.
\end{minipage}

\noindent\begin{minipage}{\columnwidth}
\noindent\includegraphics[width=\columnwidth]{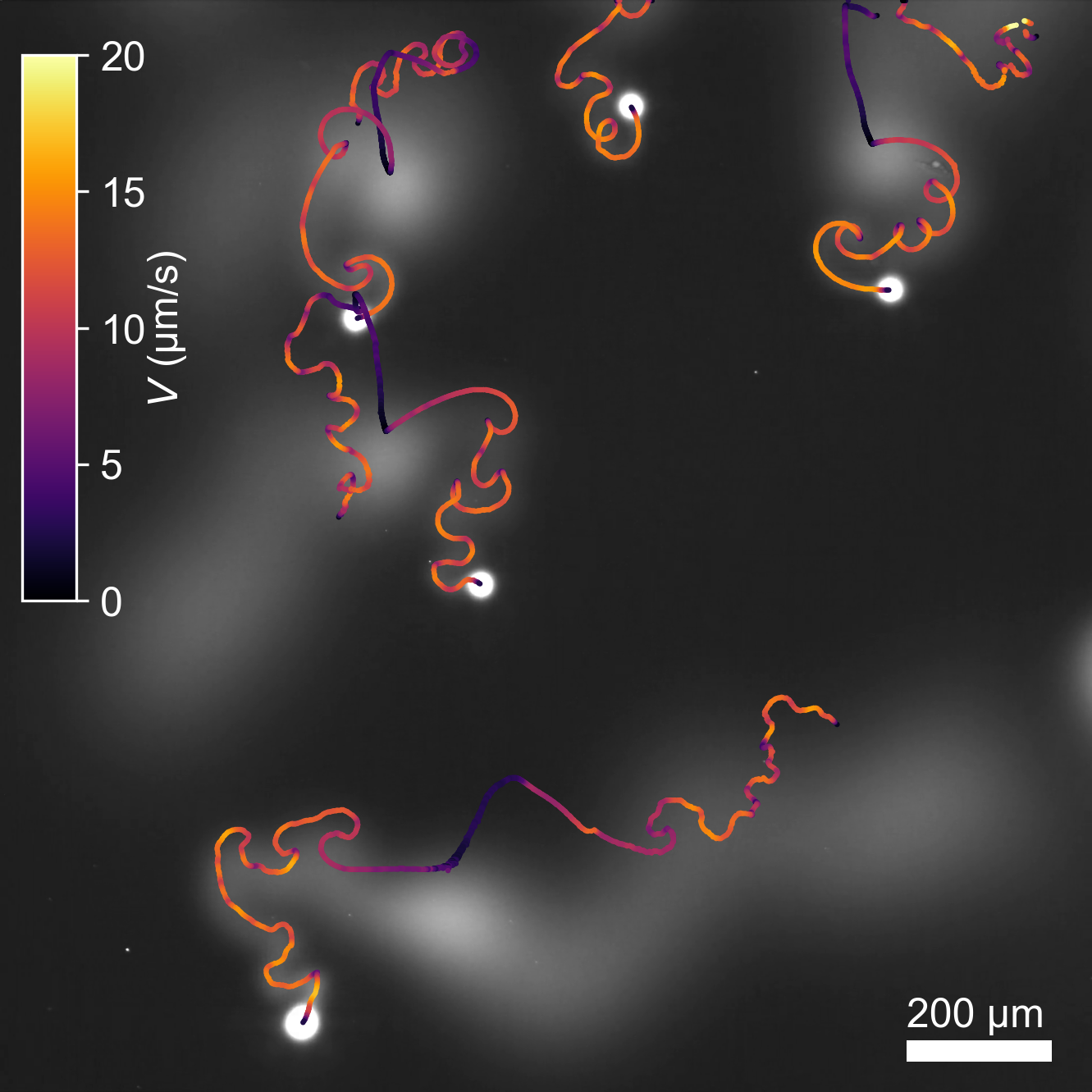}
\textbf{Movie S4.}  Fluorescent microscopy visualizing droplet chemical trails upon heating and subsequent cooling ramp with a set rate of \SI{5}{\kelvin/\minute}.
\end{minipage}

\noindent\begin{minipage}{\columnwidth}
\noindent\includegraphics[width=\columnwidth]{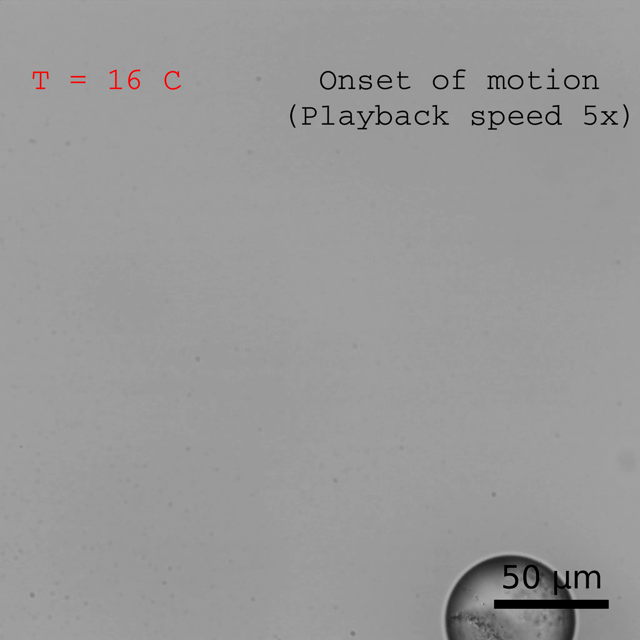}
\textbf{Movie S5.}Internal flow visualization with increasing temperature, starting at a mixed dipolar/quadrupolar mode (meandering), over a purely dipolar mode (straight) that recedes to the anterior (slowdown) and onset of motion during cooling. Excerpts recorded during one continuous experiment, playback (at 40 fps) corresponding to real-time during heating, sped up by 5x during cooling.
\end{minipage}

\noindent\begin{minipage}{\columnwidth}
\noindent\includegraphics[width=\columnwidth]{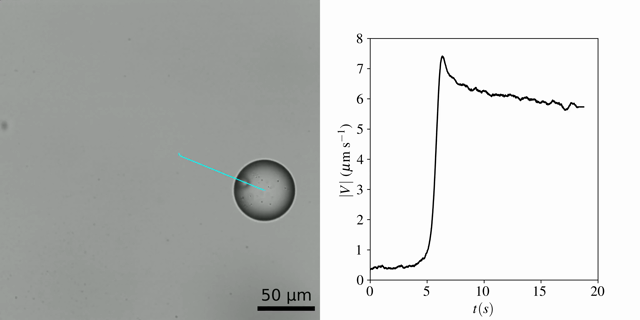}
\textbf{Movie S6.} Onset of active motion during re-cooling in a droplet containing tracer colloids, with a superimposed trajectory and the recorded droplet speed $V$. Experimental duration 18 seconds.
\end{minipage}

\clearpage

\end{document}